\documentclass[11pt]{article}
\usepackage{graphicx}
\usepackage{amssymb}
\usepackage{amsmath}
\usepackage{subfigure}
\usepackage{booktabs}
\usepackage{multirow}
\usepackage{color}
\usepackage[title]{appendix}
\usepackage{lineno}
\usepackage{multirow}
\usepackage[tableposition=top]{caption}

\newcommand{\pp}      {$K^+\rightarrow\pi^+\pi^0$~}

\newcommand{\pppc}    {$K^+\rightarrow\pi^+\pi^+\pi^-$~}
\newcommand{\kethree} {$K^+\rightarrow\pi^0e^+\nu$~}

\newcommand{\mn}      {$K^+\rightarrow\mu^+\nu$~}

\newcommand{\pnnc}    {$K^+\rightarrow\pi^+\nu\bar{\nu}$~}
\newcommand{\pnnz}    {$K_L\rightarrow\pi^0\nu\bar{\nu}$~}

\newcommand{\pic}     {$\pi^+$~}

\newcommand{\kp}      {$K^+$~}

\def\mmis{m^2_\mathrm{miss}}

\begin{document}
\pagenumbering{arabic}
\centerline{\LARGE EUROPEAN ORGANIZATION FOR NUCLEAR RESEARCH}

\begin{flushright}
CERN-EP-2020-132\\
July 14, 2020 \\
\end{flushright}
\vspace{15mm}

\begin{center}
\Large{\bf An investigation of  the  very rare \boldmath \pnnc  decay \\
\vspace{5mm}
}
The NA62 Collaboration
\end{center}
\vspace{10mm}

\begin{abstract}
The NA62 experiment reports an investigation of the \pnnc mode from a sample of  $K^+$ decays collected in 2017 at the CERN SPS. 
The experiment has achieved a single event sensitivity of $(0.389\pm 0.024)\times10^{-10}$, corresponding to 2.2 events assuming the Standard Model branching ratio of $(8.4\pm1.0)\times10^{-11}$.  
Two signal candidates are observed with an expected background of 1.5 events. 
Combined with the result of a similar analysis conducted by NA62 on a smaller data set recorded in 2016,  the collaboration now reports an upper limit of $1.78\times10^{-10}$ for the \pnnc branching ratio at 90\%\,CL. 
This, together with the corresponding 68\%\,CL measurement of $(0.48^{+0.72 }_{-0.48})\times10^{-10}$, are currently the most precise results worldwide, and are able to constrain some New Physics models that predict large enhancements still allowed by previous measurements.
\end{abstract}

\vspace{20mm}
\begin{center}
{\it Accepted for publication in JHEP}
\end{center}
\clearpage
\begin{center}
{\Large The NA62 Collaboration$\,$\renewcommand{\thefootnote}{\fnsymbol{footnote}}%
\footnotemark[1]\renewcommand{\thefootnote}{\arabic{footnote}}}\\
\end{center}
 \vspace{3mm}
\begin{raggedright}
\noindent

{\bf Universit\'e Catholique de Louvain, Louvain-La-Neuve, Belgium}\\
 E.~Cortina Gil,
 A.~Kleimenova,
 E.~Minucci$\,$\footnotemark[1]$^,\,$\footnotemark[2],
 S.~Padolski$\,$\footnotemark[3],
 P.~Petrov,
 A.~Shaikhiev$\,$\footnotemark[4],
 R.~Volpe$\,$\footnotemark[5]\\[2mm]

{\bf TRIUMF, Vancouver, British Columbia, Canada}\\
 T.~Numao,
 B.~Velghe\\[2mm]

{\bf University of British Columbia, Vancouver, British Columbia, Canada}\\
 D.~Bryman$\,$\footnotemark[6],
 J.~Fu$\,$\footnotemark[7]\\[2mm]

{\bf Charles University, Prague, Czech Republic}\\
 T.~Husek$\,$\footnotemark[8],
 J.~Jerhot$\,$\footnotemark[9],
 K.~Kampf,
 M.~Zamkovsky\\[2mm]

{\bf Institut f\"ur Physik and PRISMA Cluster of excellence, Universit\"at Mainz, Mainz, Germany}\\
 R.~Aliberti$\,$\footnotemark[10],
 G.~Khoriauli$\,$\footnotemark[11],
 J.~Kunze,
 D.~Lomidze$\,$\footnotemark[12],
 R.~Marchevski$\,$\renewcommand{\thefootnote}{\fnsymbol{footnote}}%
\footnotemark[1]$^,$\renewcommand{\thefootnote}{\arabic{footnote}}%
 \footnotemark[13],
 L.~Peruzzo,
 M.~Vormstein,
 R.~Wanke\\[2mm]

{\bf Dipartimento di Fisica e Scienze della Terra dell'Universit\`a e INFN, Sezione di Ferrara, Ferrara, Italy}\\
 P.~Dalpiaz,
 M.~Fiorini,
 I.~Neri,
 A.~Norton,
 F.~Petrucci,
 H.~Wahl\\[2mm]

{\bf INFN, Sezione di Ferrara, Ferrara, Italy}\\
 A.~Cotta Ramusino,
 A.~Gianoli\\[2mm]

{\bf Dipartimento di Fisica e Astronomia dell'Universit\`a e INFN, Sezione di Firenze, Sesto Fiorentino, Italy}\\
 E.~Iacopini,
 G.~Latino,
 M.~Lenti,
 A.~Parenti\\[2mm]

{\bf INFN, Sezione di Firenze, Sesto Fiorentino, Italy}\\
 A.~Bizzeti$\,$\footnotemark[14],
 F.~Bucci\\[2mm]

{\bf Laboratori Nazionali di Frascati, Frascati, Italy}\\
 A.~Antonelli,
 G.~Georgiev$\,$\footnotemark[15],
 V.~Kozhuharov$\,$\footnotemark[15],
 G.~Lanfranchi,
 S.~Martellotti,
 M.~Moulson,
 T.~Spadaro\\[2mm]

{\bf Dipartimento di Fisica ``Ettore Pancini'' e INFN, Sezione di Napoli, Napoli, Italy}\\
 F.~Ambrosino,
 T.~Capussela,
 M.~Corvino$\,$\footnotemark[13],
 D.~Di Filippo,
 P.~Massarotti,
 M.~Mirra,
 M.~Napolitano,
 G.~Saracino\\[2mm]

{\bf Dipartimento di Fisica e Geologia dell'Universit\`a e INFN, Sezione di Perugia, Perugia, Italy}\\
 G.~Anzivino,
 F.~Brizioli,
 E.~Imbergamo,
 R.~Lollini,
 R.~Piandani$\,$\footnotemark[16],
 C.~Santoni\\[2mm]

{\bf INFN, Sezione di Perugia, Perugia, Italy}\\
 M.~Barbanera$\,$\footnotemark[17],
 P.~Cenci,
 B.~Checcucci,
 P.~Lubrano,
 M.~Lupi$\,$\footnotemark[18],
 M.~Pepe,
 M.~Piccini\\[2mm]

{\bf Dipartimento di Fisica dell'Universit\`a e INFN, Sezione di Pisa, Pisa, Italy}\\
 F.~Costantini,
 L.~Di Lella,
 N.~Doble,
 M.~Giorgi,
 S.~Giudici,
 G.~Lamanna,
 E.~Lari,
 E.~Pedreschi,
 M.~Sozzi\\[2mm]

{\bf INFN, Sezione di Pisa, Pisa, Italy}\\
 C.~Cerri,
 R.~Fantechi,
 L.~Pontisso,
 F.~Spinella\\[2mm]
\newpage
{\bf Scuola Normale Superiore e INFN, Sezione di Pisa, Pisa, Italy}\\
 I.~Mannelli\\[2mm]

{\bf Dipartimento di Fisica, Sapienza Universit\`a di Roma e INFN, Sezione di Roma I, Roma, Italy}\\
 G.~D'Agostini,
 M.~Raggi\\[2mm]

{\bf INFN, Sezione di Roma I, Roma, Italy}\\
 A.~Biagioni,
 E.~Leonardi,
 A.~Lonardo,
 P.~Valente,
 P.~Vicini\\[2mm]

{\bf INFN, Sezione di Roma Tor Vergata, Roma, Italy}\\
 R.~Ammendola,
 V.~Bonaiuto$\,$\footnotemark[19],
 A.~Fucci,
 A.~Salamon,
 F.~Sargeni$\,$\footnotemark[20]\\[2mm]

{\bf Dipartimento di Fisica dell'Universit\`a e INFN, Sezione di Torino, Torino, Italy}\\
 R.~Arcidiacono$\,$\footnotemark[21],
 B.~Bloch-Devaux,
 M.~Boretto$\,$\footnotemark[13],
 E.~Menichetti,
 E.~Migliore,
 D.~Soldi\\[2mm]

{\bf INFN, Sezione di Torino, Torino, Italy}\\
 C.~Biino,
 A.~Filippi,
 F.~Marchetto\\[2mm]

{\bf Instituto de F\'isica, Universidad Aut\'onoma de San Luis Potos\'i, San Luis Potos\'i, Mexico}\\
 J.~Engelfried,
 N.~Estrada-Tristan$\,$\footnotemark[22]\\[2mm]

{\bf Horia Hulubei National Institute of Physics for R\&D in Physics and Nuclear Engineering, Bucharest-Magurele, Romania}\\
 A. M.~Bragadireanu,
 S. A.~Ghinescu,
 O. E.~Hutanu\\[2mm]

{\bf Joint Institute for Nuclear Research, Dubna, Russia}\\
 A.~Baeva,
 D.~Baigarashev,
 D.~Emelyanov,
 T.~Enik,
 V.~Falaleev,
 V.~Kekelidze,
 A.~Korotkova,
 L.~Litov$\,$\footnotemark[15],
 D.~Madigozhin,
 M.~Misheva$\,$\footnotemark[23],
 N.~Molokanova,
 S.~Movchan,
 I.~Polenkevich,
 Yu.~Potrebenikov,
 S.~Shkarovskiy,
 A.~Zinchenko$\,$\renewcommand{\thefootnote}{\fnsymbol{footnote}}\footnotemark[2]\renewcommand{\thefootnote}{\arabic{footnote}}\\[2mm]

{\bf Institute for Nuclear Research of the Russian Academy of Sciences, Moscow, Russia}\\
 S.~Fedotov,
 E.~Gushchin,
 A.~Khotyantsev,
 Y.~Kudenko$\,$\footnotemark[24],
 V.~Kurochka,
 M.~Medvedeva,
 A.~Mefodev\\[2mm]

{\bf Institute for High Energy Physics - State Research Center of Russian Federation, Protvino, Russia}\\
 S.~Kholodenko,
 V.~Kurshetsov,
 V.~Obraztsov,
 A.~Ostankov$\,$\renewcommand{\thefootnote}{\fnsymbol{footnote}}\footnotemark[2]\renewcommand{\thefootnote}{\arabic{footnote}},
 V.~Semenov$\,$\renewcommand{\thefootnote}{\fnsymbol{footnote}}\footnotemark[2]\renewcommand{\thefootnote}{\arabic{footnote}},
 V.~Sugonyaev,
 O.~Yushchenko\\[2mm]

{\bf Faculty of Mathematics, Physics and Informatics, Comenius University, Bratislava, Slovakia}\\
 L.~Bician$\,$\footnotemark[13],
 T.~Blazek,
 V.~Cerny,
 Z.~Kucerova\\[2mm]

{\bf CERN,  European Organization for Nuclear Research, Geneva, Switzerland}\\
 J.~Bernhard,
 A.~Ceccucci,
 H.~Danielsson,
 N.~De Simone$\,$\footnotemark[25],
 F.~Duval,
 B.~D\"obrich,
 L.~Federici,
 E.~Gamberini,
 L.~Gatignon,
 R.~Guida,
 F.~Hahn$\,$\renewcommand{\thefootnote}{\fnsymbol{footnote}}\footnotemark[2]\renewcommand{\thefootnote}{\arabic{footnote}},
 E. B.~Holzer,
 B.~Jenninger,
 M.~Koval$\,$\footnotemark[26],
 P.~Laycock$\,$\footnotemark[3],
 G.~Lehmann Miotto,
 P.~Lichard,
 A.~Mapelli,
 K.~Massri,
 M.~Noy,
 V.~Palladino$\,$\footnotemark[27],
 M.~Perrin-Terrin$\,$\footnotemark[28]$^,\,$\footnotemark[29],
 J.~Pinzino$\,$\footnotemark[30]$^,\,$\footnotemark[31],
 V.~Ryjov,
 S.~Schuchmann$\,$\footnotemark[32],
 S.~Venditti\\[2mm]

{\bf University of Birmingham, Birmingham, United Kingdom}\\
 T.~Bache,
 M. B.~Brunetti$\,$\footnotemark[33],
 V.~Duk$\,$\footnotemark[34],
 V.~Fascianelli$\,$\footnotemark[35],
 J. R.~Fry,
 F.~Gonnella,
 E.~Goudzovski,
 L.~Iacobuzio,
 C.~Lazzeroni,
 N.~Lurkin$\,$\footnotemark[9],
 F.~Newson,
 C.~Parkinson$\,$\footnotemark[9],
 A.~Romano,
 A.~Sergi,
 A.~Sturgess,
 J.~Swallow\\[2mm]
\newpage
{\bf University of Bristol, Bristol, United Kingdom}\\
 H.~Heath,
 R.~Page,
 S.~Trilov\\[2mm]

{\bf University of Glasgow, Glasgow, United Kingdom}\\
 B.~Angelucci,
 D.~Britton,
 C.~Graham,
 D.~Protopopescu\\[2mm]

{\bf University of Lancaster, Lancaster, United Kingdom}\\
 J.~Carmignani,
 J. B.~Dainton,
 R. W. L.~Jones,
 G.~Ruggiero\renewcommand{\thefootnote}{\fnsymbol{footnote}}%
\footnotemark[1]\renewcommand{\thefootnote}{\arabic{footnote}}\\[2mm]

{\bf University of Liverpool, Liverpool, United Kingdom}\\
 L.~Fulton,
 D.~Hutchcroft,
 E.~Maurice$\,$\footnotemark[36],
 B.~Wrona\\[2mm]

{\bf George Mason University, Fairfax, Virginia, USA}\\
 A.~Conovaloff,
 P.~Cooper,
 D.~Coward$\,$\footnotemark[37],
 P.~Rubin  

\end{raggedright}
%
%
\setcounter{footnote}{0}
\renewcommand{\thefootnote}{\fnsymbol{footnote}}
\footnotetext[1]{Corresponding authors: G.~Ruggiero, R.~Marchevski, \\
email:giuseppe.ruggiero@cern.ch, radoslav.marchevski@cern.ch}
\footnotetext[2]{Deceased}
\renewcommand{\thefootnote}{\arabic{footnote}}

\footnotetext[1]{Present address: Laboratori Nazionali di Frascati, I-00044 Frascati, Italy}
\footnotetext[2]{Also at CERN,  European Organization for Nuclear Research, CH-1211 Geneva 23, Switzerland}
\footnotetext[3]{Present address: Brookhaven National Laboratory, Upton, NY 11973, USA}
\footnotetext[4]{Also at Institute for Nuclear Research of the Russian Academy of Sciences, 117312 Moscow, Russia}
\footnotetext[5]{Present address: Faculty of Mathematics, Physics and Informatics, Comenius University, 842 48, Bratislava, Slovakia}
\footnotetext[6]{Also at TRIUMF, Vancouver, British Columbia, V6T 2A3, Canada}
\footnotetext[7]{Present address: UCLA Physics and Biology in Medicine, Los Angeles, CA 90095, USA}
\footnotetext[8]{Present address: IFIC, Universitat de Val\`encia - CSIC, E-46071 Val\`encia, Spain}
\footnotetext[9]{Present address: Universit\'e Catholique de Louvain, B-1348 Louvain-La-Neuve, Belgium}
\footnotetext[10]{Present address: Institut f\"ur Kernphysik and Helmholtz Institute Mainz, Universit\"at Mainz, Mainz, D-55099, Germany}
\footnotetext[11]{Present address: Universit\"at W\"urzburg, D-97070 W\"urzburg, Germany}
\footnotetext[12]{Present address: Universit\"at Hamburg, D-20146 Hamburg, Germany}
\footnotetext[13]{Present address: CERN,  European Organization for Nuclear Research, CH-1211 Geneva 23, Switzerland}
\footnotetext[14]{Also at Dipartimento di Fisica, Universit\`a di Modena e Reggio Emilia, I-41125 Modena, Italy}
\footnotetext[15]{Also at Faculty of Physics, University of Sofia, BG-1164 Sofia, Bulgaria}
\footnotetext[16]{Present address: Institut f\"ur Experimentelle Teilchenphysik (KIT), D-76131 Karlsruhe, Germany}
\footnotetext[17]{Present address: INFN, Sezione di Pisa, I-56100 Pisa, Italy}
\footnotetext[18]{Present address: Institut am Fachbereich Informatik und Mathematik, Goethe Universit\"at, D-60323 Frankfurt am Main, Germany}
\footnotetext[19]{Also at Department of Industrial Engineering, University of Roma Tor Vergata, I-00173 Roma, Italy}
\footnotetext[20]{Also at Department of Electronic Engineering, University of Roma Tor Vergata, I-00173 Roma, Italy}
\footnotetext[21]{Also at Universit\`a degli Studi del Piemonte Orientale, I-13100 Vercelli, Italy}
\footnotetext[22]{Also at Universidad de Guanajuato, Guanajuato, Mexico}
\footnotetext[23]{Present address: Institute of Nuclear Research and Nuclear Energy of Bulgarian Academy of Science (INRNE-BAS), BG-1784 Sofia, Bulgaria}
\footnotetext[24]{Also at National Research Nuclear University (MEPhI), 115409 Moscow and Moscow Institute of Physics and Technology, 141701 Moscow region, Moscow, Russia}
\footnotetext[25]{Present address: DESY, D-15738 Zeuthen, Germany}
\footnotetext[26]{Present address: Charles University, 116 36 Prague 1, Czech Republic}
\footnotetext[27]{Present address: Physics Department, Imperial College London, London, SW7 2BW, UK}
\footnotetext[28]{Present address: Aix Marseille University, CNRS/IN2P3, CPPM, F-13288, Marseille, France}
\footnotetext[29]{Also at Universit\'e Catholique de Louvain, B-1348 Louvain-La-Neuve, Belgium}
\footnotetext[30]{Present address: Department of Physics, University of Toronto, Toronto, Ontario, M5S 1A7, Canada}
\footnotetext[31]{Also at INFN, Sezione di Pisa, I-56100 Pisa, Italy}
\footnotetext[32]{Present address: Institut f\"ur Physik and PRISMA Cluster of excellence, Universit\"at Mainz, D-55099 Mainz, Germany}
\footnotetext[33]{Present address: Department of Physics, University of Warwick, Coventry, CV4 7AL, UK}
\footnotetext[34]{Present address: INFN, Sezione di Perugia, I-06100 Perugia, Italy}
\footnotetext[35]{Present address: Dipartimento di Psicologia, Universit\`a di Roma La Sapienza, I-00185 Roma, Italy}
\footnotetext[36]{Present address: Laboratoire Leprince Ringuet, F-91120 Palaiseau, France}
\footnotetext[37]{Also at SLAC National Accelerator Laboratory, Stanford University, Menlo Park, CA 94025, USA}

\clearpage
\tableofcontents
\clearpage

\section{Introduction}\label{sec:intro}
The \pnnc decay is a flavour-changing Neutral-Current process that proceeds through electroweak box and penguin diagrams in the Standard Model (SM), allowing an exploration of its flavour structure
thanks to unique theoretical cleanliness.
A quadratic GIM mechanism and the transition of the top quark to the down quark make this process extremely rare. 
The SM prediction for the \pnnc branching ratio (BR) 
can be written as~\cite{pnntheo1}:

\begin{equation}
 \text{BR}(K^+ \to \pi^+ \nu \bar{\nu})= \kappa_+(1+\Delta_{\rm EM})\left[ \left(\frac{Im \lambda_t}{\lambda^5} X(x_t)\right)^2  + \left(\frac{Re \lambda_c}{\lambda}P_c(X)+\frac{Re \lambda_t}{\lambda^5}X(x_t)\right)^2  \right],
\end{equation}
where $\Delta_{EM}=-0.003$ accounts for  
the electromagnetic radiative corrections; $x_t= m_t^2/M_W^2$; $\lambda = |V_{us}|$ and $\lambda_i=V^{*}_{is}V_{id}$  $(i= c, t)$ are  
combinations of  Cabibbo-Kobayashi-Maskawa (CKM) matrix elements; $X$ and  $P_c(X)$ are the loop functions for the top and charm quark respectively; and 

\begin{equation}
\kappa_+ = (5.173 \pm 0.025)\times 10^{-11}\left[\frac{\lambda}{0.225}\right]^8 
\end{equation}
parameterizes hadronic matrix elements.
It is worth noting that  $\text{BR}(K^+ \to \pi^+ \nu \bar{\nu})$ depends on the sum of the square of the imaginary part of the top loop, which is CP-violating, and the square of the sum of the charm contribution and the real part of 
the top loop. 
Numerically, the branching ratio can be written as an explicit function of the CKM parameters, $V_{cb}$ and the angle $\gamma$, as follows :

\begin{equation}
\text{BR}(K^+ \to \pi^+ \nu \bar{\nu})=(8.39 \pm 0.30) \times 10^{-11} \Big{[} \frac{|V_{cb}|}{40.7 \times 10^{-3}}\Big{]}^{2.8}\Big{[} \frac{\gamma}{73.2^\circ}\Big{]}^{0.74},
\end{equation}
where the numerical uncertainty is 
due to theoretical uncertainties in the NLO (NNLO) QCD 
corrections to the top (charm) quark contribution~\cite{pnntheo2,pnntheo3} and NLO electroweak corrections~\cite{pnntheo4}.  The intrinsic theoretical accuracy is at the level of 3.6\%.
Uncertainties in the hadronic matrix element largely cancel when it is evaluated from  the precisely-measured branching ratio of the $K^+\rightarrow\pi^0e^+\nu$ decay, including isospin-breaking and non-perturbative effects calculated in detail~\cite{pnntheo4,pnntheo5,pnntheo6}.
Using tree-level elements of the CKM matrix as external inputs~\cite{pdg}, averaged over  exclusive and inclusive determinations, namely 
$|V_{cb}|=(40.7 \pm 1.4 ) \times 10^{-3}$  
and $\gamma = (73.2^{+6.3}_{-7.0})$ degrees,  
the SM prediction of the branching ratio is $(8.4\pm1.0)\times10^{-11}$~ \cite{pnntheo1}.
The current precision of the CKM parameters dominates the BR uncertainty.

The \pnnc decay is sensitive to 
currently proposed SM extensions and probes higher mass scales than other 
rare meson decays.
This arises because of the absence of tree-level contributions and the quadratic GIM suppression at loop level in the SM, which together lead to a very small BR.  
 Moreover, the absence of long-distance contributions  enables the accurate BR calculation.
The largest deviations from SM predictions are expected in models with new sources of flavour violation, where constraints from B physics are weaker~\cite{pnnlh,pnnrs}. 
Models with currents of defined chirality produce specific correlation patterns between the branching ratios of \pnnc and \pnnz decay modes, which are constrained by the value of the CP-violating parameter  $\varepsilon_K$~\cite{pnnzz,pnnzp}. 
Present experimental constraints limit the range of variation within supersymmetric models~\cite{pnnsusy1,pnnsusy3,pnnsusy4}.
The \pnnc decay is also sensitive to some aspects of lepton flavour non-universality~\cite{pnnlu} and can constrain leptoquark models~\cite{pnnlq,pnnlq2} that aim to explain the measured CP-violating ratio $\varepsilon'/\varepsilon$~\cite{pdg}.

The E787 and E949 experiments at the Brookhaven National Laboratory (BNL) studied the \pnnc decay using a kaon decay-at-rest technique,
reaching an overall single event sensitivity of about $0.8\times10^{-10}$ and measuring the BR to be $(17.3^{+11.5}_{-10.5})\times10^{-11}$~\cite{bnl1,bnl2}. 
The NA62 experiment at the CERN SPS will measure  more precisely the BR of the \pnnc decay using a decay-in-flight technique and  data recorded from 2016 to 2018.
The first NA62 result  was based on the analysis of the data collected in 2016 and proved the feasibility of the technique to study the \pnnc decay~\cite{na62pnn1}.
In the following sections, NA62 reports the investigation of the \pnnc decay, based on data recorded in 2017, 
corresponding to about 30\% of the total data set collected in 2016--18.


\section{Principles of the experiment and analysis method}
\label{sec:principle}
The NA62 experiment is designed to reconstruct charged kaons and their 
daughter particles, when the kaons decay in flight inside a defined fiducial volume.
The \pnnc  decay presents two main challenges: the extremely low value of the SM signal branching ratio of order $10^{-10}$  and the open kinematics of the  
final state, as neutrinos remain undetected. 
These challenges require both the production of a sufficient number of \pnnc decays,  as can be achieved by exploiting the  high-intensity 75\,GeV$/c$ secondary $K^+$ beam produced by the CERN SPS;  and the reduction of the contribution of the dominant $K^+$ decay modes by at least eleven orders of magnitude to bring the background to a level lower than the signal. 

\begin{figure}[t]
  \begin{center}
  \includegraphics[width=25pc]{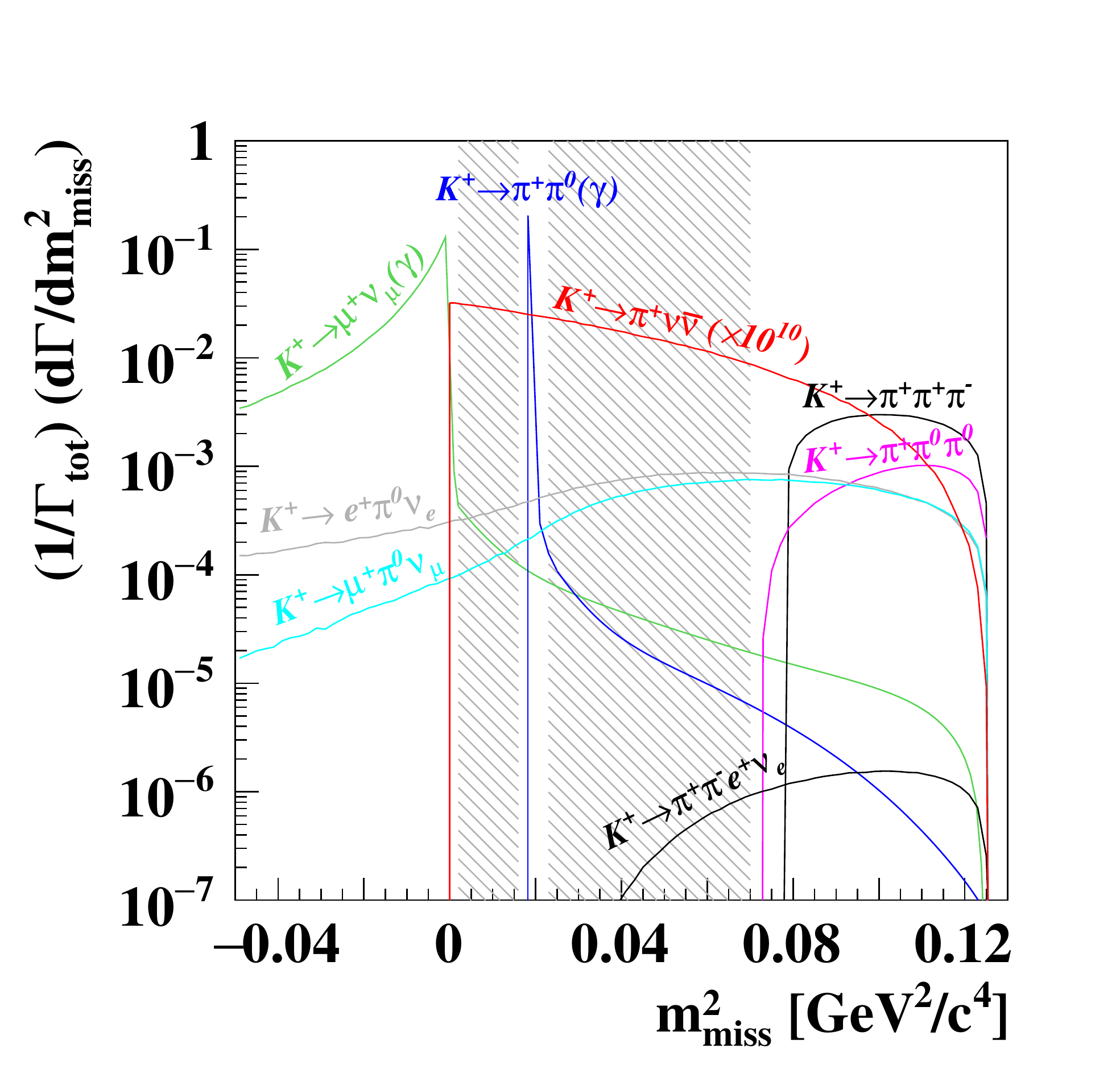}
  \caption{\label{fig:m2missth}Expected theoretical  distributions  of the $\mmis$ variable 
  relevant to the \pnnc measurement, before applying acceptance and resolution effects.
                The $\mmis$ is computed under the hypothesis that the charged particle in the final state is a $\pi^+$.
                The \pnnc signal (red line) is multiplied by $10^{10}$ for visibility. 
                The hatched areas include the signal  
                regions.}
  \end{center}
\end{figure}

The signature of the \pnnc decay is a single $\pi^+$ and missing energy. 
The squared missing mass,  $\mmis = (P_K-P_{\pi^+})^2$, where $P_K$ and $P_{\pi^+}$ indicate the 4-momenta of the $K^+$ and $\pi^+$,  
describes the kinematics of the one-track final state.
In particular,  the presence of two neutrinos makes the signal broadly distributed over the $\mmis$ range, as illustrated in Figure~\ref{fig:m2missth}.
The dominant $K^+$ decay modes $K^+\rightarrow\mu^+\nu$, $K^+\rightarrow\pi^+\pi^0$ and $K^+\rightarrow\pi^+\pi^{+(0)}\pi^{-(0)}$ have different 
$\mmis$ distributions; it is therefore possible to define 
regions, either side of the $K^+\rightarrow\pi^+\pi^0$ peak, qualitatively indicated in Figure~\ref{fig:m2missth}, where 
the search for the signal is performed, also called {\em signal regions}.

The  $K^+\rightarrow\mu^+\nu$, $K^+\rightarrow\pi^+\pi^0$ and $K^+\rightarrow\pi^+\pi^{+(0)}\pi^{-(0)}$  decays enter the signal regions through  radiative and/or  resolution tails of the reconstructed $\mmis$.
The signal selection, based on kinematics only, relies on the accurate measurement of the $\mmis$ quantity, i.e. of the $K^+$ and $\pi^+$ momenta and directions. 
In contrast, $K^+\rightarrow\pi^0 \ell^+\nu$ or rarer decays, like $K^+\rightarrow\pi^+\pi^-\ell^+\nu$, span over the signal  regions because of the presence of undetected neutrinos; however,
these background decay modes include a lepton in the final state and exhibit extra activity in the form of photons or charged particles.  
A particle identification system must  therefore  separate $\pi^+$ from $\mu^+$ and $e^+$.
Photons and additional charged particles in final state must be vetoed as efficiently as possible.

The above conditions translate into the following experimental requirements:
\begin{itemize}
\item the detection of incident $K^+$  and outgoing $\pi^+$  signals with 100\,ps time resolution to mitigate the impact of the pile-up effect due to the high particle rates; 
\item a low-mass $K^+$ and $\pi^+$ tracking system, which reconstructs  precisely the kinematics 
to suppress $K^+\rightarrow\pi^+\pi^0$ and $K^+\rightarrow\mu^+\nu$ backgrounds by at least three orders of magnitude, while keeping the background from hadronic interactions low;
\item a system of calorimeters and a Ring Imaging Cherenkov counter (RICH) to suppress decays with positrons and muons by seven to eight orders of magnitude;
\item a set of electromagnetic calorimeters, to detect photons and reduce the number of $K^+\rightarrow\pi^+\pi^0$ decays by eight orders of magnitude; and
\item 
an experimental design which guarantees the geometric acceptance for negatively charged particles in at least two detectors. 
\end{itemize}

The decay-in-flight configuration has two main advantages:
\begin{itemize}
\item 
the selection of  \pnnc decays with a $\pi^+$ momentum lower than 35\,GeV/$c$ to facilitate the background rejection by ensuring at least 40\,GeV of missing energy, and to exploit the capability of the RICH
for $\pi^+ / \mu^+$ separation; and
\item 
the achievement of  sufficient $\pi^0$ suppression by using photon detection coverage up to 50\,mrad with respect to the \kp direction, and by efficiently detecting photons of energy above 1~GeV.
\end{itemize}

The experimental layout  and the data-taking conditions are reviewed in section~\ref{sec:detector}.
The reconstruction algorithms are described in section~\ref{sec:reco}. After the \pnnc selection (section~\ref{sec:evsel}), the analysis proceeds through the evaluation of the single event sensitivity, defined as the branching ratio equivalent 
to the observation of one SM signal event (section~\ref{sec:pnnses}).
The number of signal decays is normalized to the number of~\pp decays, whose branching ratio is accurately  known~\cite{pdg}.
This allows the precise determination of the single event sensitivity without relying on the absolute measurement of the total number of $K^+$ decays.
The final step of the analysis is the evaluation of the expected background in the signal regions (section~\ref{sec:pnnbckg}). 
To avoid biasing the selection of \pnnc events,   the analysis  
follows a ``blind" procedure, with signal regions 
 kept masked until completion of all the analysis steps. Finally, the result is presented in section~\ref{sec:result}.


\section{Experimental setup and data taking}
\label{sec:detector}
\begin{figure}[t]
  \begin{center}
 \includegraphics[width=1.08\textwidth]{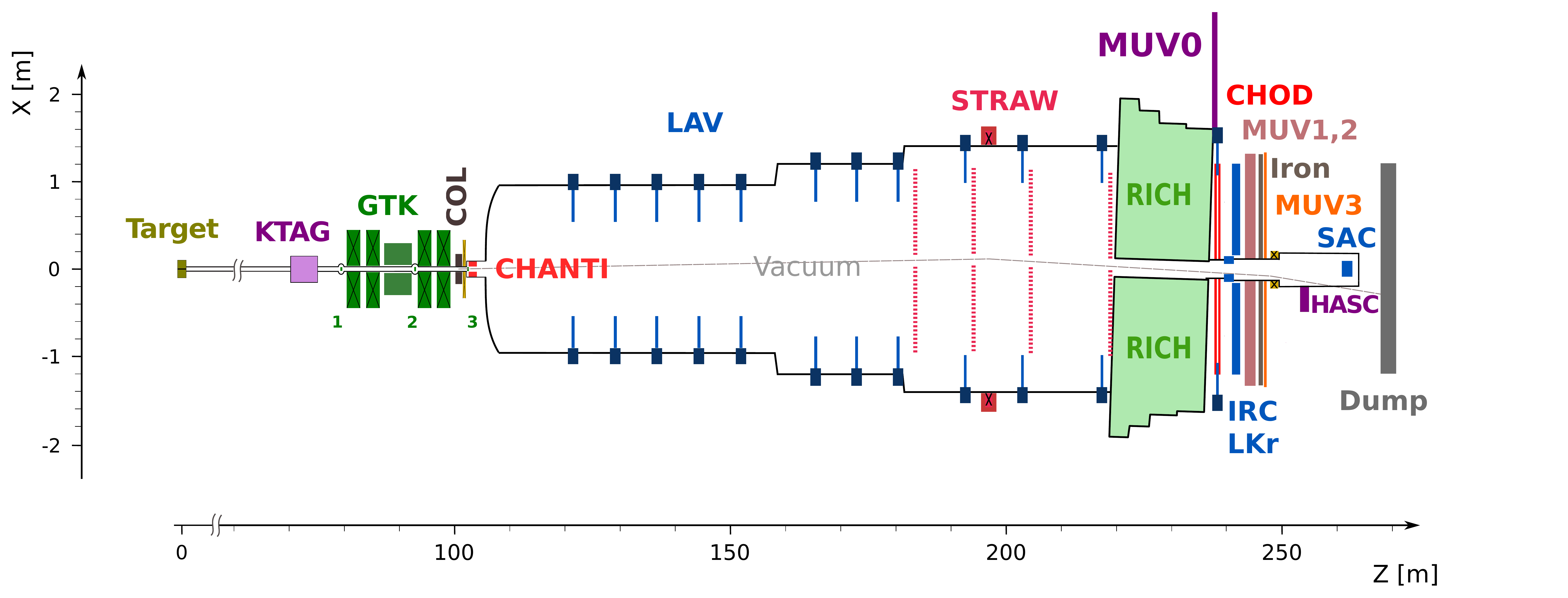}
    \caption{Schematic top view of the NA62 beam line and detector. Dipole magnets are displayed as boxes with superimposed crosses. The label ``COL'' denotes the collimator named ``final collimator'' in the text.
    The label ``CHOD'' refers both to the CHOD and NA48-CHOD detectors.
    Also shown is the trajectory of a beam particle in vacuum which crosses all the detector apertures, thus avoiding interactions with material. A dipole magnet between MUV3 and SAC deflects the beam particles out of the SAC acceptance.}
    \label{fig:layoutxz17}
  \end{center}
\end{figure}

The NA62 beam line and detector are sketched in Figure~\ref{fig:layoutxz17}.
A detailed description of them can be found in~\cite{na62det}.  The beam line defines
the Z-axis of the experiment's right-handed laboratory coordinate system.  The
origin is the kaon production target, and beam particles travel in the
positive Z-direction.  The Y-axis is vertical (positive up), and the X-axis is
horizontal (positive left).

The kaon production target is a 40~cm long beryllium rod.  
A 400~GeV proton beam extracted from the CERN Super Proton Synchrotron (SPS) 
impinges on the target in spills of three seconds effective duration.    Typical
intensities during data taking range from  $1.7$ to $1.9\times10^{12}$
protons per pulse ($ppp$).  The resulting secondary hadron beam of positively
charged particles  consists of 70\% $\pi^+$, 23\% protons, and 6\% $K^+$,
with a nominal momentum of 75~GeV/$c$ (1\% rms momentum bite).

Beam particles are characterized by a differential Cherenkov counter (KTAG)
and a three-station silicon pixel matrix (Gigatracker, GTK,  with pixel size of $300 \times 300~\mu{\rm m}^2$).  The KTAG uses N$_2$ gas 
at 1.75~bar pressure (contained in a 5~m long vessel) and is read out by photomultiplier tubes grouped in eight
sectors.  It tags incoming kaons with 70~ps time-resolution.  The GTK stations
are located before, between, and after two pairs of dipole magnets (a beam
achromat), forming a  spectrometer that measures beam particle momentum, direction,
and time with resolutions of  0.15~GeV/$c$, 16~$\mu$rad, and 100~ps,
respectively.  

The last GTK station (GTK3) is immediately preceded by a 1~m thick, variable
aperture steel collimator (final collimator).  Its inner aperture is typically set at
66~mm $\times$ 33~mm, and its outer dimensions are about 15~cm.  It serves as
a partial shield against hadrons produced by upstream $K^+$ decays.

GTK3 marks the beginning of a 117~m long vacuum tank. The first 80~m of the
tank define a  volume  in which 13\% of the kaons decay.  The beam
has a  rectangular transverse profile of 52 $\times$ 24 mm$^2$ and a divergence
of 0.11~mrad (rms) in each plane at the decay volume entrance.

The time, momentum, and direction of charged daughters of kaon decays-in-flight
are measured by a magnetic spectrometer (STRAW), a ring-imaging Cherenkov
counter (RICH), and two scintillator hodoscopes (CHOD and NA48-CHOD).
The STRAW, consisting of two pairs of straw chambers on either side of a
dipole magnet,  measures momentum-vectors with a resolution,
$\sigma_p / p$, between 0.3\% and 0.4\%.  The RICH, filled with neon at
atmospheric pressure, tags the decay particles with a timing precision of
better than 100~ps and provides particle identification.  The CHOD, a matrix of
tiles read out by SiPMs, and the NA48-CHOD, comprising two orthogonal planes
of scintillating slabs reused from the NA48 experiment, are used for
triggering and timing, providing a time measurement with 200~ps resolution.
 
Other sub-detectors suppress decays into photons or into multiple charged
particles (electrons, pions or muons) or provide complementary particle
identification.  Six stations of plastic scintillator bars (CHANTI) detect,
with 1~ns time resolution, extra activity, including
inelastic interactions in  GTK3.  Twelve stations of ring-shaped
electromagnetic calorimeters (LAV1 to LAV12), made of lead-glass blocks,
surround the vacuum tank and downstream sub-detectors to achieve hermetic
acceptance for photons emitted by $K^+$ decays in the decay volume at polar angles
between 10 and 50~mrad.  A 27 radiation-length thick, quasi-homogeneous liquid
krypton electromagnetic calorimeter (LKr) detects photons from $K^+$  decays
emitted at angles between 1 and 10~mrad.  The LKr also complements the RICH
for particle identification.  Its energy resolution in NA62 conditions is
$\sigma_E / E = 1.4\%$ for energy deposits of 25~GeV.  Its spatial and time
resolutions are 1~mm and between 0.5 and 1~ns, respectively, depending on the
amount and type of energy released.  Two hadronic iron/scintillator-strip
sampling calorimeters (MUV1,2) and an array of scintillator tiles located
behind 80 cm of iron (MUV3)  supplement the pion/muon identification system. 
MUV3 has a time resolution of 400~ps.  A lead/scintillator shashlik calorimeter
(IRC) located in front of the LKr, covering an annular region between 65 and
135~mm from the Z-axis, and a similar detector (SAC) placed on the Z-axis at
the downstream end of the apparatus, ensure the detection of photons down to
zero degrees in the forward direction.  Additional counters (MUV0, HASC)
installed at optimized locations provide nearly hermetic coverage for charged
particles produced in multi-track kaon decays.

All detectors are read out with TDCs, except for LKr and MUV1, 2, which are
read out with 14-bit FADCs.  The IRC and SAC are read out with both.  All
TDCs are mounted on custom-made (TEL62) boards, except for GTK and STRAW,
which each have specialized TDC boards.  TEL62 boards both read out data and
provide trigger information.  A dedicated processor interprets calorimeter
signals for triggering. 
A dedicated board (L0TP) combines logical signals (primitives) from the RICH, CHOD,
NA48-CHOD, LKr, LAV, and MUV3 into a low-level trigger (L0) whose decision is
dispatched to sub-detectors for data readout ~\cite{tdaq}.  
A software trigger (L1) exploits reconstruction
algorithms similar to those used offline with data from KTAG, LAV, and
STRAW to further cull the data before storing it on disk~\cite{na62det}.

The data come from $3\times10^5$ SPS spills accumulated during a four-month data-taking period in 2017,  recorded at an average  beam intensity of 450 MHz.
The instantaneous beam intensity is measured event-by-event 
using the number of signals recorded out-of-time in the GTK detector.
The average beam intensity per spill was stable within $\pm$10\% throughout the data-taking period, while the instantaneous beam intensity showed fluctuations up to a factor of two around the average value.

The data have been collected using a trigger specifically setup  for the \pnnc measurement, called {\it PNN trigger}, concurrently with  a minimum-bias  trigger.
The PNN trigger is defined as follows. 
The L0 trigger requires a signal in the RICH to tag a charged particle. 
The time of this signal,  called trigger time, is used as a reference to define a coincidence within 6.3\,ns of:
a signal in  one to four CHOD tiles; 
no signals in opposite CHOD quadrants to suppress \pppc decays;
no signals in MUV3 to reject \mn decays;  
less than 30\,GeV energy deposited in LKr and 
no more than one cluster to reject \pp decays. 
The L1 trigger requires: 
a kaon identified in KTAG;
signals within 10 ns of the trigger time in at most two blocks of each LAV station;
at least one STRAW track corresponding to a particle with momentum below 50\,GeV$/c$  
and forming a vertex with the nominal beam axis upstream of the first STRAW chamber. 
Events collected by the PNN trigger are referred to as {\it PNN} events or data.
The minimum-bias trigger is based on NA48-CHOD information downscaled by a factor of 400.
The trigger time is the time of the NA48-CHOD signal.
Data collected by the minimum-bias trigger are used at analysis level to determine the $K^+$ flux, to measure efficiencies, and to estimate backgrounds. 
These data are called {\it minimum-bias} events or data.

Acceptances and backgrounds are evaluated using Monte Carlo (MC) simulation based on the
{\tt GEANT4} toolkit~\cite{geant4} to describe detector geometry and response.
The $K^+$ decays are generated in the kaon rest frame using the appropriate matrix elements and form factors.
The simulation also includes a description of the 
collimators and dipole and quadrupole magnets in the beam line, necessary to accurately simulate the beam shape.  
Certain aspects of the simulation are tuned using input from data, namely signal formation and 
readout detector inefficiencies. 
Accidental activity is added to the KTAG Cherenkov counter and to the GTK beam tracker assuming 450~MHz  beam intensity, and using a library of pileup beam particles built from data. 
No accidental activity is simulated in the detectors downstream of the last station of the beam tracker. 
Simulated data are subjected to the same reconstruction and calibration procedures as real data.

\section{Data reconstruction and calibration}
\label{sec:reco}
The channels of the Cherenkov beam counter KTAG are time-aligned with the trigger time, and signals are grouped within 2\,ns wide windows to define KTAG candidates.  
A $K^+$ KTAG candidate must have signals in at least five of  eight sectors.

The arrival time of the pulses measured in each of the GTK pixels is aligned to the trigger time and corrected for pulse-amplitude slewing.  
Signals from the three GTK stations grouped within 10\,ns of the trigger time form a beam track.
A track must have pulses in all  three stations, therefore it is made of at least three hit pixels.
Nevertheless, a particle can leave a signal in more than one adjacent pixel in the same station if hitting  the edge of a pixel or because of $\delta$-rays.
In this case, pulses in neighbouring pixels form a cluster that is used to reconstruct the track.
Fully reconstructed \pppc decays in the STRAW spectrometer are used to align the GTK stations transversely to a precision of better than 100~$\mu$m and to tune the GTK momentum scale.

The STRAW reconstruction relies on the trigger time as a reference to determine the 
drift time. 
A track is defined by space-points in the chambers describing a path compatible with magnetic bending.
A Kalman-filter fit provides the track parameters.
The $\chi^2$ fit value and the number of space-points characterize the track quality. 
Straight tracks collected with the magnet off serve to align the straw tubes to 30\,$\mu$m accuracy. 
The average value of the \kp mass reconstructed for \pppc decays provides fine tuning of the momentum scale to a  part per thousand precision.  

Two  algorithms reconstruct  RICH ring candidates, both grouping signals from photomultipliers (PM) in time around the trigger time. 
The first one, called {\it track-seeded ring}, makes use of a STRAW track as a seed to build a RICH ring and compute a likelihood for several mass hypotheses ($e^+$, $\mu^+$, $\pi^+$ and $K^+$). 
The second one, called  {\it single ring}, fits the signals to a ring assuming that they are produced by a single particle, with the fit $\chi^2$ characterizing the quality of this hypothesis. 
Positrons are used to calibrate the RICH response and align the twenty RICH mirrors to a precision of $30$\,$\mu$rad~\cite{richperf}. 

The CHOD candidates are defined by the response of  two silicon-photomultipliers (SiPM) reading out the same tile. 
Signals in crossing horizontal and vertical slabs compatible with the passage of a charged particle form NA48-CHOD candidates.
Each slab is time-aligned to the trigger time. 
Time offsets depending on the intersection position account for the effect of light propagation along a slab.

Groups of LKr cells with deposited energy within 100\,mm of a seed form LKr candidates (clusters). 
A seed is defined by a cell in which an energy of at least 250\,MeV is released. 
Cluster energies, positions, and times are reconstructed taking into account energy calibration, non-linearity, energy sharing for nearby clusters and noisy cells. 
The final calibration is performed using positrons from \kethree decays. 
An additional reconstruction algorithm is applied to maximise the reconstruction efficiency. 
This is achieved by defining candidates as sets of cells with at least 40\,MeV energy, closer than 100\,mm and in time within 40\,ns of each other. 

The reconstruction of MUV1(2) candidates relies on the track impact point. 
Signals in fewer than 8 (6) nearby scintillator strips around the track are grouped to form a candidate.
The energy of a candidate is defined as the sum of the energies in the strips, calibrated using weighting factors extracted from dedicated simulations and tested on samples of $\pi^+$ and $\mu^+$.

Candidates in MUV3 are defined by time coincidences of the response of the two PMs reading the same tile. 
The time of a candidate is defined by the later of the two PM signals, to avoid the effect of the time spread 
due to the early Cherenkov light produced by particles traversing the PM window.

CHANTI candidates are defined by signals clustered in time and belonging either to adjacent parallel bars or to intersecting orthogonal bars. 

Two threshold settings discriminate the CHANTI, LAV, IRC and SAC TDC signals~\cite{na62det}.
Thus up to four time measurements are associated with each signal, corresponding to the leading and trailing edge times of the high and low thresholds. 
The relation between the amplitude of the IRC and SAC pulses provided by the FADC readout, and the energy release is 
calibrated for each channel after baseline subtraction using a sample of \pp decays. 

Signal times measured by GTK, KTAG, CHOD, RICH and LKr are further aligned to the trigger time for each spill, resulting in a better than 10\,ps stability through the whole data sample. 


\section{Selection of signal and  normalization decays} 
\label{sec:evsel}
The selection of both \pnnc signal  and $K^+\rightarrow\pi^+\pi^0$ normalization decays requires 
the identification of the downstream charged particle  as a $\pi^+$ and the parent beam particle as a $K^+$.
Further  specific criteria are applied to separate signal and normalization events.

\subsection{Downstream charged particle }
\label{sec:pdef}
A downstream charged particle is defined as a track reconstructed in the STRAW spectrometer (downstream track) and matching signals in the two hodoscopes CHOD and NA48-CHOD, in the electromagnetic calorimeter LKr, and in the RICH counter.

The downstream track must include space-points reconstructed in all  four chambers of the STRAW spectrometer, satisfy suitable quality criteria, and be consistent with a positively charged particle. 
The extrapolation of this track to any downstream detector defines the expected position of the charged particle's impact point on that detector.
These positions must lie within 
 the geometric acceptance of the  corresponding downstream detectors and outside the acceptance of the large  and small angle calorimeters LAV and IRC.
The impact points of the charged particles are used to match 
the downstream tracks with signals in the hodoscopes and the electromagnetic calorimeter.

Two discriminant variables are built  using the difference of time and spatial coordinates between each hodoscope candidate and the track.
The NA48-CHOD candidate with the lowest discriminant value and the CHOD candidate closest in space to the particle impact point are matched to the track. 
The latter candidate must be within $\pm5$\,ns of the assigned NA48-CHOD candidate.
Cuts on maximum allowed values of the discriminant variables are also implemented to avoid fake or accidental signals in the hodoscopes.

A  LKr cluster  is matched to a charged particle if its distance from the particle impact point is smaller than 100\,mm.
The energy released by the track in the calorimeter is defined as the energy of the associated cluster.
The time of the associated cluster is the time of the most energetic cell 
of the cluster. 
A 2\,ns time coincidence is required between the cluster and the NA48-CHOD candidate associated with  the track.
 
The association between the track and a single ring of the RICH counter exploits the relationship between the slope of the track and the position of the ring center. 
A track-seeded ring is also considered  for particle identification purposes (section~\ref{sec:pid}).  Both types of RICH rings must be in time within $\pm3$\,ns of the NA48-CHOD candidate associated to the track.  The time of the downstream charged particle is defined as the time of the associated RICH single ring.
  
Track-matching with a CHOD, NA48-CHOD, RICH and LKr candidate is mandatory.

\subsection{Parent  beam particle}  
\label{sec:kdef}
The parent $K^+$ of a selected downstream charged particle is defined by:
the $K^+$ candidate in KTAG closest in time and within $\pm2$\,ns of the downstream particle;
a beam track  in GTK associated in time with the KTAG candidate and in space with the downstream track in the STRAW.

\begin{figure}[t]
  \begin{center}
  \begin{minipage}{18pc}
  \includegraphics[width=18pc]{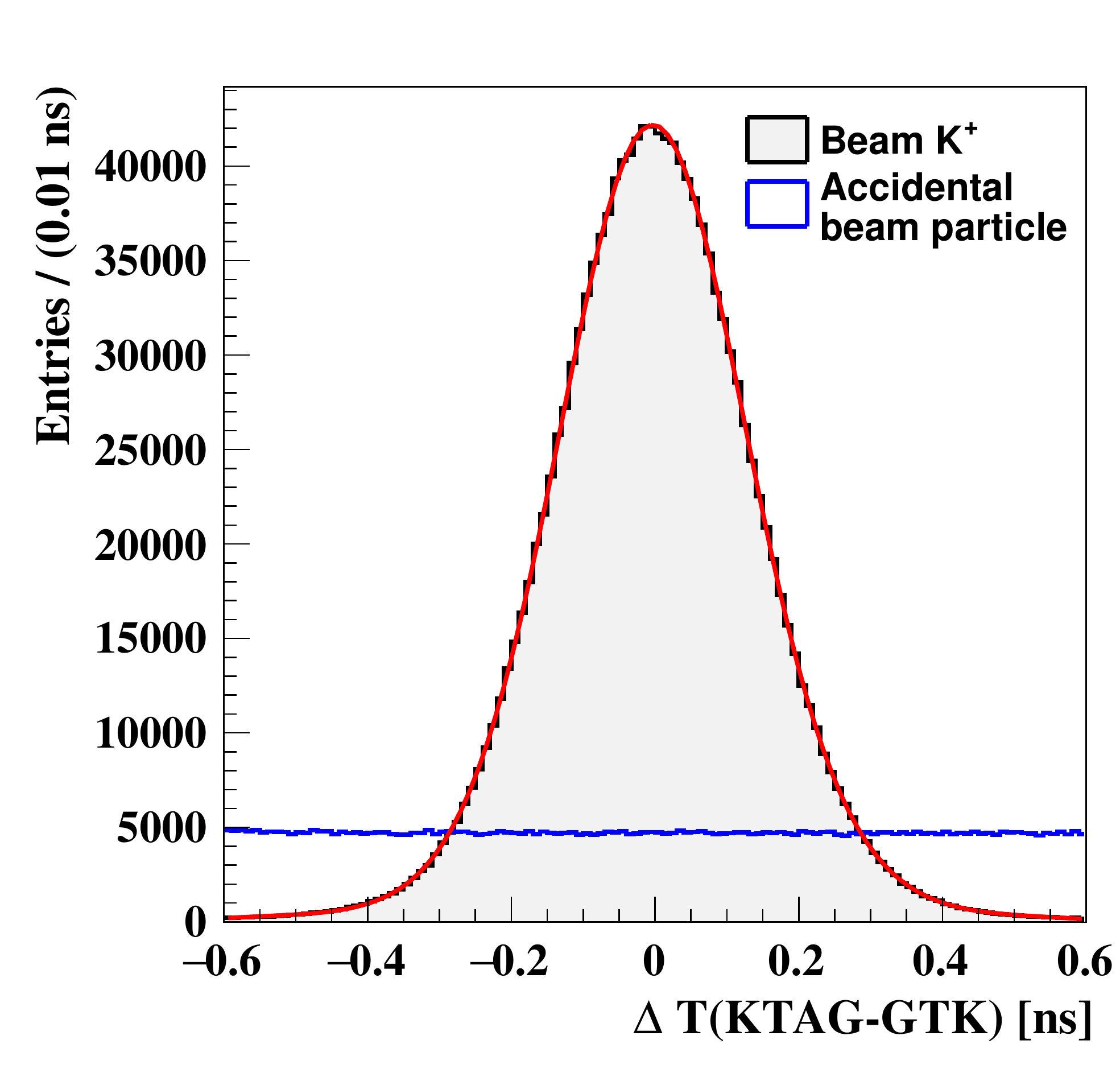}   
  \end{minipage}\hspace{1pc}
  \begin{minipage}{18pc}
  \includegraphics[width=18pc]{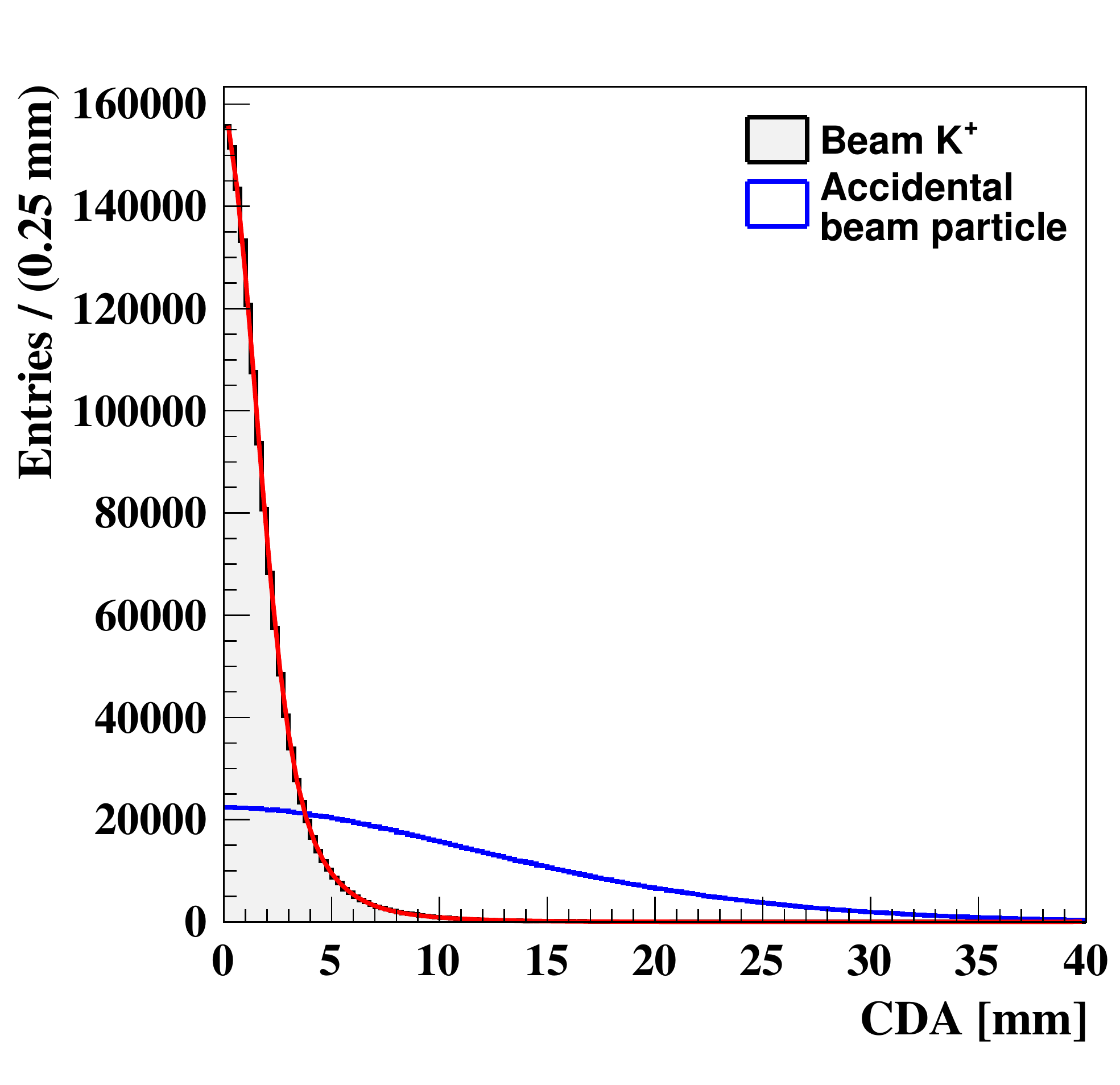}   
  \end{minipage}
  \caption{\label{fig:kpitemplate}Distributions of $\Delta$T(KTAG-GTK) (left) and CDA (right) for events with beam $K^+$ (shaded histogram) and accidental beam particle  (empty histogram),  as obtained from fully reconstructed $K^+\rightarrow\pi^+\pi^+\pi^-$ decays in the data. The red curves superimposed on the histograms describe the functions 
  used to model  the time and CDA distributions of the beam $K^+$.}
  \end{center}
\end{figure}

The association between GTK, KTAG and STRAW candidates relies on a likelihood discriminant built from two variables:
the time difference between the KTAG candidate and the beam track ($\Delta$T(KTAG-GTK)); and
the closest distance of approach of the beam track to the downstream charged particle (CDA) computed  taking into account bending of the particle trajectory in the stray magnetic field in the vacuum tank.
The templates of the $\Delta$T(KTAG-GTK) and CDA distributions of the parent $K^+$ are derived from a sample of $K^+\rightarrow\pi^+\pi^+\pi^-$ selected on data.
In this case, the clean three-pion final state signature tags the $K^+$ track in the GTK and one of the positively charged pions is chosen to be the downstream charged particle.
The resulting distributions are shown in Figure~\ref{fig:kpitemplate}, together with the corresponding distributions for events including a random GTK track instead of the $K^+$ track.
In contrast with the parent $K^+$, the shape of the CDA distribution in the presence of random beam tracks depends on the size and divergence of the beam, and the emission angle of the $\pi^+$.
The beam track with the largest 
discriminant value is, by construction, the parent $K^+$;
its momentum and direction must be consistent with the nominal beam properties.

Because of the high particle rate in the beam tracker, several beam particles may overlap with the $K^+$ within $\pm1$\,ns; they are referred to as  {\it pileup} (or {\it accidental}) particles and the corresponding GTK track is called a pileup (or accidental) track.
A {\it wrong association} occurs when a pileup track leads to a likelihood discriminant value larger than that of the actual $K^+$ track.
An {\it accidental association} occurs when the $K^+$ track is not reconstructed in the beam tracker and a pileup track is associated to the downstream charged particle.
A sharp cut on the minimum allowed value of the likelihood discriminant reduces the probabilities of wrong and accidental association.
Events are also rejected if more than 5 pileup tracks are reconstructed or if the likelihood discriminant values of different beam tracks matching the same downstream charged particle are similar.
Finally, a cut is applied on 
a discriminant computed using the time difference between the beam track and the downstream charged particle, instead of $\Delta$T(KTAG-GTK). 

The $K^+\rightarrow\pi^+\pi^+\pi^-$ decays allow the 
performance of the beam-track matching to be monitored.  
The probabilities of wrong and accidental association depend on the instantaneous beam intensity and are about 1.3\% and 3.5\% on average, respectively.
The latter includes also the probability that a pileup track time  is within $\pm$1\,ns  of the KTAG time.
Both probabilities depend on the type of process under study.  

\subsection{Kaon decay} 
\label{sec:fv}
A downstream charged particle and its parent $K^+$ define the kaon decay.
The mid-point between the beam and downstream track at the closest distance of approach defines the position of the $K^+$ decay, called {\it decay vertex}.

Several downstream charged particles may be reconstructed in the same event as a result of overlapping accidental charged particles in the downstream detectors.
In particular, this occurs in the STRAW spectrometer which makes use of a large 200\,ns readout window.
If two downstream charged particles are reconstructed 
and both match a parent $K^+$, the one closer to the trigger time is accepted. 
The same trigger time requirement is applied independently to each detector
signal matched with beam and downstream tracks.
Further conditions are applied to suppress $K^+$ decay like $K^+\rightarrow\pi^+\pi^+\pi^-$:
no more than two tracks reconstructed in the STRAW are allowed in total; 
if there are two tracks, both must be positively charged and should not form a vertex with a $Z$-position between GTK3 and the first STRAW station.

\begin{figure}[t]
  \begin{center}
  \begin{minipage}{18pc}
  \includegraphics[width=18pc]{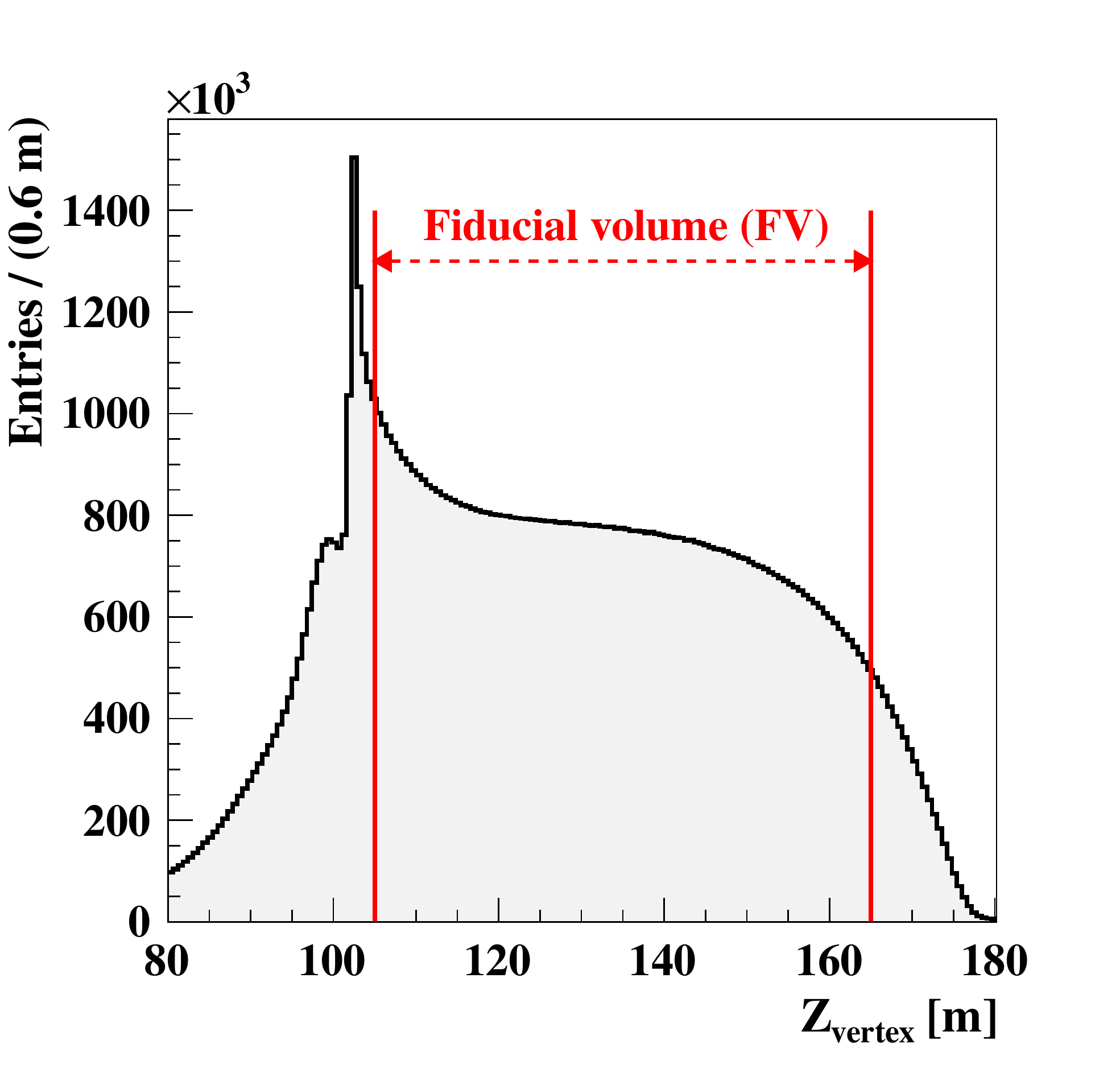}
  \end{minipage}\hspace{1pc}
  \begin{minipage}{18pc}
  \includegraphics[width=18pc]{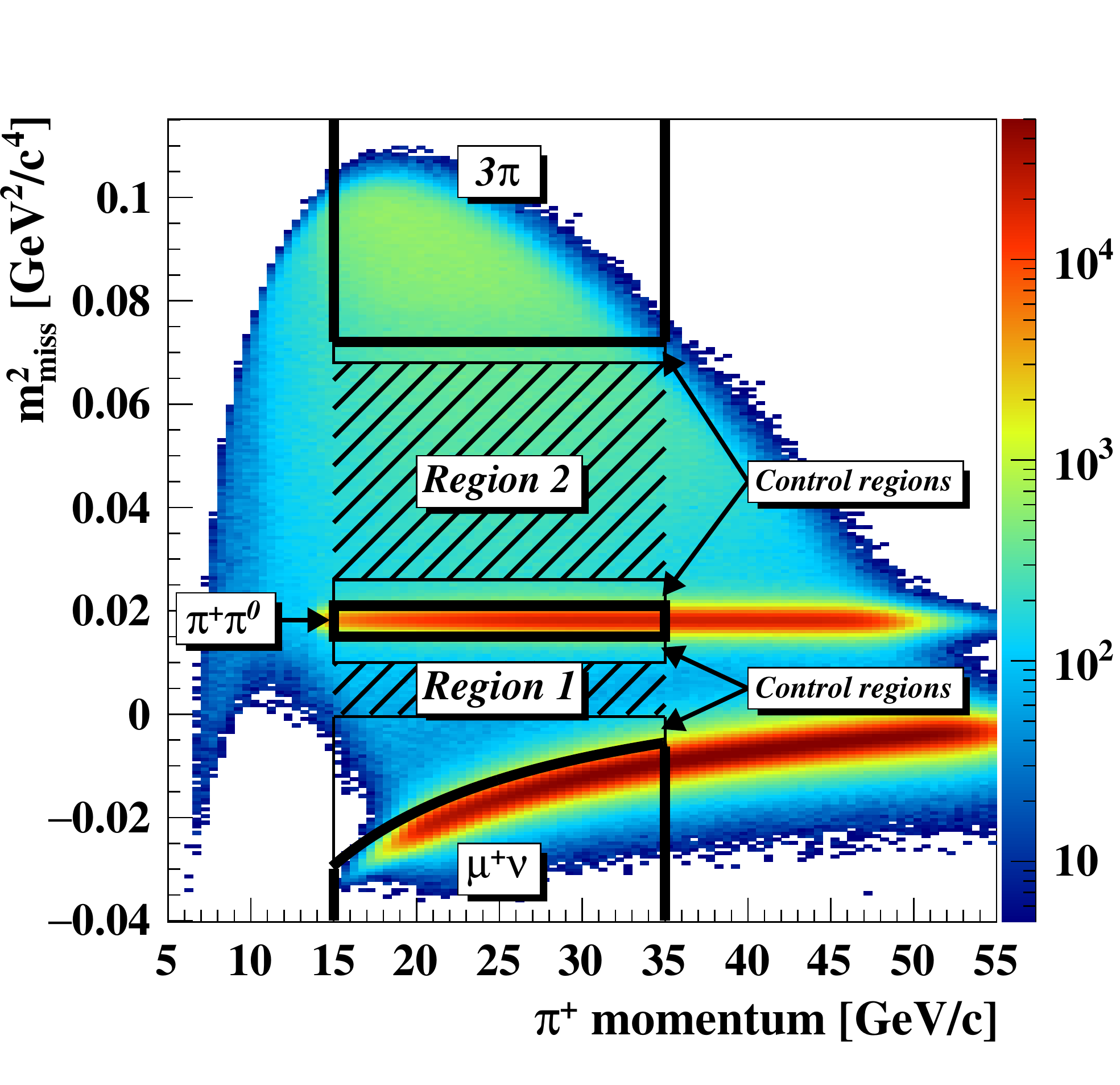}
  \end{minipage}\hspace{1pc}
  \caption{\label{fig:zvtx}{\bf Left}: distribution of the longitudinal position of the reconstructed decay vertex.
 The FV is defined between 105 and 165\,m (vertical red lines).
                                    {\bf Right}: reconstructed $\mmis$ as a function of the decay particle momentum
                                    for minimum-bias events selected without applying \pic identification and photon rejection, assuming the \kp and \pic mass for the parent and decay particle, respectively.
                                      Signal regions 1 and 2 (hatched areas), as well as  $3\pi$, $\pi^+\pi^0$, and $\mu^+\nu$ background regions  (solid thick contours) are shown. 
                                      The control regions are located between the signal and background regions.} 
\end{center}
\end{figure}
Figure~\ref{fig:zvtx} (left) displays the distribution of the longitudinal position ($Z_{vertex}$) of the reconstructed decay vertex of a $K^+$ decay. 
The events with $Z_{vertex}<100$\,m mostly originate from $K^+$ decays upstream of the final collimator.
The peaking structure starting 
at about 100\,m is due to nuclear interactions of beam particles grazing the edges of the final collimator or passing through the last station of the beam tracker located at 103\,m.
Charged particles created by decays upstream of the final collimator or by nuclear interactions can reach the detectors downstream and create fake $K^+$ decays.
To mitigate this effect, the decay vertex  
 is required to lie within a fiducial volume (FV) defined as 105\,m to 165\,m from the target.
The coordinates of this vertex must also 
be consistent with the beam envelope.
Wrong or accidental associations or mis-reconstruction of the $Z_{vertex}$ can shift the origin of these events within the FV,  imitating a $K^+$ decay.

Cuts on the direction of the decay particles as a function of $Z_{vertex}$ are applied to reduce the number of events reconstructed within the FV, but which actually originated upstream  (section~\ref{sec:ups}).
These cuts are also useful against
 $K^+\rightarrow\pi^+\pi^+\pi^-$ decays with only one $\pi^+$ reconstructed.
The CHANTI detector further protects the FV against nuclear interactions by vetoing events 
with CHANTI  signals within 3\,ns of the decay particle candidate.
Extra pulses in at least two GTK stations in time with the $K^+$ candidate may indicate that the $K^+$ has decayed before entering the decay region.
In this case the event is rejected if at least one pileup track is reconstructed in the beam tracker in addition to the $K^+$ candidate.
Finally, events are also discarded if the  decay particle track points back to the active area of GTK3.

\subsection{Kinematic regions}
\label{sec:kindef}
Figure~\ref{fig:zvtx}~(right) shows the $\mmis$ distribution as a function of the decay particle momentum
for $K^+$ decays selected as above 
from minimum-bias data.
Here, the $\mmis$ quantity is computed  using the three-momenta measured by the beam tracker and the STRAW spectrometer, assuming   $K^+$  and $\pi^+$ masses. 
Events from $K^+\rightarrow\pi^+\pi^0$ and $K^+\rightarrow\mu^+\nu$ decays accumulate at $\mmis=m^2_{\pi^0}$ and $\mmis<0$, respectively.
Events above $\mmis=4m^2_{\pi^+}(4m^2_{\pi^0})$ are mostly $K^+\rightarrow\pi^+\pi^+\pi^-(\pi^+\pi^0\pi^0)$ decays.
The shape of the region at low momentum arises from the $Z_{vertex}$ cuts.

The  $\mmis$ resolution varies with  $\mmis$ and is about $10^{-3}\,\text{GeV}^2/c^4$ at the $K^+\rightarrow\pi^+\pi^0$ peak.
This sets the definition of the boundaries of signal region 1 and 2:   

{\bf Region 1}: $0<\mmis<0.01\,\text{GeV}^2/c^4$;

{\bf Region 2}: $0.026<\mmis<0.068\,\text{GeV}^2/c^4$.\\
Additional momentum-dependent constraints supplement this definition by selecting 
$\mmis$ values computed using either
the decay particle momentum measured by the RICH under the $\pi^+$ mass hypothesis instead of the STRAW momentum,
or the nominal beam momentum and direction instead of those measured by the GTK tracker.
These requirements are intended to reduce the probability of wrong reconstruction of the $\mmis$ quantity due to a mis-measurement of the momenta of the decay particle or $K^+$ candidate. 

The momentum of the decay particle in the range  $15-35$\,GeV$/c$ 
 complements the definition of the signal regions.
The $\pi^+$ Cherenkov threshold of the RICH sets the lower boundary at 15\,GeV$/c$.
The $K^+\rightarrow\mu^+\nu$ kinematics and the requirement of  
a large missing energy drive the choice of the 35\,GeV$/c$ upper boundary.
The two signal regions are kept masked (blind) until the completion of the analysis.

In addition to the signal regions, three exclusive {\it background regions} are defined:

{\bf The \boldmath $\mu\nu$ region}: $-0.05< \mmis<  m^2_{\mu-kin}+3\sigma$, where $m^2_{\mu-kin}$ is the $\mmis$ of the $K^+\rightarrow\mu^+\nu$ decays under the $\pi^+$ mass hypothesis and $\sigma$ its resolution;

{\bf The \boldmath $\pi^+\pi^0$ region}: $0.015 <\mmis< 0.021\,\text{GeV}^2 / c^4$;

{\bf The \boldmath $3\pi$ region}: $0.072 < \mmis < 0.150\,\text{GeV}^2 / c^4$.\\
Once photons, muons and positrons are rejected (sections~\ref{sec:pid} and \ref{sec:sig}), simulations show that solely $K^+\rightarrow\mu^+\nu$, $K^+\rightarrow\pi^+\pi^0$ and
 $K^+\rightarrow\pi^+\pi^+\pi^-$ decays populate these regions, respectively.

Regions of the $\mmis$  distribution between signal and background regions, referred to as {\it control regions}, are masked until backgrounds are estimated and then used to validate the estimates.  Two regions around the $\pi^+\pi^0$ peak, for the $\pi^+\pi^0$  background, and one region each for the $K^+\rightarrow\mu^+\nu$ and $K^+\rightarrow\pi^+\pi^+\pi^-$  backgrounds, are identified.
Both background and control regions are restricted to the $15-35$\,GeV/$c$ $\pi^+$ momentum range
for consistency with the definition of the signal regions.

\subsection{Pion identification} 
\label{sec:pid}

The PNN trigger (section~\ref{sec:detector}) discards kaons decaying to muons by vetoing events with a signal in the MUV3 detector. 
A similar requirement applied offline reinforces the trigger condition, recovering possible online veto inefficiencies and makes the $\pi^+$ identification in minimum-bias and PNN data identical.
Muons may fail to be detected by MUV3 because of inefficiency or catastrophic interaction in the calorimeter, or if they decay upstream.

Pions can be distinguished from muons and positrons using information from the LKr calorimeter and, should any be present, from the MUV1 and MUV2 hadronic calorimeters.
A multivariate classifier resulting from a Boosted Decision Tree algorithm, BDT,  combines 13 variables  characterizing the calorimetric energy depositions.
A first group of variables consists of the ratios between the calorimetric energy deposited and the particle  momentum measured in the STRAW.
The energy in the LKr is used alone,  and in combination with the hadronic energies.
A second group of variables describes the longitudinal and transverse development of the calorimetric showers. 
The energy sharing between LKr, MUV1 and MUV2 provides information about the longitudinal shape of the energy deposition,  and 
the shape of the clusters characterizes the transverse size of the shower.
Finally, the BDT makes use of the distance between the particle impact point and the reconstructed cluster position.
The BDT training is performed using  samples of $\mu^+$, $\pi^+$ and $e^+$ selected from minimum-bias data recorded in 2016 and not used in the present analysis. 
The BDT returns the probability for a particle to be a $\pi^+$, a $\mu^+$, or a positron.
Pion identification requires the $\pi^+$ probability to be  larger than a minimum value that depends on the particle  momentum and is optimised with data.

\begin{figure}[t]
\begin{center}
 \begin{minipage}{18pc}
  \includegraphics[width=18pc]{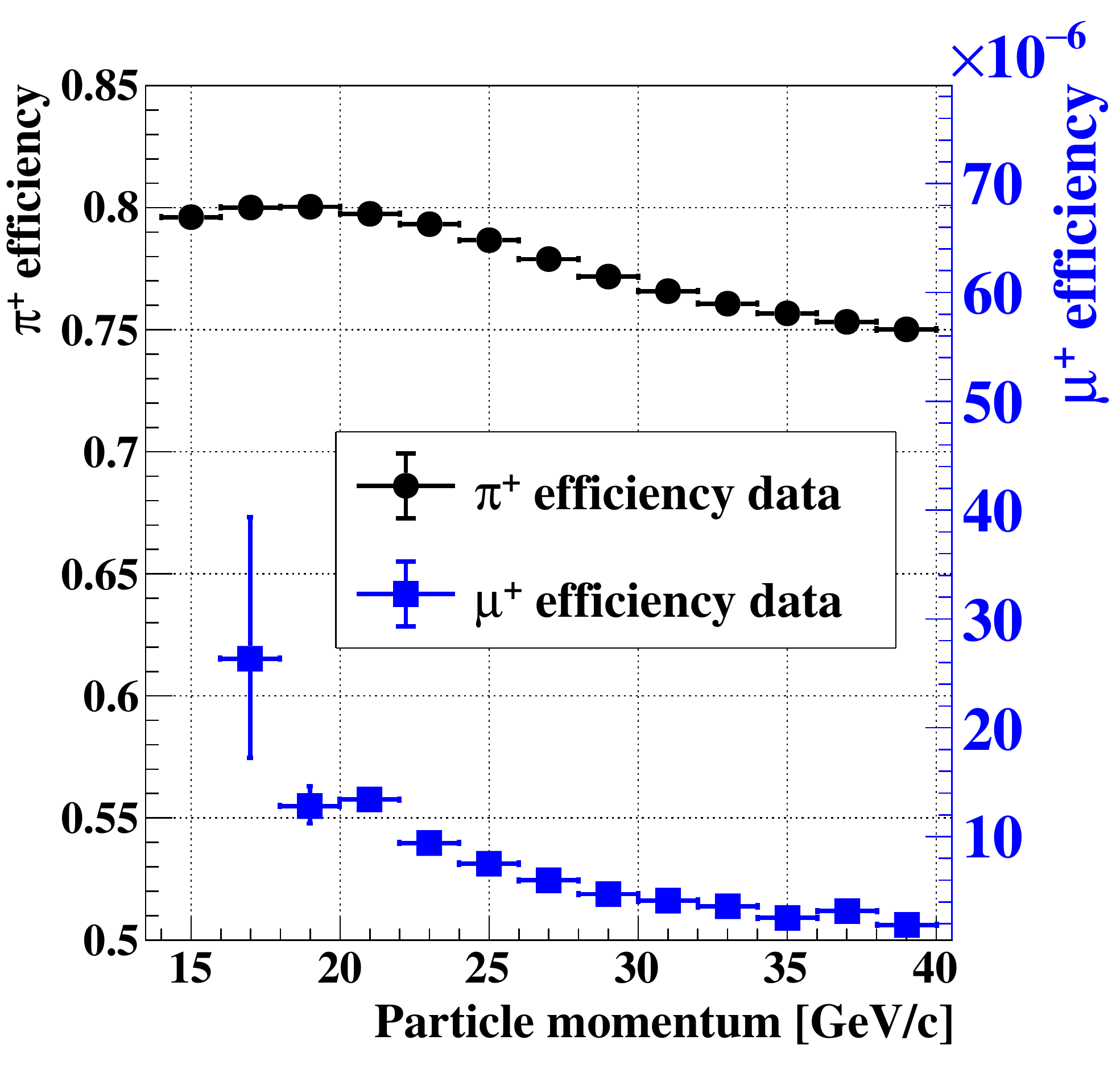} 
 \end{minipage}\hspace{1pc}
 \begin{minipage}{18pc}
  \includegraphics[width=18pc] {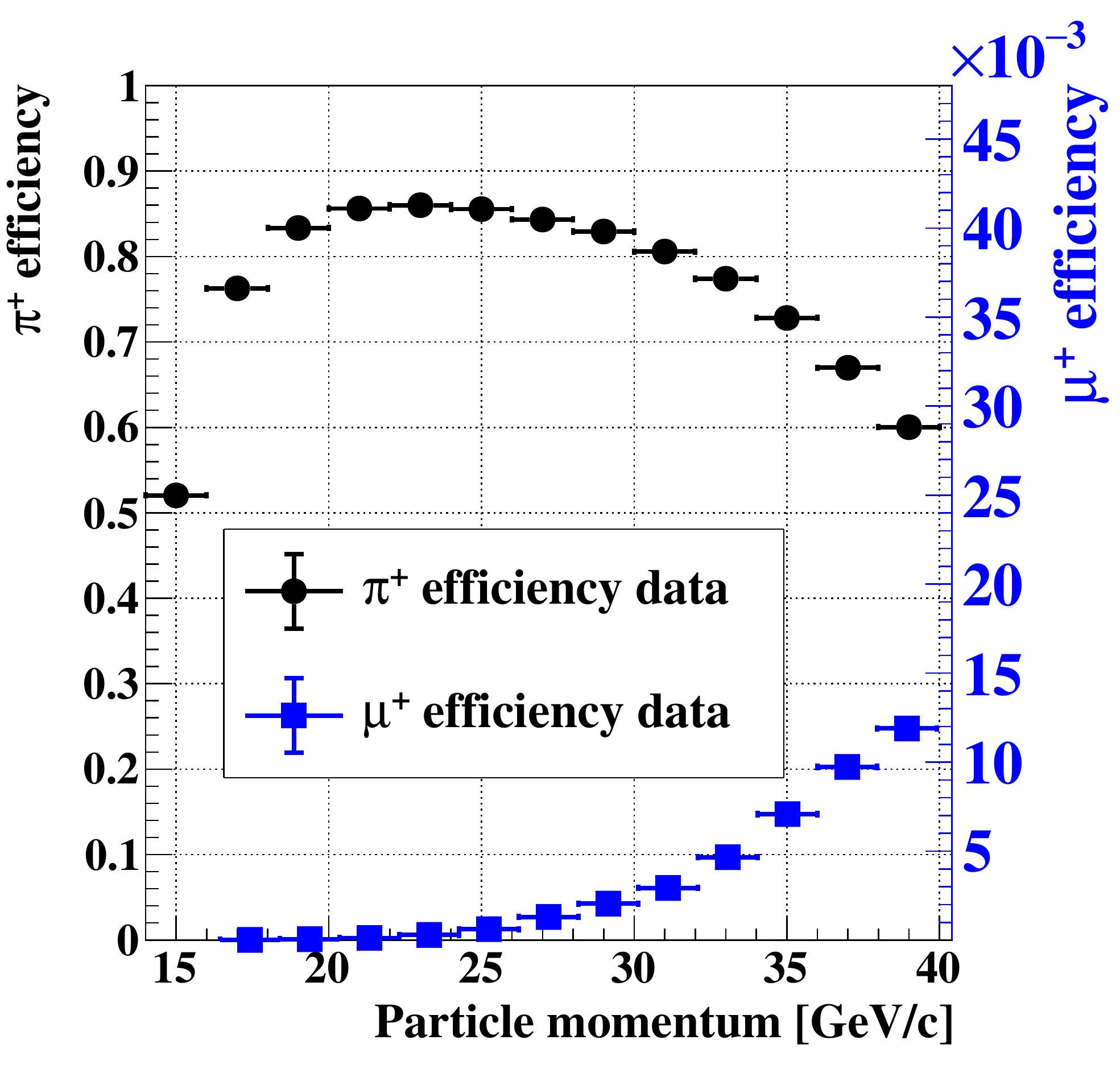} 
 \end{minipage}
  \caption{\label{fig:pid}Performance of the $\pi^+$ identification using calorimeters (left) and RICH (right) measured on data.
                Performance is quantified in terms of $\pi^+$ and $\mu^+$ efficiency, defined as the fraction of pions and muons passing the pion identification criteria, respectively.
                These criteria  include the corresponding RICH and calorimeter reconstruction efficiency.
                On each plot, the $\pi^+$ efficiency scale is shown on the left (black) vertical axis, the  $\mu^+$  
efficiency (misidentification) scale is shown on the right (blue) vertical axis.}
  \end{center}
\end{figure}

Samples of $K^+\rightarrow\pi^+\pi^0$ and $K^+\rightarrow\mu^+\nu$ decays selected from minimum-bias data are used to monitor the performance of the  $\pi^+$ identification efficiency 
and resulting $\mu^+$ misidentification probability,  
shown in Figure~\ref{fig:pid} (left).

Finally, the RICH separates $\pi^+$, $\mu^+$ and $e^+$ independently of the calorimeter responses. 
The reconstructed mass and the likelihood of the particle must be consistent with the $\pi^+$ hypothesis.
Figure~\ref{fig:pid} (right) shows the performance of the $\pi^+ / \mu^+$ separation using the RICH as a function of the particle momentum,  evaluated using data. 

The $\pi^+$ identification is required for both  signal $K^+ \to \pi^+ \nu \bar{\nu}$ and normalization $K^+ \to \pi^+ \pi^0$ selections.

\subsection{Signal selection} 
\label{sec:sig}
Additional requirements are applied to PNN data
to reject events with in-time  photons or non-accidental additional charged particles in the final state that are compatible with a physics process producing the downstream $\pi^+$.

Photon rejection discriminates against partially reconstructed $K^+\rightarrow\pi^+\pi^0$ decays.
An extra in-time photon in the LKr calorimeter is defined as a cluster located at least 100\,mm away from the $\pi^+$ impact point and 
within a cluster energy-dependent time coincidence with the $\pi^+$ time that ranges from $\pm5$\,ns below 1\,GeV to $\pm$50\,ns above 15\,GeV.
Pileup clusters can overlap in space with the photon to be rejected, spoiling the time of such a photon 
by as much as several tens of ns.
The choice of a broad timing window at high energy keeps the detection inefficiency below $10^{-5}$.

An extra  in-time photon in the LAV detector is defined as any signal in a LAV station within $\pm3$\,ns of the $\pi^+$ time.                 
Appropriate combinations of the TDC leading and trailing edges of the high and low threshold channels define a LAV signal \cite{na62det}.
A similar method identifies photons in the small angle calorimeters IRC and SAC, using a time-window of $\pm7$\,ns around the $\pi^+$ time.
In addition to the signals from the TDC readout, photon rejection in IRC and SAC exploits the FADC readout; here,  a photon signal is defined as an energy deposit larger than 1\,GeV in a $\pm7$\,ns time window.
 
Multiplicity rejection discriminates against tracks produced by photons interacting in the material before reaching the calorimeters, and against  tracks from $K^+\rightarrow\pi^+\pi^+\pi^-$ decays partially reconstructed in the STRAW.
The first category of charged particles is expected to leave signals in the detectors downstream of the STRAW. 
The rejection criteria exploit the time and spatial coincidence of isolated signals reconstructed in at least two of the CHOD, NA48-CHOD and LKr detectors.
In-time signals in the peripheral detectors MUV0 and HASC are also included. 
The second category of charged particles is  characterized by the presence of  
track segments, defined as  pairs of signals in the first-second or third-fourth STRAW stations and consistent with a particle coming from the FV.
\begin{figure}[t]
  \begin{center}
  \begin{minipage}{18pc}
   \vspace{-25mm}
 \includegraphics[width=18pc]{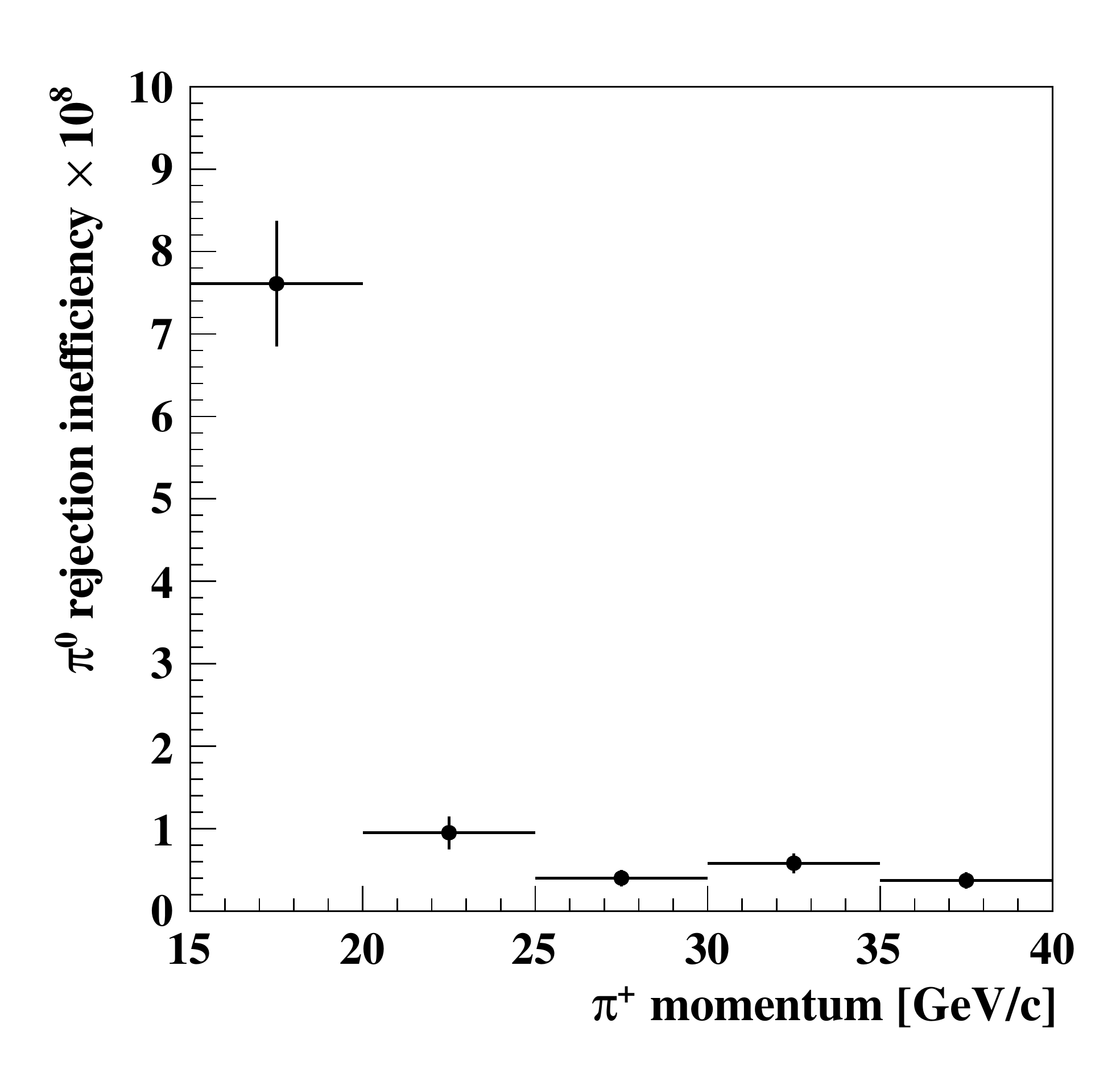} 
 \caption{\label{fig:pp0rej} 
  Rejection inefficiency of the $\pi^0$ from $K^+\rightarrow\pi^+\pi^0$ decays as a function of the $\pi^+$ momentum. The quoted uncertainties are statistical only.}
   \end{minipage}\hspace{1pc}
  \begin{minipage}{18pc}
  \includegraphics[width=18pc]{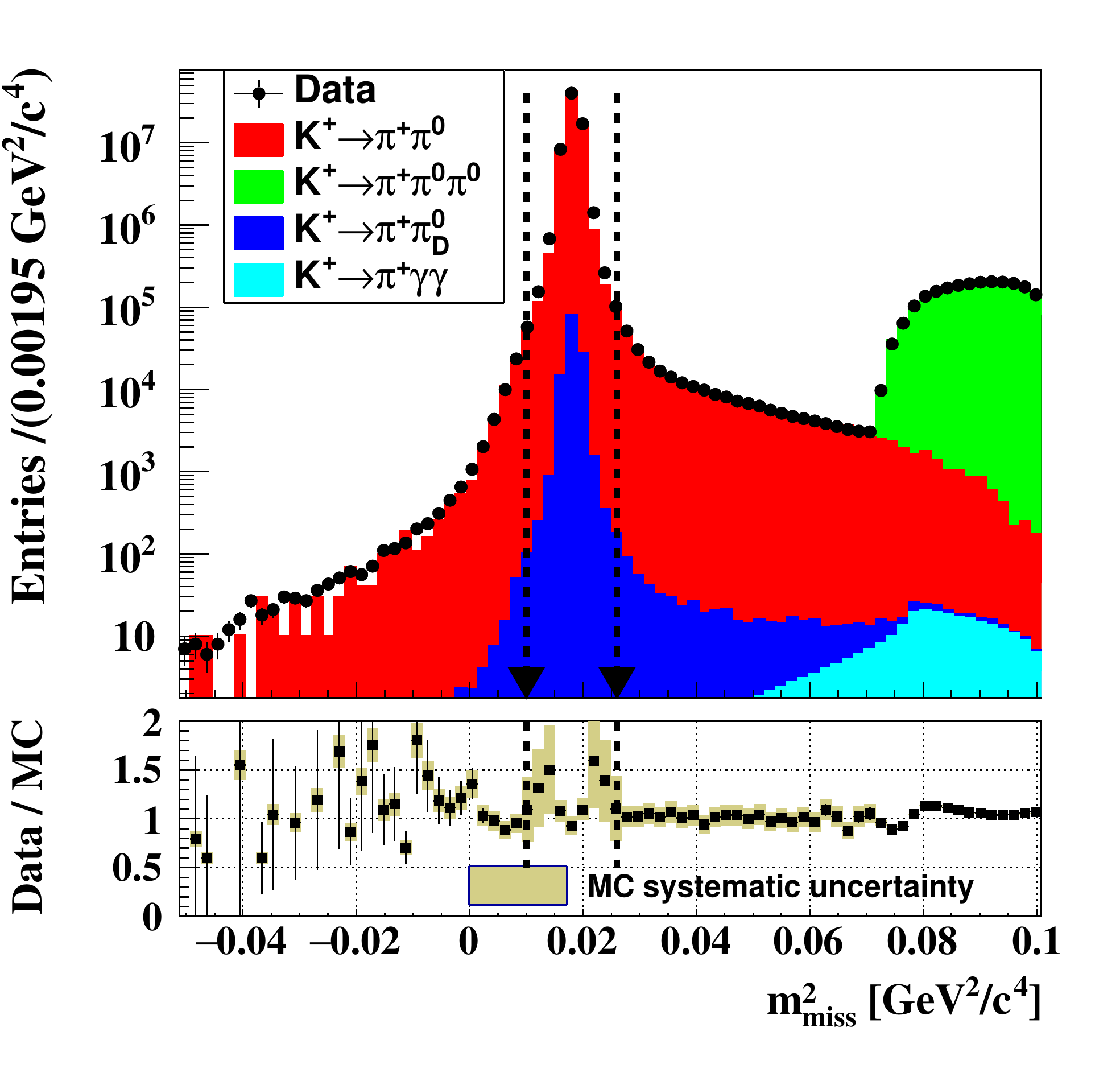} 
  \caption{\label{fig:pp0sel}
 Distribution of the $\mmis$ of events selected from minimum-bias data for normalization. 
 Data and  MC simulation are superimposed. 
 The bottom insert shows the data/MC ratio.
  The error bars correspond to the statistical uncertainty of the ratio, the yellow band is the systematic uncertainty due to the imperfect simulation of the detector response.}
  \end{minipage}
  \end{center}
\end{figure}

The reduction of reconstructed $K^+\rightarrow\pi^+\pi^0$ decays 
quantifies the performance of the photon and multiplicity rejection.
The number of PNN events 
in the  $\pi^+\pi^0$ region
remaining after rejection is compared to the number of minimum-bias events 
in the same region before rejection.
The ratio of these two numbers, corrected for the minimum-bias downscaling factor (section~\ref{sec:pnnses}) and trigger efficiency (section~\ref{sec:trigeff}),  is  
the rejection inefficiency of the $\pi^0$ produced in $K^+\rightarrow\pi^+\pi^0$ decays.
This inefficiency depends on the $\pi^+$ momentum and is about $1.3\times10^{-8}$ on average, as shown in Figure~\ref{fig:pp0rej}. 
The measured $\pi^0$ rejection can be explained in terms of single-photon detection inefficiencies in the LKr, LAV, IRC and SAC calorimeters which are measured from a sample of minimum-bias 
$K^+ \to \pi^+ \pi^0$ data using a tag-and-probe method. 
The estimated $\pi^0$ efficiency stems from the measured single-photon detection efficiencies convoluted with simulated $K^+ \to \pi^+ \pi^0$ decays 
and is in agreement with the measured $\pi^0$ efficiency within the statistical uncertainty.
The rise at low $\pi^+$ momentum is a consequence of lower detection efficiency for photons travelling close to the beam axis 
and interacting with the beam pipe.

In addition to the photon and multiplicity rejection, the \pnnc selection enforces specific requirements   
against particles entering the FV from upstream.
The $\pi^+$ track is extrapolated back to the $Z$-position of the final collimator and the $X,Y$ transverse 
coordinates are required  to be 
outside of a box with $|X| < 100$~mm and $|Y| < 500$~mm.
This cut removes a region with weaker shielding against particles coming from upstream and 
corresponds to the central aperture of the last dipole magnet of the beam line (section~\ref{sec:ups}).
This condition is referred to as the {\it box cut} in the following sections.

Finally, signal selection requires the $\mmis$  value 
to be within the signal regions defined in section~\ref{sec:kindef}.
The set of criteria described  in this section   
is  called {\it PNN selection} in the following.

\subsection{Normalization selection}  
\label{sec:norm}
The $K^+\rightarrow\pi^+\pi^0$ decays  used for normalization are selected from minimum-bias data, as defined in sections~\ref{sec:fv} and \ref{sec:pid},  and 
their $\mmis$ value must be in the 0.01--0.026 GeV$^2/c^4$ range.
Figure~\ref{fig:pp0sel}   shows the $\mmis$ spectrum of these events before the 
$\mmis$  cut,  together with the simulated distribution.
The shape of the $K^+\rightarrow\pi^+\pi^0$ peak depends on the resolution of the STRAW spectrometer, on multiple scattering in the tracker material, on the rate of  pileup tracks,
and on the calibration of the beam and STRAW trackers. 
The uncertainty in the simulation of these effects affects 
the data/MC agreement in the peak region only, and is taken into account in the evaluation of the $SES$ (section~\ref{sec:pnnses}).
The overall background under the peak is at the 
one part per thousand level and stems from $K^+\rightarrow\pi^+\pi^0$ decays with $\pi^0\rightarrow e^+e^-\gamma$.


\section{Single Event Sensitivity determination}
\label{sec:pnnses}
Denoting $N_{K^+}$ the number of kaon decays occurring in the FV, the single-event sensitivity ($SES$) of the present data sample to \pnnc can be written as 
\begin{equation}
 \label{eq:ses2}
   SES=\frac{1}{N_{K^+}\cdot\epsilon_{\pi\nu\nu}\cdot\epsilon_{trig}^{\scriptscriptstyle PNN}}= \frac{\text{BR}(K^+\rightarrow\pi^+\pi^0)}{D\cdot N_{\pi\pi}}
   \frac{\epsilon_{\pi\pi}\cdot\epsilon_{trig}^{\scriptscriptstyle MB}}
    {\epsilon_{\pi\nu\nu}\cdot\epsilon_{trig}^{\scriptscriptstyle PNN}}~\cdot
\end{equation}
Here $N_{\pi\pi}$ is the number of $K^+\rightarrow\pi^+\pi^0$ events reconstructed in the FV from minimum-bias data (section~\ref{sec:norm}), also called normalization events;
$D$ is the reduction, or down-scaling, factor applied online to reduce the
minimum-bias contribution to the total trigger rate;
$\epsilon_{\pi\nu\nu}$ and $\epsilon_{\pi\pi}$ are the efficiencies to identify a \pnnc and a $K^+\rightarrow\pi^+\pi^0$ decay in the FV, also called signal and normalization efficiencies, respectively;
$\epsilon_{trig}^{\scriptscriptstyle PNN}$ and
$\epsilon_{trig}^{\scriptscriptstyle MB}$,  are the trigger efficiencies that account for the data loss after
the event selection due to the PNN and minimum-bias triggers.
The efficiencies and $N_{\pi\pi}$ depend on the $\pi^+$ momentum, $p_{\pi}$,
and on the instantaneous beam intensity, $I$.  The $SES$ is consequently
computed in bins of $p_{\pi}$ and $I$:
the momentum range 15--35~GeV/$c$ is subdivided into  four bins of 5~GeV/$c$ width and the instantaneous beam intensity into  five bins of approximately the same statistics of \pp  normalization events.

\subsection{Number of  {\boldmath $K^+\rightarrow\pi^+\pi^0$} decays}
\label{sec:npp0ses}
The number of events satisfying the conditions described in
section~\ref{sec:norm} is $N_{\pi\pi}=68\times10^6$.

The $\pi^0$ mainly decays to $\gamma\gamma$, but in about 1\% of cases it
decays to $\gamma e^+e^-$, called a Dalitz decay ($\pi^0_D$).  The relative impact of
$\pi^0_D$ decays on the $SES$ is estimated to be less than 0.3\% and is
assigned as systematic uncertainty.  In the following sections
$K^+\rightarrow\pi^+\pi^0$ refers only to $\pi^0\rightarrow\gamma\gamma$ decays.

\subsection{Signal and normalization efficiencies}
\label{sec:seleff}
The efficiencies $\epsilon_{\pi\nu\nu}$ and $\epsilon_{\pi\pi}$ quantify the
effects of reconstruction and selection (section~\ref{sec:evsel}) on the
counting of signal and normalization channels.  Event losses can be grouped
into 6 classes:
\begin{enumerate} 
\item geometric and kinematic acceptances;
\item reconstruction of the $K^+$ and of the downstream charged particle;
\item matching the $K^+$ with the downstream charged particle;
\item $\pi^+$ identification by the RICH and calorimeters; 
\item decay region definition; and
\item selection criteria unique to the $K^+\rightarrow\pi^+\nu\bar{\nu}$ mode.
\end{enumerate}
The impact of these effects on $\epsilon_{\pi\nu\nu}$ and $\epsilon_{\pi\pi}$
depends on the kinematics of the decay, detector resolutions and efficiencies,
and the accidental presence of unassociated particles in an event.

The kinematics of the decays are studied with simulations, while detector
performance is studied with data and either reproduced by simulation or
factored out from $\epsilon_{\pi\nu\nu}$ and $\epsilon_{\pi\pi}$.

Accidental particles have a twofold effect.
They affect detector response and therefore the reconstruction of $K^+$ decays
and kinematic resolution.  They also randomly satisfy conditions in the
GTK, CHANTI, STRAW, MUV3, calorimeters, CHOD, and NA48-CHOD that lead to an
event being rejected, referred to here as a random veto.  The first effect is
modelled with simulation.  The second effect, which is independent of decay mode
topology, is measured directly with data as a function of the instantaneous beam intensity and factored out of
$\epsilon_{\pi\nu\nu}$ and $\epsilon_{\pi\pi}$.

As a consequence, signal and normalization efficiencies may take the form:
\begin{equation}
	\epsilon_{\pi\nu\nu}=\epsilon^{MC}_{\pi\nu\nu}\cdot
	\epsilon^{Random}_{\pi\nu\nu}~~~~~~~~\epsilon_{\pi\pi}=
	\epsilon^{MC}_{\pi\pi}\cdot\epsilon^{Random}_{\pi\pi}.
\end{equation}
The Monte Carlo efficiency, $\epsilon^{MC}_{decay}$, quantifies the effects of
the factors listed above, except for random losses, 
and the random efficiency, $\epsilon^{Random}_{decay}$, quantifies the fraction of
events randomly lost because of 
the accidental presence of at least one veto condition.

The $SES$ depends only on the ratio of the $\epsilon_{\pi\nu\nu}$ and $\epsilon_{\pi\pi}$  
efficiencies. As both signal and normalization channels contain a $\pi^+$ in the final state, 
the ratio effectively cancels significant components of the two efficiencies, decreasing the dependence of the $SES$ on their magnitude and reducing significantly their contribution to the SES uncertainty.

\subsubsection{Monte Carlo efficiencies}
\label{sec:deteff}
The Monte Carlo efficiency, $\epsilon^{MC}_{decay}$, is the ratio of the
number of simulated events passing signal or normalization selection to the
corresponding number of generated events in the FV. 

\begin{figure}[t]
  \begin{center}
  \begin{minipage}{18pc}
  \includegraphics[width=18pc]{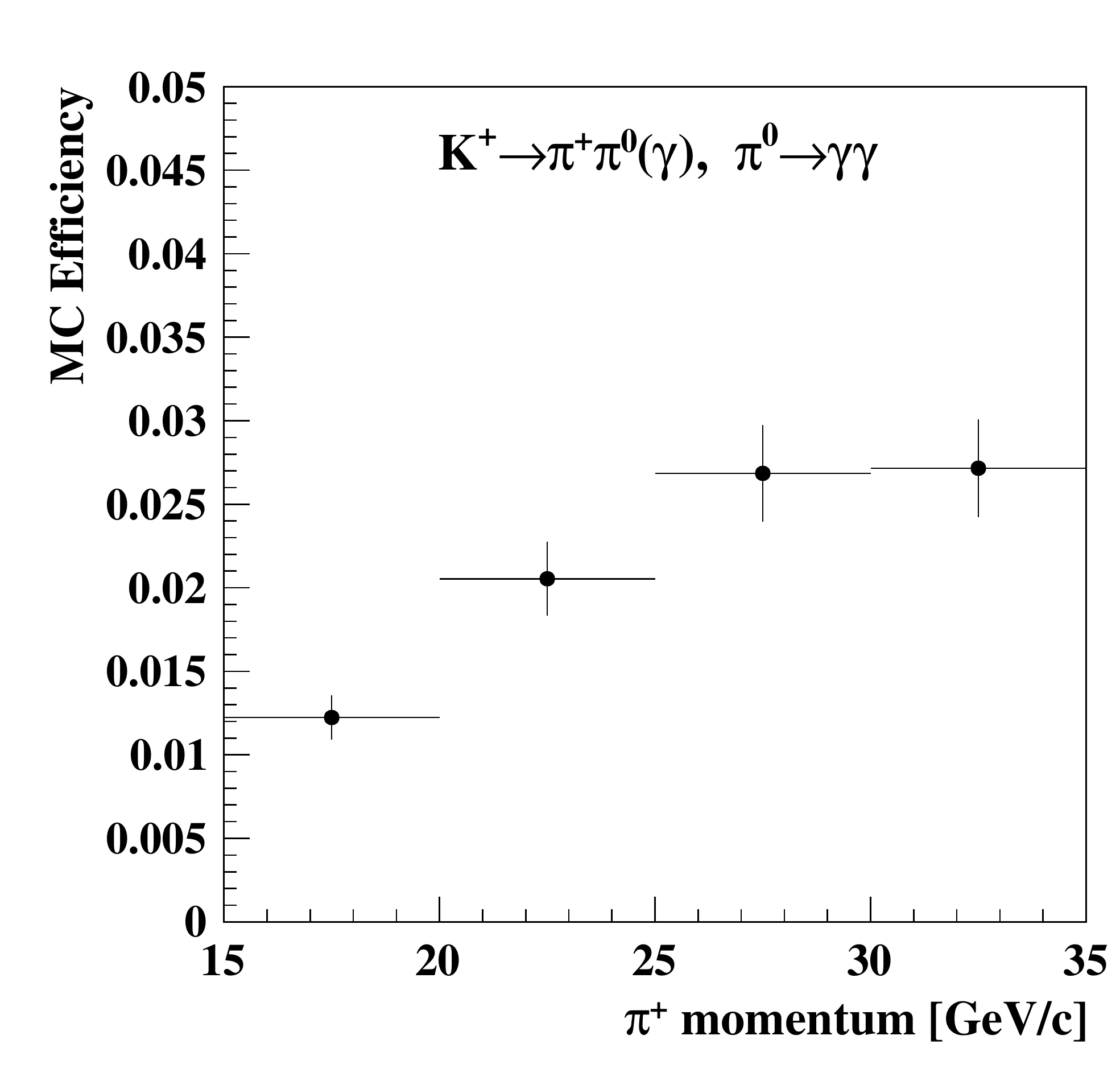}
  \end{minipage}\hspace{1pc}
  \begin{minipage}{18pc}
  \includegraphics[width=18pc]{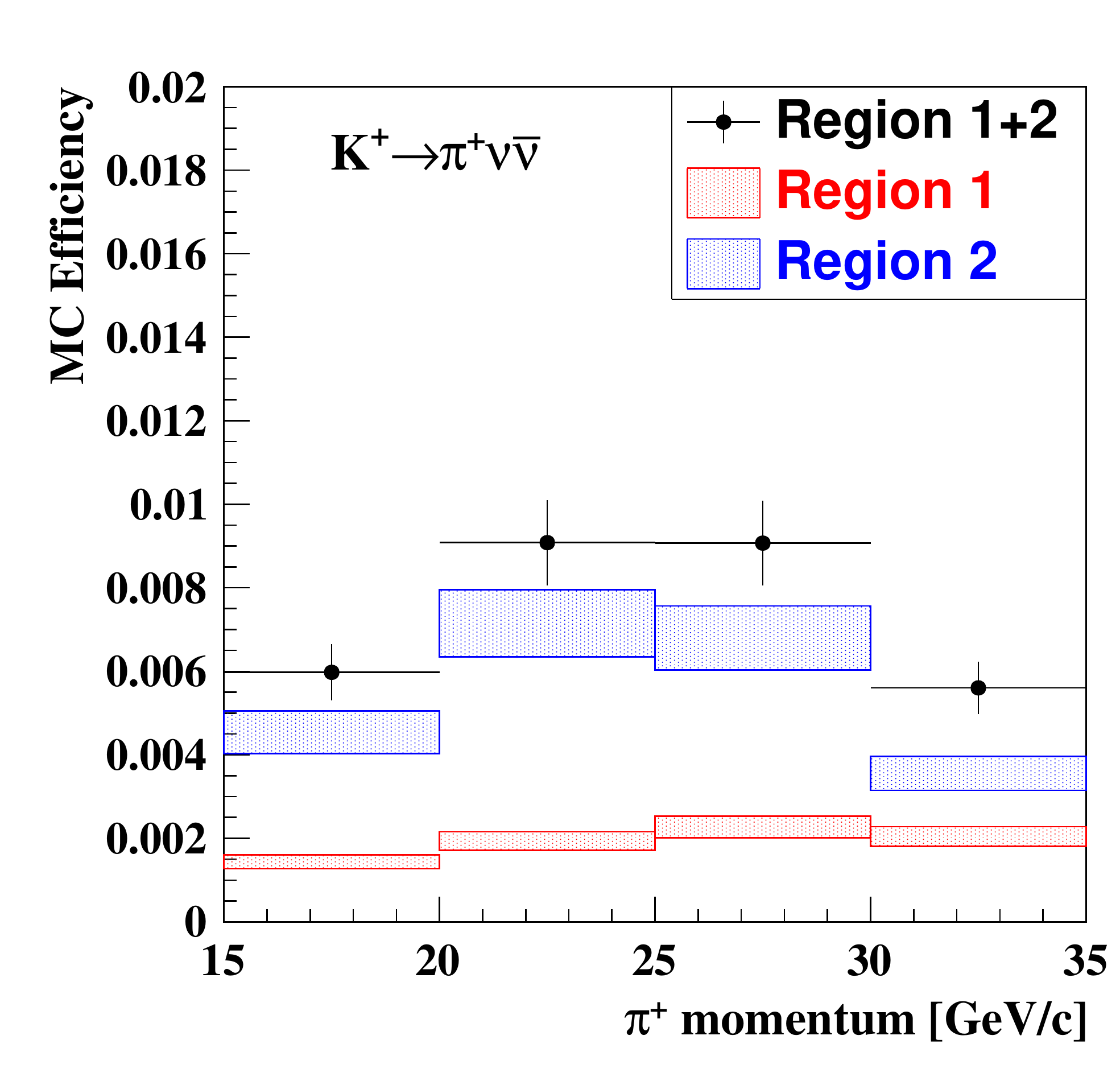}
  \end{minipage}
  \caption{\label{fig:accept}{\bf Left}: $K^+\rightarrow\pi^+\pi^0$ MC
	  efficiency in independent bins of $\pi^+$ momentum.
	  {\bf Right}:
	  $K^+\rightarrow\pi^+\nu\bar{\nu}$ MC efficiency  
	  in independent bins of $\pi^+$ momentum.  The efficiencies in regions 1 and 2
	  are shown separately and summed (full symbols).  The width of the coloured bands represents the
	  uncertainties in the measured values.}
  \end{center}
\end{figure}

Figure~\ref{fig:accept}  shows the values of
$\epsilon^{MC}_{\pi\pi}$ and $\epsilon^{MC}_{\pi\nu\nu}$ in bins of
$\pi^+$ momentum.  The sums over all bins are $0.087\pm 0.009$ and
$0.030\pm 0.003$, respectively.  The uncertainties are systematic, due mainly to the accuracy
of $\pi^+$ identification and $K/\pi$ track matching in the simulation.

Table~\ref{tab:effic} shows estimates of the contributions to
$\epsilon^{MC}_{\pi\nu\nu}$ and $\epsilon^{MC}_{\pi\pi}$ of the components
listed in section~\ref{sec:seleff}.  The values in the table are
approximated, due to correlations among the components.  A 10\% relative
uncertainty is assigned to each component and conservatively considered as 100\% correlated.  The difference between
$\epsilon^{MC}_{\pi\nu\nu}$ and $\epsilon^{MC}_{\pi\pi}$ is attributable to
differences in acceptance, particle reconstruction, and cuts specific to the
signal channel.  The accuracy with which these factors are simulated is the
primary source of uncertainty in the $SES$.
\begin{table}[!htbp]
  \begin{center}
  \caption{\label{tab:effic}Monte Carlo efficiencies for normalization
	  and signal decay modes.  
	  The uncertainties in the total efficiencies are
	  systematic and reflect the accuracy of the simulation.}  
    \begin{tabular}{l|c|c}
     \toprule
      Source              & $K^+\rightarrow\pi^+\pi^0$ & \pnnc \\\midrule
      Acceptance                   & $0.27$ & $0.16$ \\
      Particle reconstruction & $0.64$ & $0.70$ \\
      $K^+$ matching           & $0.84$ & $0.84$ \\
      $\pi^+$ identification   & $0.72$ & $0.72$ \\
      Decay region selection  & $0.83$ & $0.81$ \\
      \pnnc selection             & $-$ & $0.55$ \\\midrule
      Total                       & $0.087\pm 0.009$ & $0.030\pm 0.003$ \\
      \bottomrule
    \end{tabular}
  \end{center}
\end{table}
The next paragraphs focus on the contributions from each component of the Monte Carlo efficiencies listed in Table~\ref{tab:effic}. \\\\

{\bf \noindent Acceptance}\\
Events fail to be selected because of detector geometry as well as restrictions on the $\pi^+$ 
momentum and  $\mmis$ ranges. 
 The effects of these three factors are different for
signal and normalization selection efficiencies and are therefore a potential source of SES uncertainty.

The impact of the limited accuracy of the simulated $\mmis$ distribution has been quantified by recalculating 
the $SES$ with $K^+\rightarrow\pi^+\pi^0$ decays  in a smaller $\mmis$ region,  (0.015,0.021)\,$\text{GeV}^2/c^4$, 
where data and MC marginally agree (section 5.7). The corresponding variation of the SES is approximately 1\% and
 assigned as systematic uncertainty due to the simulation of the $\mmis$. 

Detector illumination and the momentum spectrum contribute to a lesser
extent to the difference between signal and normalization acceptances.
The accuracy of the simulation with respect to these is ascertained by
measuring the branching ratio of $K^+\rightarrow\mu^+\nu$ normalized to
$K^+\rightarrow\pi^+\pi^0$.  A systematic uncertainty is assigned after
comparing the result of this measurement to the accepted value (section~\ref{sec:brkm2}).
\\\\
{\bf Particle reconstruction}\\
The particle reconstruction efficiency is the product of
the KTAG and GTK efficiencies for reconstructing the parent $K^+$, and
the STRAW, RICH, CHOD, NA48-CHOD and LKr efficiencies for
reconstructing the daughter $\pi^+$.  The RICH, CHOD, NA48-CHOD and LKr
efficiencies include detector signal association with a STRAW track.

The effect of local inefficiencies due to detector readout or to accidental activity cancels 
at first order in the ratio of efficiencies, as signal and normalization decays are recorded
simultaneously. Nonetheless, these effects are measured with data and added to the simulation.
Table~\ref{tab:deteff} details the impact of the various subdetectors on the
reconstruction efficiency.  The numbers are averages over $\pi^+$ momentum
between 15 and 35\,GeV$/c$ and instantaneous beam intensity.
 \begin{table}[t]
  \begin{center}
  \caption{Average detector efficiencies over $\pi^+$ momentum and instantaneous beam intensity.  The uncertainties are estimated
  by comparing data and simulation, 
  and with systematic studies, such as checks of time stability.  The values for the RICH efficiency refer to pion identification efficiencies from
 \pnnc and $K^+\rightarrow\pi^+\pi^0$,  respectively.
	  All  other detector efficiencies are equal for signal and normalization.}  \label{tab:deteff}
    \begin{tabular}{l|l}
      \toprule
      Source & Efficiency \\\midrule
      KTAG    & $0.97\pm 0.01$ \\
      GTK detector    & $0.92\pm 0.02$ \\
      GTK reconstruction & $0.94\pm0.02$ \\
      STRAW & $0.95\pm 0.01$ \\
      RICH     & $0.95\pm 0.03$ $(0.87\pm 0.03)$ \\
      CHOD   & $>0.99$  \\
      NA48-CHOD & $>0.99$ \\
      LKr        & $0.95\pm 0.03$ \\
      \bottomrule
    \end{tabular}
  \end{center}
\end{table}

KTAG and GTK efficiencies refer to $K^+$ detection  and 
are equal for signal and normalization.  Both efficiencies are measured with data,
using $K^+\rightarrow\pi^+\pi^+\pi^-$ decays.  KTAG inefficiencies come mainly
from the readout.  
GTK inefficiencies arise from geometric acceptance and identified readout malfunctioning  (5\%) and  
from the detector (3\%)~\cite{gtkdet}.  
The GTK reconstruction efficiency is due to the conditions applied to identify a track of good quality.

The efficiency to reconstruct a $\pi^+$ track with the STRAW is measured with
$K^+\rightarrow\pi^+\pi^0$ decays in the data.  In the $15-35$\,GeV/$c$
momentum range, the efficiency depends only on the instantaneous beam intensity,
which is directly related to accidental activity in the detectors.

The RICH efficiency for reconstructing a $\pi^+$ with momentum between 15 and
35 GeV$/c$ is measured with data using $K^+\rightarrow\pi^+\pi^0$ decays.
It is directly related to the statistics of Cherenkov photons and depends
only on the $\pi^+$ momentum.  Simulation reproduces this efficiency with a
relative accuracy of about 3\%.  The simulation indicates that this efficiency
is about 7\% higher for \pnnc decays than for $K^+\rightarrow\pi^+\pi^0$
decays.  This difference is attributable to extra hits created when photons
from $\pi^0$ decay in $K^+\rightarrow\pi^+\pi^0$ events convert in RICH material and
spoil the charged pion ring shape.  Therefore, the RICH reconstruction efficiency
does not cancel in the ratio of Equation~(\ref{eq:ses2}).
A sample of $K^+\rightarrow\mu^+\nu$ decays in the data is used to test the
accuracy of RICH particle reconstruction in the simulation.  The resulting
ratio of data to MC agrees with that of $K^+\rightarrow\pi^+\pi^0$ to
within 1.5\%.  This value is assigned as a relative systematic uncertainty in
the $SES$ due to the simulation of the RICH reconstruction efficiency.

Measurements with data show that the CHOD and NA48-CHOD detectors are highly
efficient.  An overall 0.99 efficiency is assigned to account for small losses
in the association of detector signals with STRAW tracks that define downstream
charged particles.

The LKr calorimeter detects signals from minimum ionizing particles with an
efficiency  greater than 99\%, as measured with data.  
In the case of  $\pi^+$  inelastic hadronic interactions, an additional inefficiency may arise
 in associating LKr clusters with STRAW tracks.
\\\\
{\bf $\boldsymbol{K^+}$ matching}\\
The efficiency for matching a $K^+$ with a downstream charged particle is 0.84 and depends
on the GTK efficiency and on time and CDA resolutions.  The simulation
reproduces the matching performance measured with data to within 5\% relative
accuracy, once accidental pileup in the GTK and GTK efficiency are 
simulated.  This measurement of the accuracy is taken as a systematic
uncertainty in the magnitudes of both $\epsilon^{MC}_{\pi\nu\nu}$ and
$\epsilon^{MC}_{\pi\pi}$.  However, the effect of $K^+$ matching is equal
for signal and normalization, and therefore no corresponding uncertainty is
assigned to the $SES$.  As a cross check, the $SES$ is found to be nearly
insensitive to the simulated level of GTK inefficiency.
\\\\
{\bf $\boldsymbol{\pi^+}$ identification}\\
Not every  $\pi^+$ is identified due to the intrinsic efficiencies of
the RICH and calorimeters and to $\pi^+$ decays in flight.

The RICH efficiency for identifying undecayed $\pi^+$s from
$K^+\rightarrow\pi^+\pi^0$ events is measured with data and found to be about
0.95.  Simulation reproduces this number with 3\% accuracy and indicates that
$\pi^+$s from \pnnc decays are identified with a comparable efficiency.

Simulation reproduces the measured efficiency for the RICH to reconstruct and
identify a $\pi^+$ with an accuracy of about 6\%.  This value is assigned as a
relative uncertainty to $\epsilon^{MC}_{\pi\nu\nu}$ and
$\epsilon^{MC}_{\pi\pi}$.  However, no additional uncertainty is assigned to
the $SES$ beyond that from the RICH reconstruction efficiency, because the
RICH identification algorithm treats signal and normalization modes the same.

The average efficiency of $\pi^+$ identification with the calorimeters is
about 0.80, as measured with data.  Simulation reproduces this result with
2\% accuracy.  This degree of accuracy is propagated as a relative uncertainty
to $\epsilon^{MC}_{\pi\nu\nu}$ and $\epsilon^{MC}_{\pi\pi}$.  Simulation also
shows 
that the efficiencies to identify charged pions with the
calorimeters are the same for signal and normalization modes.  Therefore, the
accuracy of calorimeter simulations does not affect the $SES$ measurement.

The $\pi^+$ identification efficiencies reported in Table~\ref{tab:effic}
include an additional factor of 0.95 to account for the probability of $\pi^+$ decay.
\\\\
{\bf Decay region}\\
In addition to the definition of the 105--165~m FV, 
the decay region is shaped by the cuts on the $\pi^+$ direction as a function of 
$Z_{vertex}$ as discussed in section~\ref{sec:fv}.
These selection criteria reject a slightly different number of signal and normalization events.
The simulation accounts for the corresponding effect in the $SES$ together with the kinematic and geometric acceptances, as the various contributions are correlated. 
\\\\
{\bf  Signal $\boldsymbol{K^+\rightarrow\pi^+\nu\bar{\nu}}$ selection}\\
Photon and multiplicity rejection and the box cut are applied only to signal
events.  These selection criteria, therefore, directly impact the measurement of
the $SES$.

In the absence of random activity, the box cut alone rejects about 40\% of signal events. 
The GTK, CHANTI, STRAW, and MUV3 veto conditions do not affect $\epsilon^{MC}_{\pi\nu\nu}$ or
$\epsilon^{MC}_{\pi\pi}$.  On the other hand, because charged pions may interact
in RICH material, vetoing photons and extra charged particles can inadvertently
reject \pnnc events.  The accuracy with which the simulation models this
effect is studied by selecting from data $K^+\rightarrow\pi^+\pi^0$ events in
which both photons from the $\pi^0$ decay are detected in LAV stations.
The loss of events because of $\pi^+$ interactions is measured on these data and 
compared with simulation, leading to about 6\% discrepancy.
The efficiency
$\epsilon^{MC}_{\pi\nu\nu}$ is 
corrected for half of this difference. 
An uncertainty of 100\% is assigned to this correction factor,
resulting in about 3\% relative uncertainty in the $SES$.

\subsubsection{{\boldmath $K^+\rightarrow\mu^+\nu$} branching ratio measurement}
\label{sec:brkm2}
The measurement of the branching ratio of the $K^+\rightarrow\mu^+\nu$ decay
provides a test of the accuracy of the MC simulation of the
kinematic and geometric acceptances.

The measurement follows a procedure similar to that adopted for the $SES$.
The decay $K^+\rightarrow\pi^+\pi^0$ is used for normalization and the
branching ratio can be expressed as:
\begin{equation}
	\text{BR}(K^+\rightarrow\mu^+\nu)=\text{BR}(K^+\rightarrow\pi^+\pi^0)~\frac{N_{\mu2}}{\hat{N}_{\pi\pi}}
	~\frac{\hat{\epsilon}_{\pi\pi}}{\epsilon_{\mu2}}~\cdot
\end{equation}
Here $N_{\mu2}$ and $\hat{N}_{\pi\pi}$ are the number of selected
$K^+\rightarrow\mu^+\nu$ and $K^+\rightarrow\pi^+\pi^0$ events,
$\epsilon_{\mu2}$ and $\hat{\epsilon}_{\pi\pi}$ are the efficiencies for
selecting them. 

Event selection for both modes differs slightly from the procedures described
in sections~\ref{sec:pdef} and \ref{sec:kdef}.  Both modes require a RICH ring
associated with the STRAW track, but $\pi^+$ identification for the
$K^+\rightarrow\pi^+\pi^0$ decay relies on  the electromagnetic and
hadronic calorimeters only. 
MUV3 provides positive identification of the $\mu^+$ from  $K^+\rightarrow\mu^+\nu$ decay. 

The kinematic range $0.01<\mmis<0.026\,\text{GeV}^2/c^4$ defines
$K^+\rightarrow\pi^+\pi^0$ events.  The requirement that
$|\mmis(\mu)|<0.01\,\text{GeV}^2/c^4$ defines $K^+\rightarrow\mu^+\nu$
decays.  Here, $\mmis(\mu)$ is the squared missing mass computed assuming 
the particle associated with the STRAW track to be a muon.  The background in
both selected modes is of the order of $10^{-3}$.
Estimations of $\hat{\epsilon}_{\pi\pi}$ and $\epsilon_{\mu2}$ rely on
Monte Carlo simulations, as for the $SES$.  Their magnitudes are about
0.09 and 0.10.

The procedure described in section~\ref{sec:seleff} is adopted to quantify the
efficiency bias introduced by the simulation of the RICH reconstruction.
The corresponding correction factor applied to $\text{BR}(K^+\rightarrow\mu^+\nu)$ is $+0.005\pm0.005$.

The $\pi^+$ identification efficiency affects only the
$K^+\rightarrow\pi^+\pi^0$ mode.  As stated in section~\ref{sec:seleff}, this
efficiency can be measured with data, and the simulation reproduces the value
within 2\% accuracy.  Half this discrepancy is applied as a correction to
$\hat{\epsilon}_{\pi\pi}$.  Assuming that the uncertainty in this correction is
100\%, this amounts to correcting $\text{BR}(K^+\rightarrow\mu^+\nu)$
by $-0.008\pm0.008$.

\begin{figure}[t]
  \begin{center}
    \begin{minipage}{25pc} 
    \includegraphics[width=25pc]{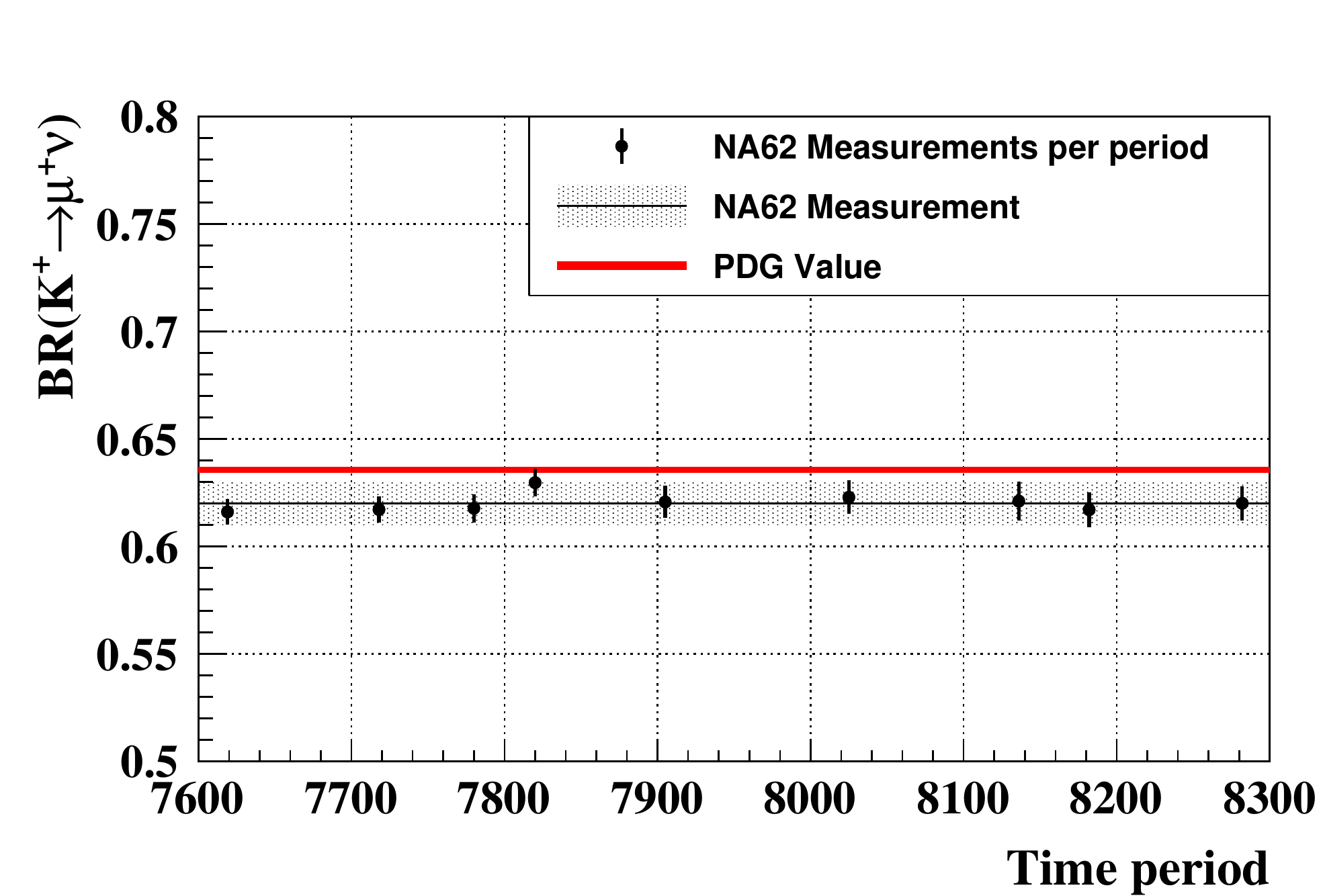}
   \end{minipage}\hspace{1pc}
    \caption{\label{fig:brkm2} Measured branching ratio of the
	  $K^+\rightarrow\mu^+\nu$ decay mode in different time periods of the
	  2017 data taking.   The overall quoted uncertainty is mostly systematic.} 
  \end{center}
\end{figure}

The $K^+\rightarrow\mu^+\nu$ branching ratio is  measured to be
\begin{equation}
\text{BR}(K^+\rightarrow\mu^+\nu)=0.62\pm0.01, 
\end{equation}
in agreement within 2.5\% with the PDG value~\cite{pdg}.  The 0.01 uncertainty is
systematic, mainly attributable to the corrections described above.  The
statistical uncertainty is negligible.  The result is stable within
uncertainties when signal and normalization selection cuts are varied.  It is
also stable throughout the 2017 data taking, as shown in Figure~\ref{fig:brkm2}.

This result relies on simulation to account for the different acceptances of the $K^+\rightarrow\mu^+\nu$ and $K^+\rightarrow\pi^+\pi^0$ decay modes, as  in the case of the $SES$ measurement.
The comparison between the measured and PDG branching ratios is used to set the level of accuracy in
the simulations, leading to a relative uncertainty of $2.5\%$ being propagated to $SES$.

\subsubsection{Random veto efficiency}
\label{sec:rv}
In both \pnnc and $K^+\rightarrow\pi^+\pi^0$ event selections, the GTK, CHANTI,
STRAW, and MUV3 are also used to veto backgrounds.  Data are used to estimate the
fraction of kaon decays rejected due to accidental activity in these detectors.
Measurements on samples of $K^+\rightarrow\pi^+\pi^0$ and
$K^+\rightarrow\mu^+\nu$ show that the fraction of events 
accepted by each of these detectors is
about 0.9, 0.97, 0.9, and 0.95.  These veto requirements are uncorrelated and, in
total, reject about 25\% of signal and normalization decays. Because the
average beam intensity of selected normalization and signal-like events is
comparable, the effects of the GTK, CHANTI, STRAW and MUV3 vetoes cancel in the
ratio of Equation~(\ref{eq:ses2}).

The criteria, collectively termed {\it photon and multiplicity rejection} (section~\ref{sec:sig}),
employed to veto $K^+$ decays with photons or more than one charged
particle in the final state also reject signal events if accidental particles
overlap the $\pi^+$ in time.  
\begin{figure}[t]
  \begin{center}
\includegraphics[width=20pc]{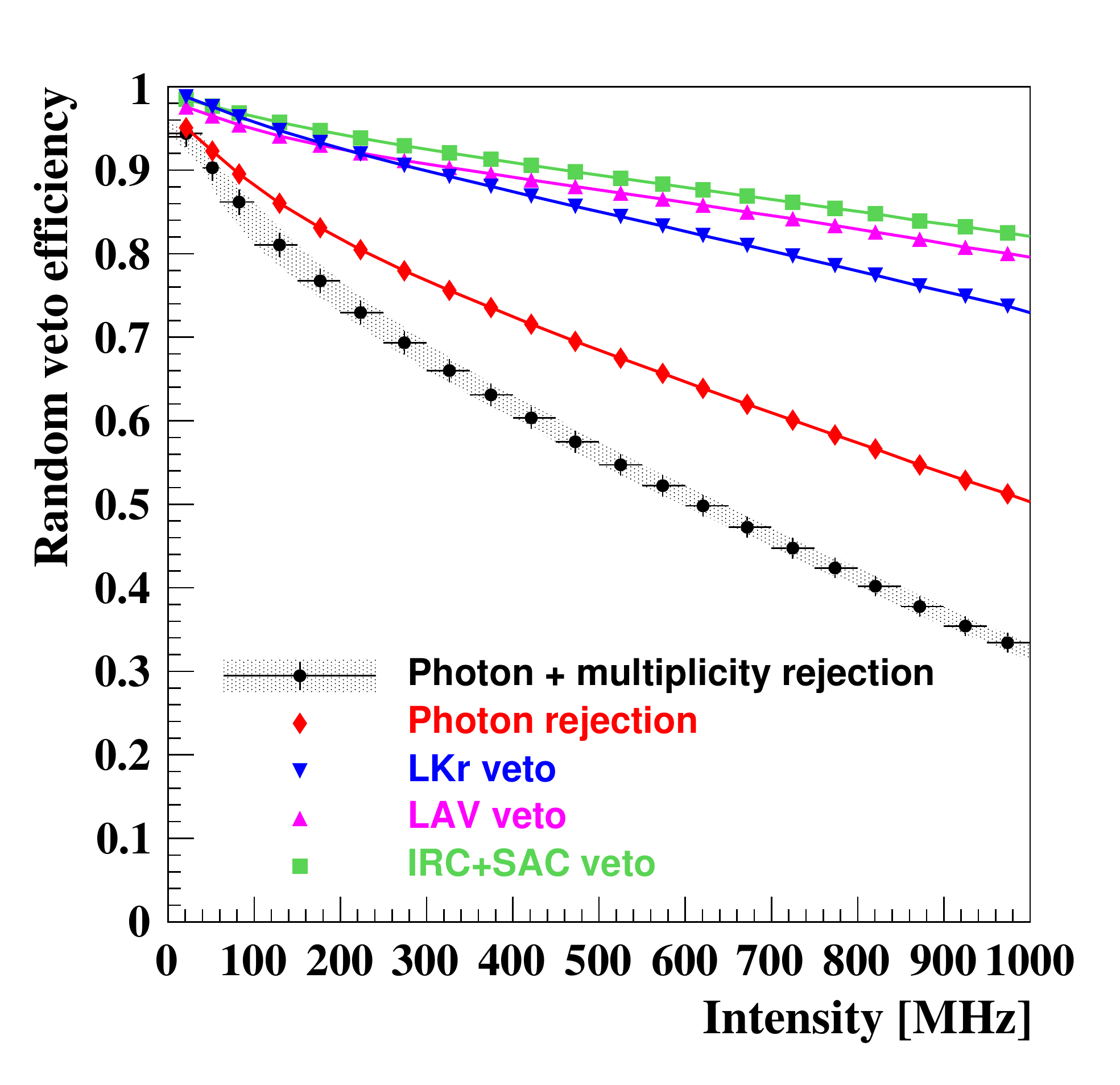}
  \caption{\label{fig:rv} Random veto efficiency $\epsilon_{RV}$ in bins of instantaneous beam
	  intensity after photon and multiplicity rejection, after photon
	  rejection, after LKr veto only, after LAV veto only, and after IRC and SAC veto only.
	 The error bars on the photon and multiplicity
	  rejection points indicates the total uncertainty.  Lines are drawn as
	  guides for the eye.} 
  \end{center}
\end{figure}

The fraction of signal events passing photon and multiplicity rejection is
denoted $\epsilon_{RV}$ and called the {\it random veto efficiency}.
A sample of $K^+\rightarrow\mu^+\nu$ decays selected from minimum-bias data
is used to estimate $\epsilon_{RV}$.  The selection closely follows that of
\pnnc, including GTK-, CHANTI-, and STRAW-based veto criteria, except that:  
$|(P_K-P_{\mu})^2|<0.006$\,GeV$^2/\text{c}^4$ replaces the missing-mass
squared regions; the calorimeters and the MUV3 are used for $\mu^+$
identification; and no box cut or the photon and multiplicity rejection
criteria are applied.  Simulation shows that the background to
$K^+\rightarrow\mu^+\nu$ is less than a part per thousand.
\begin{table}[t]
  \begin{center}
  \caption{Contributions to the uncertainty of the random veto efficiency
	  measurement.  The total uncertainty is the sum in quadrature of
	  the contributions.}  \label{tab:rv}
    \begin{tabular}{l|l}
      \toprule
      {Source} & {Uncertainty in $\epsilon_{RV}$}  \\\midrule
      $\mu^+$ interaction correction & $\pm0.011$ \\
      $\mu^+$ identification              & $\pm0.008$  \\
      Momentum dependence          & $\pm0.003$  \\
      Statistical uncertainty               & $<0.001$      \\\midrule
      Total                                         & $\pm0.014$ \\
      \bottomrule
    \end{tabular}
  \end{center}
\end{table}

The random veto efficiency is computed as the ratio between the number of
$K^+\rightarrow\mu^+\nu$ events remaining before and after photon and
multiplicity rejection.  Figure~\ref{fig:rv} displays $\epsilon_{RV}$ as a
function of the instantaneous beam intensity.  This result is corrected for the
probability of event loss induced by $\mu^+$ interactions, such as
$\delta$-ray production in the RICH material, as estimated by simulation.
This correction increases $\epsilon_{RV}$ by about 1\%.  An uncertainty of
100\% is assigned to this correction, leading to a 1\% systematic uncertainty
in $\epsilon_{RV}$. The stability of $\epsilon_{RV}$ is tested against cuts on
$(P_K-P_{\mu})^2$ and $\mu^+$ identification.  The maximum observed relative
variation is $2.4\%$ due to the cut on the calorimetric BDT probability.  Half \
this variation is used to correct the measured $\epsilon_{RV}$ and half is
assigned as a systematic uncertainty.  
A residual dependence on the
$\mu^+$ momentum is observed and added to the total systematic uncertainty. 
The final average random veto efficiency is $0.638\pm0.014$.
Table~\ref{tab:rv} summarizes the different contributions to the uncertainty.

Because the random veto affects only the signal, 
the uncertainty in $\epsilon_{RV}$ contributes linearly to the uncertainty in the $SES$.
\subsection{Trigger efficiency}
\label{sec:trigeff}
Normalization events are selected from minimum-bias data, and signal events are
selected from PNN data.  Problems in the hardware and trigger definitions in
conflict with offline cuts may cause the trigger to reject good normalization
and signal events.  Because mimimum-bias and PNN triggers differ, their
efficiencies, denoted $\epsilon_{trig}^{\scriptscriptstyle MB}$ and
$\epsilon_{trig}^{\scriptscriptstyle PNN}$ in Equation~(\ref{eq:ses2}), do not
cancel in the ratio, which therefore must be precisely evaluated.  The L0 and
L1 trigger algorithms which identify signal candidates employ different sets of
detectors, so their efficiencies can be studied separately.

\subsubsection{PNN L0 trigger efficiency}
\label{sec:l0eff}
The L0 efficiency stems from conditions in the RICH, CHOD, and MUV3, termed
L0NoCalo, and veto conditions in the LKr, called L0Calo.  A sample of
$K^+\rightarrow\pi^+\pi^0$ events selected from minimum-bias data using
PNN-like criteria allows the measurement of the L0NoCalo efficiency.  The
contributions from the RICH and CHOD are also estimated with
$K^+\rightarrow\mu^+\nu$ decays.  
The measured L0NoCalo efficiency is about 0.980 at the mean intensity of 450 MHz and varies almost linearly as a function of the instantaneous beam intensity, decreasing by about 1\% at twice the mean intensity.
The main source of inefficiency comes from the MUV3 veto criteria, because the
veto timing window is larger online than offline due to online time resolution.
The uncertainty in the measured value is at the level of 0.5\%, is mostly systematic and reflects the deviation of the efficiency from linearity.

The L0Calo efficiency is measured with a sample of $K^+\rightarrow\pi^+\pi^0$
decays in which the two photons are detected in LAV stations.  Events of this
type result in a $\pi^+$ with momentum greater than 45 GeV$/c$ in the
LKr.  The L0Calo efficiency, defined as the fraction of events passing the
L0Calo conditions, is measured as a function of the energy, $E_{LKr}$, that
the $\pi^+$ deposits in the LKr.  The dependence on $E_{LKr}$ is converted into
a dependence on the $\pi^+$ momentum, $p_{\pi^+}$, in the $15-35$\,GeV$/$c
range, with a conversion factor extracted from the $E_{LKr}/p_{\pi^+}$
distribution of a sample of $\pi^+$s selected from
$K^+\rightarrow\pi^+\pi^+\pi^-$ decays.  
The L0Calo efficiency depends on the $\pi^+$ momentum, and  decreases from 0.965 to 0.910 between the first and the last momentum bin.
The requirement that there be no more 
than 30\,GeV detected in the LKr, convoluted with the energy resolution of the LKr, is the main source of inefficiency. 
The uncertainty in the L0Calo trigger efficiency comes from the statistics of
the $K^+\rightarrow\pi^+\pi^+\pi^-$ sample used to map $E_{LKr}$ into $p_{\pi^+}$.

The overall L0 trigger efficiency is the product of the L0NoCalo and L0Calo
efficiencies as a function of $\pi^+$ and intensity.  The measured value
decreases with both increasing $\pi^+$ momentum and intensity, ranging from
0.95 to 0.9.

\subsubsection{PNN L1 trigger efficiency}
\label{sec:l1eff}
The effects of independent KTAG, LAV, and STRAW requirements in the L1 trigger
efficiency are uncorrelated, such that the overall efficiency is the product of the
individual efficiencies.  Samples of $K^+\rightarrow\mu^+\nu$ selected from
minimum-bias data and of $K^+\rightarrow\pi^+\pi^0$ selected from data 
triggered by the PNN L0 conditions and recorded irrespective of the L1 trigger
decision were used to measure these efficiencies.  
The L1 trigger algorithms were emulated offline, including the effects of resolution.   

After applying PNN selection criteria, the KTAG L1 requirements do not
introduce additional loss of signal.  On the other hand, the LAV requirements
introduce intensity-dependent losses of events which pass signal offline
selection criteria, because the online LAV time resolution requires a larger
veto timing window than that used offline.  The L1 LAV efficiency in the first
part of the 2017 data-taking period ranges from 0.965 to 0.955, depending
on the intensity.  This efficiency is about 1\% higher and exhibits less
intensity dependence in the second part of 2017 data taking as a consequence of
an optimization of the L1 LAV algorithm.  The spread of the efficiency among  
data-taking periods is used to set a systematic uncertainty for this measurement, which amounts
to about 0.4\% (1.4\%) at low (high) intensity.

The efficiency of the L1 STRAW algorithm is greater than 0.99 and independent
of intensity.  A $\pm$0.2\% uncertainty is assigned to this value to account
for an observed $\pi^+$ momentum dependence.

\subsubsection{Trigger efficiency and \boldmath$SES$}
\label{sec:toteff}
\begin{figure}[t]
  \begin{center}
\includegraphics[width=30pc]{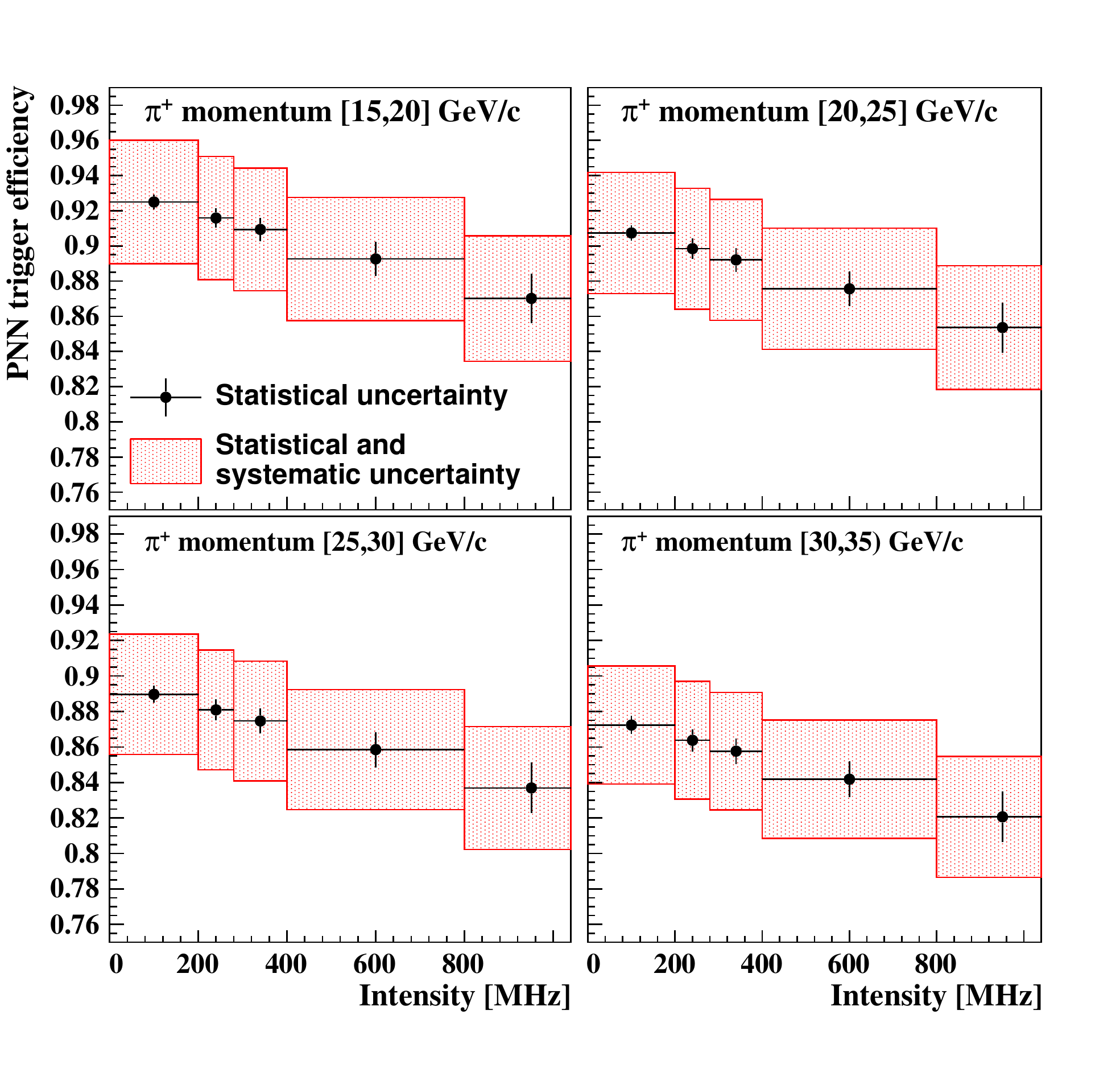}
  \caption{\label{fig:trigfinal} Measured PNN trigger efficiency as a function
	  of instantaneous beam intensity in four bins of $\pi^+$ momentum.
              The shaded band corresponds to the total uncertainty.}
  \end{center}
\end{figure}
The effect of the trigger efficiency on the $SES$ is determined using
Equation~(\ref{eq:ses2}) with the following assumptions:
the total PNN trigger efficiency is the product of the L0 and L1 efficiencies;
the efficiency of the L0TP is the same for the PNN and the minimum bias trigger 
and therefore cancels in Equation~(\ref{eq:ses2}); and
the minimum-bias trigger is 100\% efficient. 

The following test is performed to check the accuracy of these assumptions.
The PNN selection, except $\pi^+$ identification with the RICH, is applied to minimum-bias data,  
leading to $N^{MB}=701\pm26$ events in the $\mu\nu$ region of the $\mmis$ distribution.
The expected number of PNN data in this region passing the same selection can be written, under the above assumptions, as:
\begin{equation}
N^{PNN}(\text{expected})\,=\,D\cdot N^{MB}\cdot\epsilon^{PNN}_{L0}\cdot\epsilon^{PNN}_{L1}.
\end{equation}
Here $D$ is the minimum-bias reduction factor and $\epsilon^{PNN}_{L0}$  ($\epsilon^{PNN}_{L1}$) are the L0 (L1) PNN trigger efficiencies for $K^+\rightarrow\mu^+\nu$ decays in which the muon resembles a pion in the calorimeters and does not hit MUV3.
Considering that $K^+\rightarrow\mu^+\nu$ decays are fully efficient under the L0Calo condition (section~\ref{sec:l0eff}), the measured values of $\epsilon^{PNN}_{L0}$ and $\epsilon^{PNN}_{L1}$ 
 lead to $N^{PNN}(\text{expected})=(263\pm10)\times10^3$.
The number of PNN data observed in the $\mu\nu$ region of the $\mmis$ distribution after removing the RICH identification from the PNN selection is $N^{PNN}(\text{observed})=(255.6\pm0.6)\times10^3$, in agreement within $\pm\,3.8\%$ with $N^{PNN}(\text{expected})$.
This value is assigned as systematic uncertainty to the measured PNN trigger efficiency (noted  Global in Table~\ref{tab:trigerr}). 

The PNN trigger efficiency relevant to the measurement of $SES$ is shown in
Figure~\ref{fig:trigfinal} as a function of instantaneous beam intensity and $\pi^+$ momentum. 
The overall average trigger efficiency is $0.87\pm0.03$.
Table~\ref{tab:trigerr} summarizes the various contribution to the uncertainty
in the trigger efficiency.
\begin{table}[t]
  \begin{center}
  \caption{Contributions to the uncertainty of the trigger efficiency. When quoted,  the range corresponds to the efficiency dependence on the instantaneous beam intensity and the $\pi^+$  momentum.}  \label{tab:trigerr}
    \begin{tabular}{l|c}
      \toprule
      Source & Trigger efficiency uncertainty  \\\midrule
      L0NoCalo     & $\pm0.002$ to $\pm0.004$ \\ 
      L0Calo         & $\pm0.003$ to $\pm0.004$  \\
      L1 LAV         & $\pm0.004$ to $\pm0.014$  \\
      L1 Straw      & $\pm0.002$ \\
      Global          & $\pm0.035$ \\
      \bottomrule
    \end{tabular}
  \end{center}
\end{table}

\subsection{\boldmath$SES$ result}
The single event sensitivity and the total number of expected Standard Model \pnnc decays are:
\begin{align}
  {}&SES=(0.389\pm0.024_{syst})\times10^{-10},\\
  {}&N^{exp}_{\pi\nu\nu}(SM)=2.16\pm0.13_{syst}\pm0.26_{ext}.
\end{align}
The statistical uncertainty is negligible.
Table~\ref{tab:sesval} details the various contributions to the $SES$,
averaged over instantaneous beam intensity and $\pi^+$ momentum.
\begin{table}[t]
  \begin{center}
  \caption{Contributions to $SES$, averaged over instantaneous beam intensity and $\pi^+$ momentum.}  \label{tab:sesval}
    \begin{tabular}{l|l}
      \toprule
      Contribution & value \\\midrule
      $N_{\pi\pi}$                                                                       & $68\times10^6$\\
      \pnnc Monte Carlo efficiency, $\epsilon_{\pi\nu\nu}^{MC}$   & $0.030\pm0.003$\\
      $K^+\rightarrow\pi^+\pi^0$ Monte Carlo efficiency, $\epsilon_{\pi^+\pi^0}^{MC}$            & $0.087\pm0.009$\\
      Random veto efficiency (photon and multiplicity rejection)    & $0.638\pm0.014$\\
      Trigger efficiency                                                                & $0.87\pm0.03$\\
      \bottomrule
    \end{tabular}
  \end{center}
\end{table}
This list of contributions is for reference only, as the measured value of
the $SES$ comes from Equation~(\ref{eq:ses2}) in bins of  instantaneous beam intensity and $\pi^+$ momentum.

The above $SES$ corresponds to about $1.5\times10^{12}$ effective $K^+$ decays in the FV, defined as $(D\cdot N_{\pi\pi})/(\epsilon_{\pi^+\pi^0}^{MC}\cdot\text{BR}(K^+\rightarrow\pi^+\pi^0))$.
This quantity is proportional to the actual number of $K^+$ decays in the FV, although not strictly equal because $\epsilon_{\pi^+\pi^0}^{MC}$ does not include the elements which factor out and cancel in Equation~(\ref{eq:ses2}).

\begin{figure}[h]
  \begin{center}
  \begin{minipage}{18pc}
\includegraphics[width=18pc]{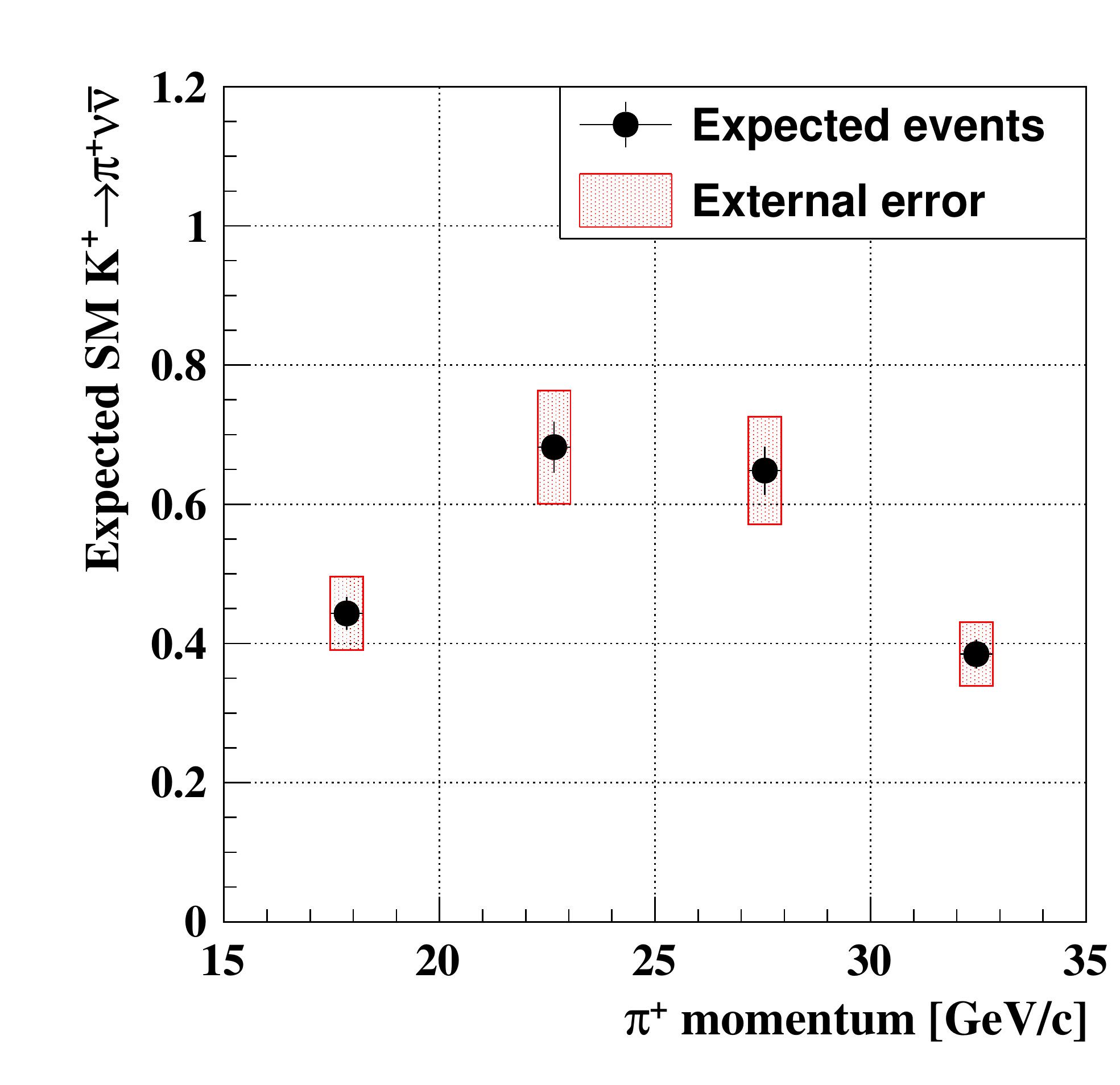}
  \end{minipage}\hspace{1pc}
  \begin{minipage}{18pc}
\includegraphics[width=18pc]{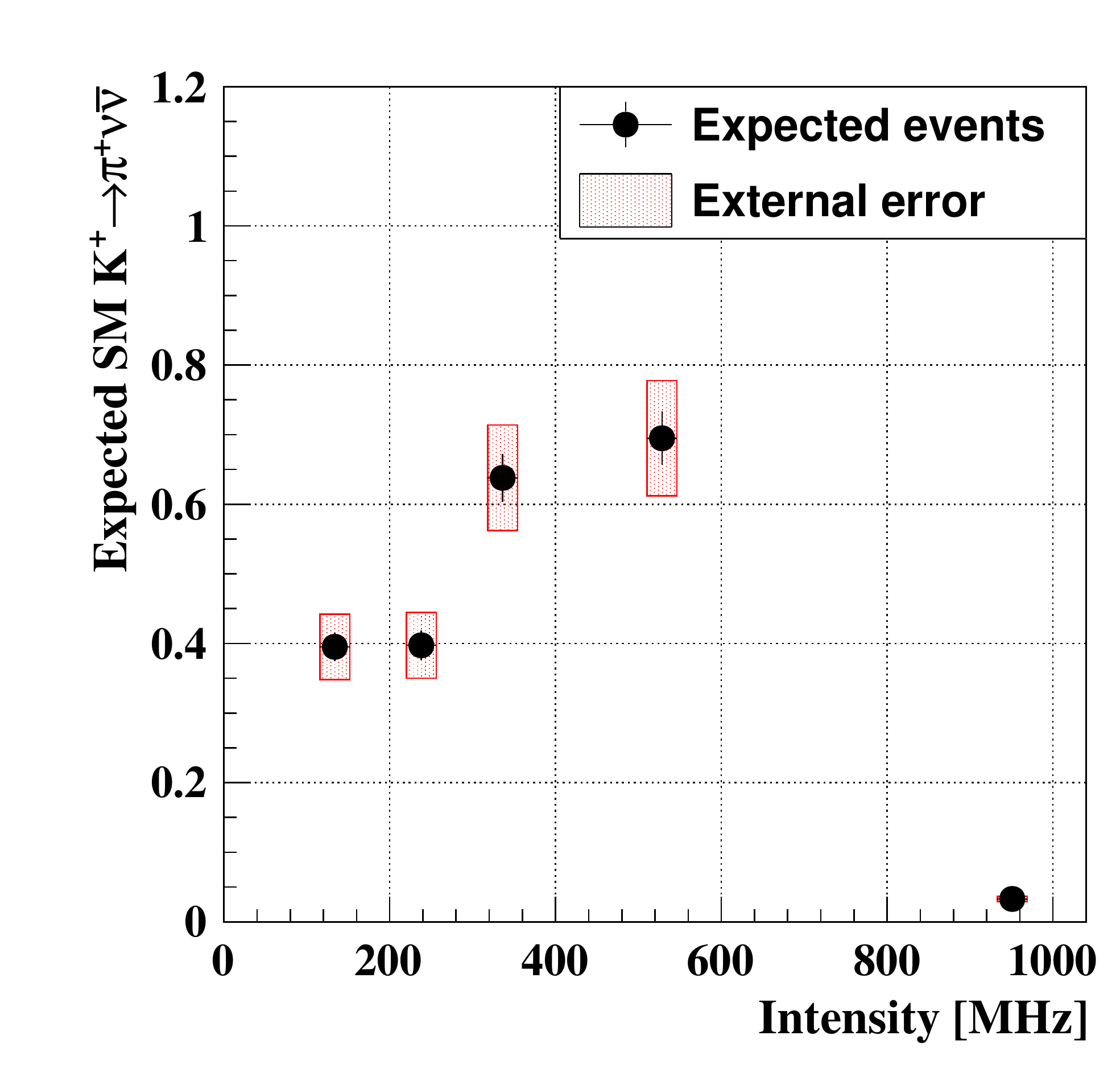}
  \end{minipage}
  \caption{\label{fig:npnn}Number of expected Standard Model \pnnc events in
	  bins of $\pi^+$ momentum (left) and average beam intensity (right).  The average beam
	  intensity per bin, obtained from the $K^+\rightarrow\pi^+\pi^0$ sample used for normalization, is plotted at the barycentre of the bin.}
  \end{center}
\end{figure}

Table~\ref{tab:seserr} lists the different sources of $SES$
uncertainty, including contributions to the uncertainty of the
Monte Carlo and trigger efficiency ratios, as discussed in
sections~\ref{sec:seleff} and \ref{sec:trigeff}, respectively.
The external error on $N^{exp}_{\pi\nu\nu}$ stems from the uncertainty in
the theoretical prediction of $\text{BR}(K^+\rightarrow\pi^+\nu\bar{\nu})$.
Figure~\ref{fig:npnn} shows $N^{exp}_{\pi\nu\nu}$ in bins of
$\pi^+$ momentum and instantaneous beam intensity.
\begin{table}[h]
  \begin{center}
  \caption{Sources contributing to the uncertainty in the $SES$ measurement.
	  ``Normalization background'' refers to the impact of
	  $K^+\rightarrow\pi^+\pi^0_D$ decays on the normalization sample.
	  The total uncertainty is the sum in quadrature of  the four contributions listed in the first column.}  \label{tab:seserr}
    \begin{tabular}{l|lc}
      \toprule
      {Source} & \multicolumn{2}{c}{Uncertainty in $SES$ ($\times 10^{10}$)}  \\\midrule
      Monte Carlo efficiency ratio & \multicolumn{2}{c}{$\pm0.017$} \\\cline{2-3}
      & $\pi^+$ interactions                                  & $\pm0.012$  \\
      & RICH reconstruction                                   & $\pm0.006$ \\
      & $\mmis$ Selection                            & $\pm0.004$  \\
      & Acceptance simulation                               & $\pm0.010$ \\\midrule
      Trigger efficiency & \multicolumn{2}{c}{$\pm0.015$} \\\cline{2-3}
      & L0 Efficiency  & $\pm0.002$ \\
      & L1 Efficiency  & $\pm0.003$ \\
      & Global           & $\pm0.015$ \\\hline                      
      Random Veto efficiency & \multicolumn{2}{c}{$\pm0.008$} \\\midrule
      Normalization Background & \multicolumn{2}{c}{$<0.001$}  \\\midrule
      Total & \multicolumn{2}{c}{$\pm0.024$} \\
      \bottomrule
    \end{tabular}
  \end{center}
\end{table}

\section{Expected background}
\label{sec:pnnbckg}
The background to \pnnc decays
can be divided into two classes. The $K^+$ decay background is due to kaon decays in the FV other than $K^+\to\pi^+\nu\bar\nu$, while the upstream background is due to $\pi^+$ particles produced either by beam particle interactions or by kaon decays upstream of the FV. To mimic a signal, a background event should have a \kp reconstructed upstream and matched to a $\pi^+$ downstream, and $\mmis$ reconstructed in the signal region. Furthermore, either the extra particles produced in association with the $\pi^+$ should escape detection, or a lepton in the final state should be mis-identified as a $\pi^+$.

\boldmath
\subsection{\kp decay background}
\unboldmath
\label{sec:kdecbckg}
The background from \kp decays in the FV is primarily due to the $K^+\to\pi^+\pi^0$, $K^+\to\mu^+\nu$, ~\pppc and $K^+\to\pi^+\pi^-e^+\nu$ decays.

The first three processes are constrained kinematically, and enter the signal regions via $\mmis$ mis-reconstruction due  to large-angle Coulomb scattering, elastic hadronic interactions in GTK and STRAW material, incorrect $K/\pi$ association, pattern recognition errors, or position mis-measurement in the spectrometers. In addition to $\mmis$ mis-reconstruction, at least one of the following should occur: photons from a $K^+\to\pi^+\pi^0$ decay are not detected by electromagnetic calorimeters; the muon from a $K^+\to\mu^+\nu$ decay is mis-identified as $\pi^+$ by the RICH counter, hadronic calorimeters and MUV3; a $\pi^+\pi^-$ pair from a \pppc decay is undetected by the STRAW and the other downstream detectors.

The background  from the three kinematically-constrained decays is evaluated with data.
Denoting by $N_{\rm decay}$ the number of events in the corresponding background region of $\mmis$  in the PNN data sample passing the PNN selection, and by $f_{\rm kin}$  the probability that $\mmis$ is reconstructed in the signal region, 
  the expected number of background events from each decay is given by
\begin{equation}
N^{\rm exp}_{\rm decay}=N_{\rm decay}\cdot f_{\rm kin} \cdot
\label{eq:bckgest}
\end{equation}
The value of $N_{\rm decay}$ is obtained directly from the PNN data, while the probability $f_{\rm kin}$ is measured with minimum-bias data. This technique does not require  knowledge of photon and charged particle rejection inefficiencies or of the $\pi^+$ mis-identification probability. Nevertheless, the precision of the method relies on three assumptions 
whose reasonableness is tested with both data and simulations:
\begin{enumerate}
\item $f_{\rm kin}$ represents the probability that an event of a given decay mode enters the signal region; 
\item $f_{\rm kin}$ and $N_{\rm decay}$ are uncorrelated; and
\item $N_{\rm decay}$ accounts only for events of the corresponding decay mode.
\end{enumerate}

Backgrounds from the $K^+\to\pi^+\pi^-e^+\nu$ decay, as well as from the rare decay $K^+\to\pi^+\gamma\gamma$ and the semileptonic decays $K^+\to\pi^0\ell^+\nu$  ($\ell=e,\mu$), are evaluated with simulations.

In the following subsections the $K^+\to\pi^+\pi^0$ and the $K^+\to\mu^+\nu$ backgrounds are shown in bins of $\pi^+$ momentum up to 40\,GeV$/c$, albeit only the 15--35~GeV/$c$ momentum
range is used to evaluate the corresponding backgrounds in the signal regions.

\subsubsection{{\boldmath $K^+\to\pi^+\pi^0$} decay}
\label{sec:pp0}

\begin{figure}[t]
  \begin{center}
 \begin{minipage}{20pc}
\includegraphics[width=20pc]{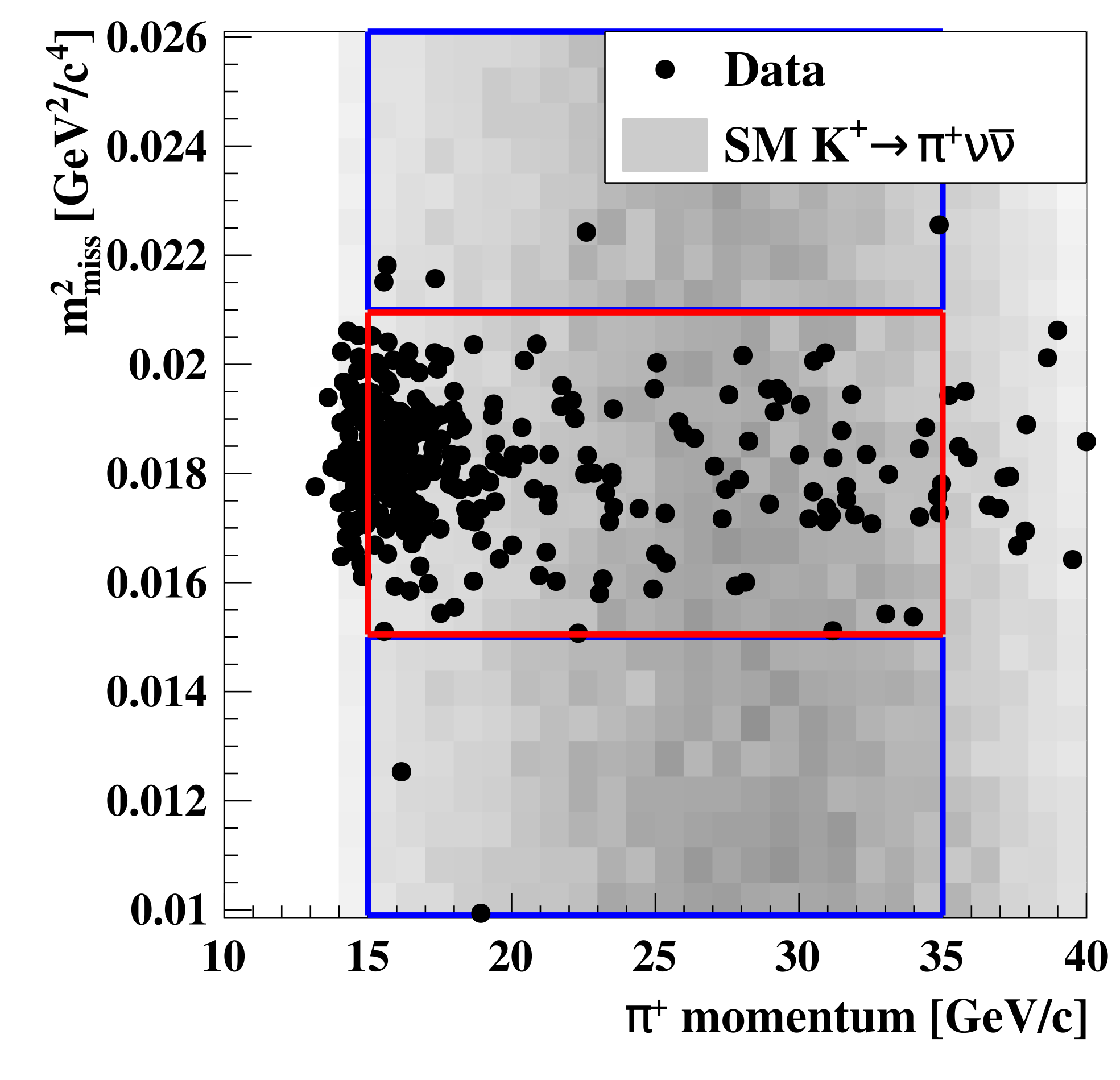}
 \end{minipage}\hspace{1pc}
\caption{\label{fig:pp0res} Distribution of the PNN  events in the ($\pi^+$ momentum, $\mmis$) plane  after the PNN selection in the $\pi^+\pi^0$ (red box) and control $\mmis$ regions (blue boxes). 
Both control regions are used only for validation of the background estimation.
The shaded grey area represents the distribution of the simulated SM $K^+\to\pi^+\nu\bar\nu$ events (arbitrarily normalized).}
\end{center}
\end{figure}

After the PNN selection, $N_{\pi\pi}=264$ events from the PNN sample remain in the $\pi^+\pi^0$ region. The distribution of these events in the ($\pi^+$ momentum, $\mmis$) plane is shown in Figure~\ref{fig:pp0res}, after 
the unblinding of the $K^+\to\pi^+\pi^0$ control regions. The $\pi^+$ momentum lies in the $15-20$\,GeV/$c$ range for 60\% of these events, due to the degradation of the $\pi^0$ detection efficiency for photons emitted at small angles (section~\ref{sec:sig}).
The measurement of $f_{\rm kin}$ is based on a $K^+\to\pi^+\pi^0$ sample selected from minimum-bias data. The selection involves the $K^+\to\pi^+$ decay definition described in sections~\ref{sec:pdef}, \ref{sec:kdef} and \ref{sec:pid}. The conditions of section~\ref{sec:fv} are applied as well, however the decay region is defined as $115<Z_{\rm vertex}<165$~m. Specific selection criteria are employed to tag the $\pi^0$ by reconstructing two photons from the $\pi^0\to\gamma\gamma$ decay in the LKr calorimeter independently of the $\pi^+$ and $K^+$ tracks. The quantity $Z_{\rm vertex}$ is evaluated 
from the coordinates of the two photon energy clusters in LKr by assuming that they  
originate from a $\pi^0$ decay on the nominal beam axis. 
The vertex is required to be within the decay region, and its position is used to reconstruct the photon and 
the $\pi^0$ momenta. Consequently, the expected $\pi^+$ trajectory is reconstructed and is required to be in the geometric acceptance of the detectors. The reconstructed squared missing mass $(P_K-P_0)^2$, where $P_K$ and $P_0$ are the four-momenta of the nominal $K^+$ and the reconstructed $\pi^0$, peaks at the squared $\pi^+$ mass for $K^+\to\pi^+\pi^0$ decays. A cut on this quantity is applied to select an almost background-free $K^+\to\pi^+\pi^0$ sample without biasing the $\mmis$ reconstruction. 

\begin{figure}[p]
\begin{center}
 \begin{minipage}{18pc}
\includegraphics[width=18pc]{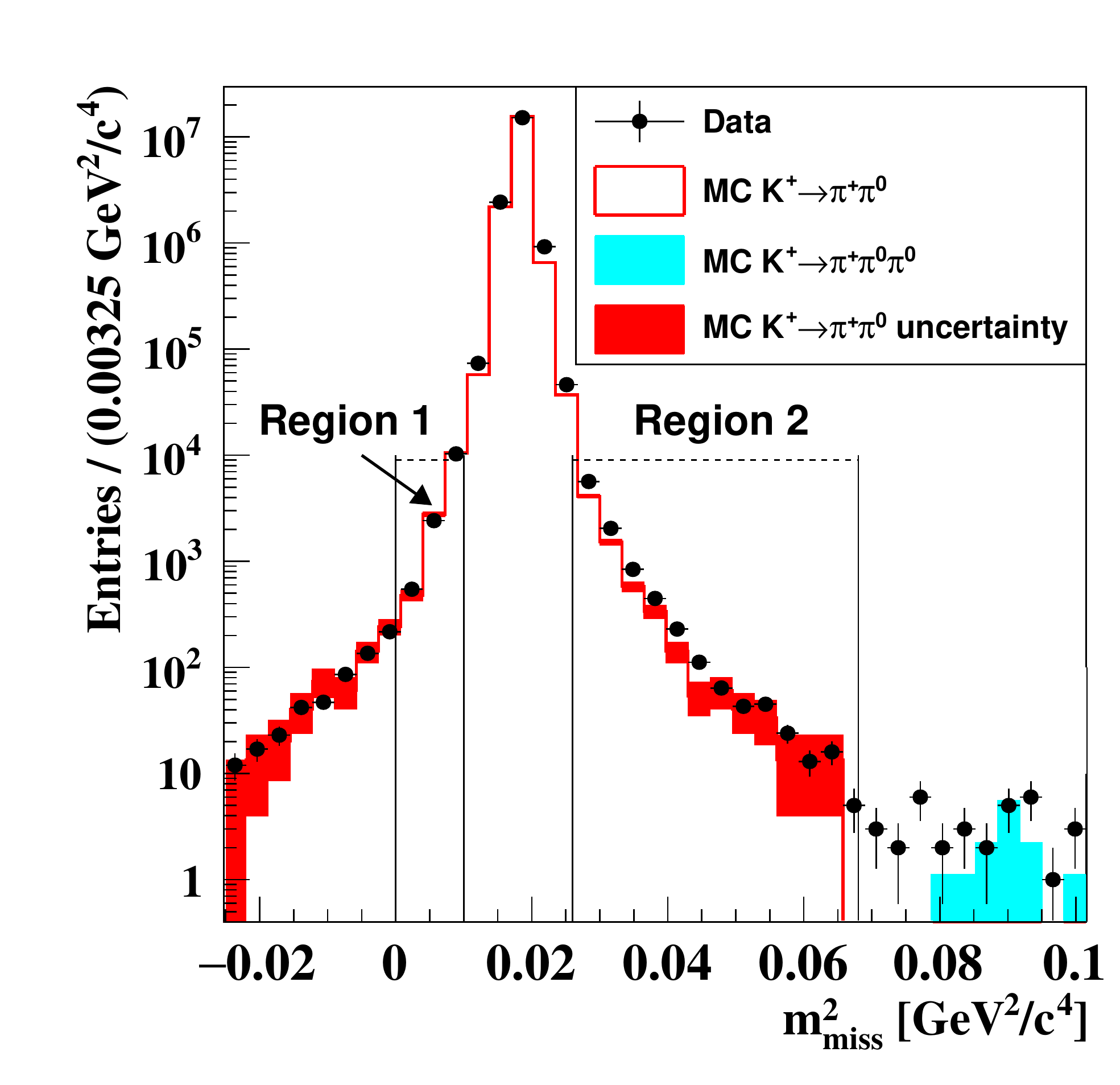}
\end{minipage}\hspace{1pc}
\begin{minipage}{18pc}
\includegraphics[width=18pc]{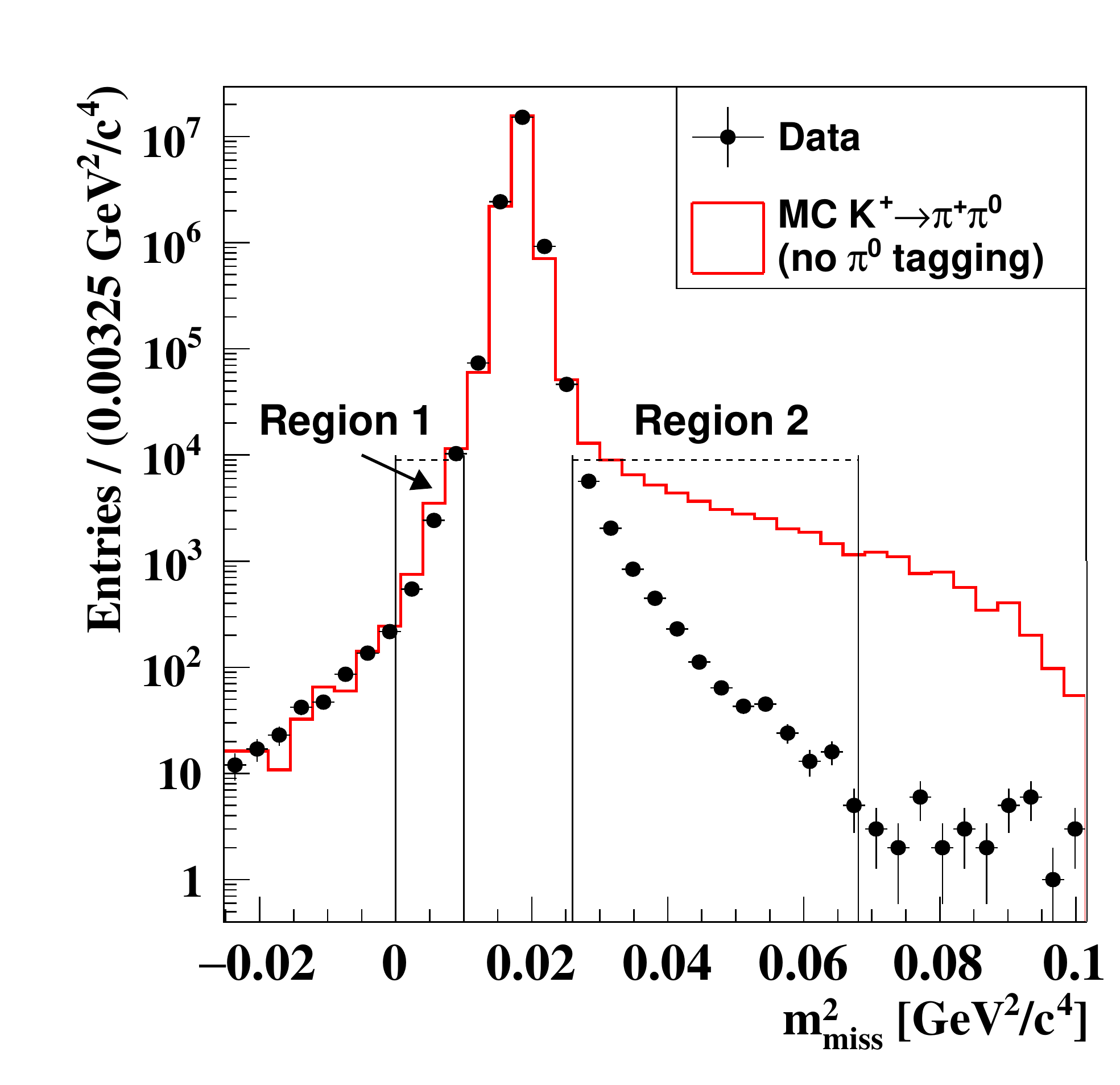}
\end{minipage}\hspace{1pc}
\begin{minipage}{18pc} 
\includegraphics[width=17pc]{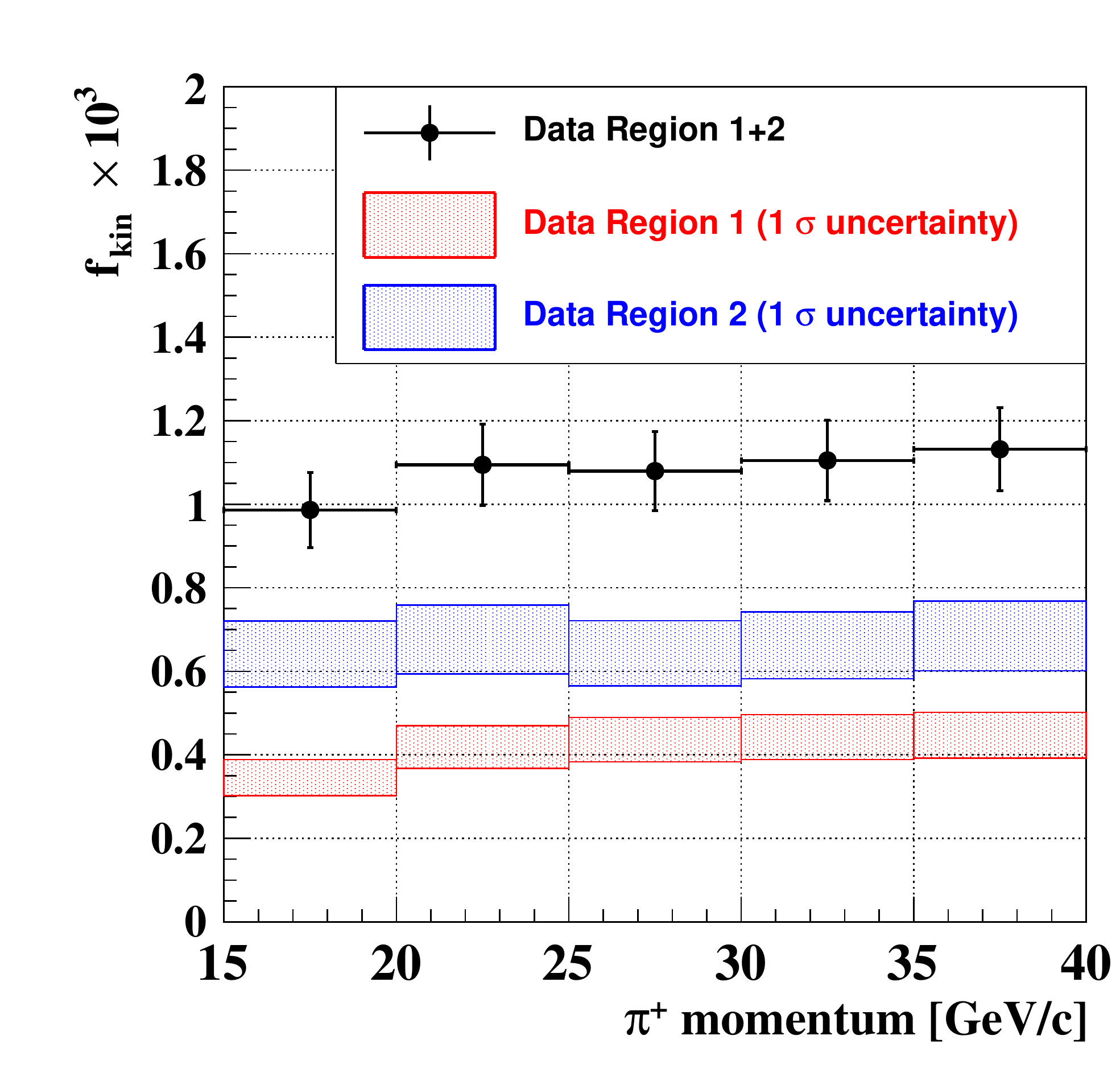}
\end{minipage} 
\begin{minipage}{18pc}\hspace{8mm}
\includegraphics[width=16pc]{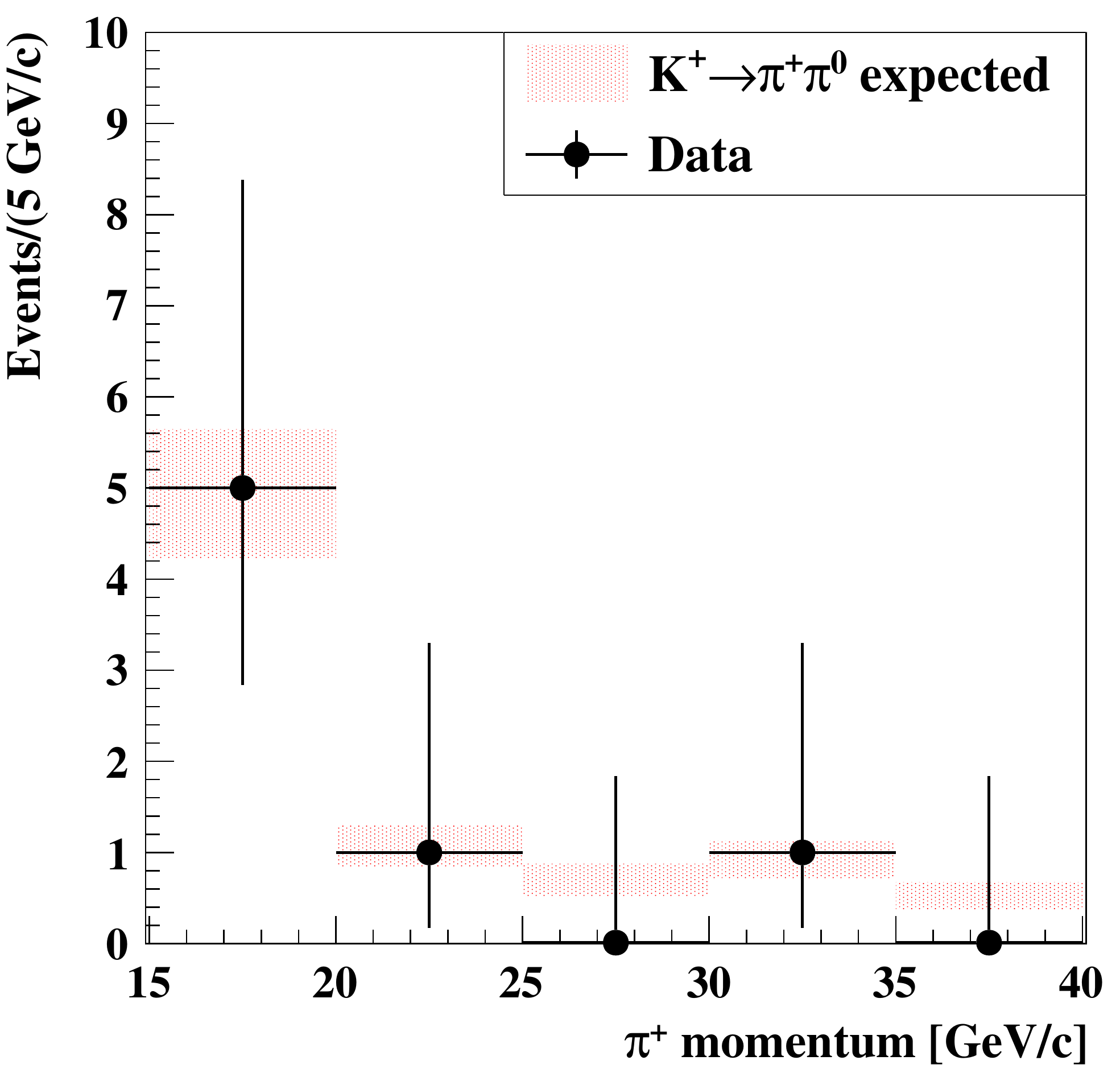}
\end{minipage}
\caption{\label{fig:pp0tail} {\bf Top left}: reconstructed $\mmis$ distribution of the~\pp minimum-bias data events selected by tagging the $\pi^0$ (full symbols, see text for details) integrated over the $15-35\,\text{GeV}/c$ momentum range.
Simulated samples of~\pp decays and backgrounds (normalized to the data in the $K^+\to\pi^+\pi^0$ region) are superimposed. Signal regions 1 and 2 are shown. The~\pp region, defined by the condition 
$0.015<\mmis<0.021\,\text{GeV}^2/c^4$, and the control regions, comprised between the signal and~\pp regions, are not shown.
{\bf Top right}: same as top-left, but the simulated~\pp sample is selected without applying the $\pi^0$ tagging; simulated backgrounds are not shown. 
{\bf Bottom left }: the probability $f_{\rm kin}$, defined in the text,  measured using the $K^+\to\pi^+\pi^0$ minimum-bias sample in bins of $\pi^+$ momentum, separately for signal region 1 and 2 and combined, with the statistical uncertainties.
{\bf Bottom right}: expected and observed numbers of background events in the $K^+\to\pi^+\pi^0(\gamma)$ decay  control regions  in $\pi^+$ momentum bins. 
The errors are statistical for the observed numbers of events, and dominated by  systematics for the expected numbers of events.}
\end{center}
\end{figure}

Figure~\ref{fig:pp0tail} (top left) displays the $\mmis$ spectrum of the $K^+\to\pi^+\pi^0$ minimum-bias sample used for $f_{\rm kin}$ measurement: $f_{\rm kin}$ is evaluated for each of the signal regions 1 and 2 as the ratio of the numbers of events in the signal region and in the $K^+\to\pi^+\pi^0$ region.
The simulation reproduces the tails within the statistical uncertainties, and the background is negligible. The measured values of $f_{\rm kin}$ in bins of $\pi^+$ momentum are shown in Figure~\ref{fig:pp0tail} (bottom left). Incorrect $K/\pi$ association due to the pileup in the GTK accounts for 50\% of the contribution to $f_{\rm kin}$ in region 1, and 30\% in region 2.

The total expected background in the signal regions, evaluated by applying Equation~(\ref{eq:bckgest}) in each $\pi^+$ momentum bin, is $N_{\pi\pi}^{\rm exp}=0.27\pm 0.026_{\rm stat}\pm 0.014_{\rm syst}$. The statistical uncertainty is mainly due to $N_{\pi\pi}$. The systematic uncertainty accounts for a possible bias to the shape of the $\mmis$ spectrum induced by the $\pi^0$ tagging used to measure $f_{\rm kin}$. It is evaluated by comparing the simulated shape of the $\mmis$ spectrum in the minimum-bias sample with that of $K^+\to\pi^+\pi^0$ decays used for normalization (Figure~\ref{fig:pp0tail}, top right). The 5\% difference between the numbers of events in region 1 in the two samples is taken as a systematic uncertainty.

Radiative decays in the simulated $K^+\to\pi^+\pi^0$ sample are modeled according to~\cite{gatti}.
Simulation studies show that decays with radiative photons energetic enough to 
shift the reconstructed $\mmis$ value to the signal regions are absent in the minimum-bias sample, 
due to the $\pi^0$ tagging suppression (Figure~\ref{fig:pp0tail} top right).
However the presence of an additional photon in the final state improves the photon veto, compensating for the weaker kinematic suppression.
The contribution of the radiative component to the $K^+\to\pi^+\pi^0$ decays is computed by applying the measured single photon detection efficiency (section~\ref{sec:sig}) to the simulated $K^+\to\pi^+\pi^0$ decays entering signal region 2
because of the presence of radiated photons.
It is concluded that the presence of an additional photon improves the rejection of $K^+\to\pi^+\pi^0$ in region 2 by a factor of almost 30 with respect to the case of the photons from the $\pi^0$ decay only.
This leads to an increase of the expected $K^+\to\pi^+\pi^0$ background of 0.02 events. 
A systematic uncertainty of 100\% is conservatively considered for this value, mainly due to the accuracy of the simulation and the modelling of the single photon detection efficiency.

\begin{table}[t]
\begin{center}
\caption{Expected numbers of $K^+\to\pi^+\pi^0$ events in the signal regions, and expected and observed numbers of events in the control regions. 
``Control region 1'' corresponds to $0.010<\mmis<0.015$\,GeV$^2/c^4$, ``Control region 2'' to $0.021<\mmis<0.026$\,GeV$^2/c^4$.
Expected events in both signal and control regions 2 are corrected for the contribution from the radiative component of the decay.
The uncertainties are the sums in quadrature of the statistical and systematic ones. Radiative decays are kinematically forbidden in both signal region 1 and control region 1.}
  \label{tab:pp0result}
  \begin{tabular}{l|c|c}
      \toprule
      {Region} & {Expected $K^+\rightarrow\pi^+\pi^0$} & Observed \\\midrule
      Signal region 1 & $0.11\pm0.01$ &  masked\\
      Signal region 2 & $0.18\pm0.04$ & masked\\
      Control region 1 & $2.6\pm0.3$ & 2 \\
      Control region 2 & $5.2\pm0.6$ & 5 \\
      \bottomrule
    \end{tabular}
 \end{center}
\end{table}

The numbers of expected $K^+\to\pi^+\pi^0$ events in the signal regions are presented in Table~\ref{tab:pp0result}. 
The overall background expected in the signal regions, including the effect of the radiative decays, is
\begin{equation}
N^{\rm exp}_{\pi\pi} = 0.29\pm0.03_{\rm stat}\pm0.03_{\rm syst}.
\end{equation}

To validate this result, the numbers of expected and observed events are compared in the two $\pi^+\pi^0$ control regions. 
The probability $f_{\rm kin}$ for the control regions is measured to be about 25 times higher than for the corresponding signal regions, and the expected background scales accordingly.
The contribution from radiative decays in control regions is found negligible.  
Table~\ref{tab:pp0result} and in Figure~\ref{fig:pp0tail} (bottom right) present the numbers of expected and observed events in the control regions, found to be in good agreement.
The uncertainties in the expected background in the control regions are mostly systematic due to the modelling of the $\mmis$ spectrum.

\subsubsection{\boldmath $K^+\to\mu^+\nu$ decay}
\label{sec:kmu2}

\begin{figure}[p]
  \begin{center}
   \begin{minipage}{18pc}\vspace{1pc}
\includegraphics[width=17pc]{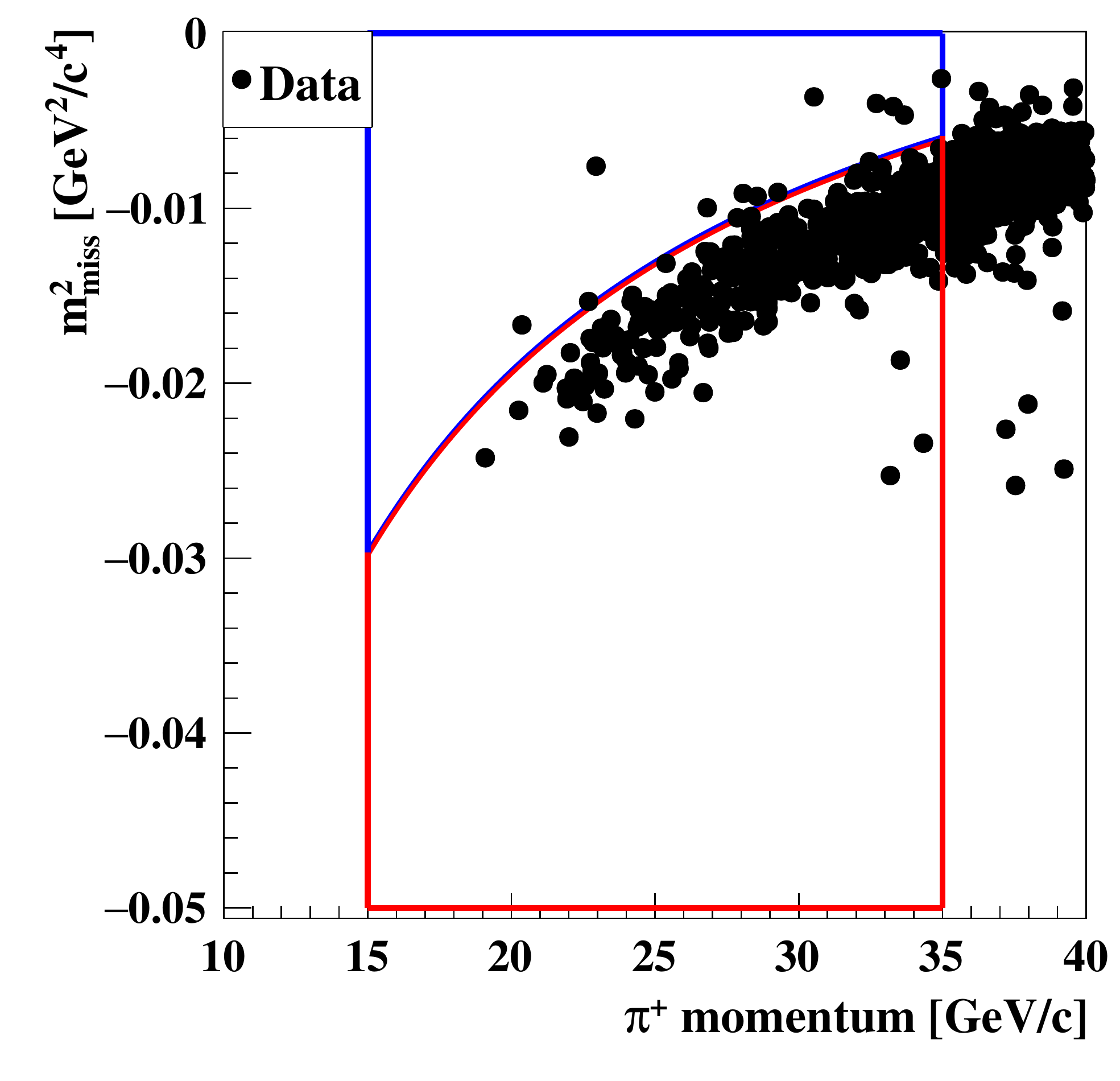} 
\end{minipage}\hspace{1pc}
 \begin{minipage}{18pc}
 \includegraphics[width=18pc]{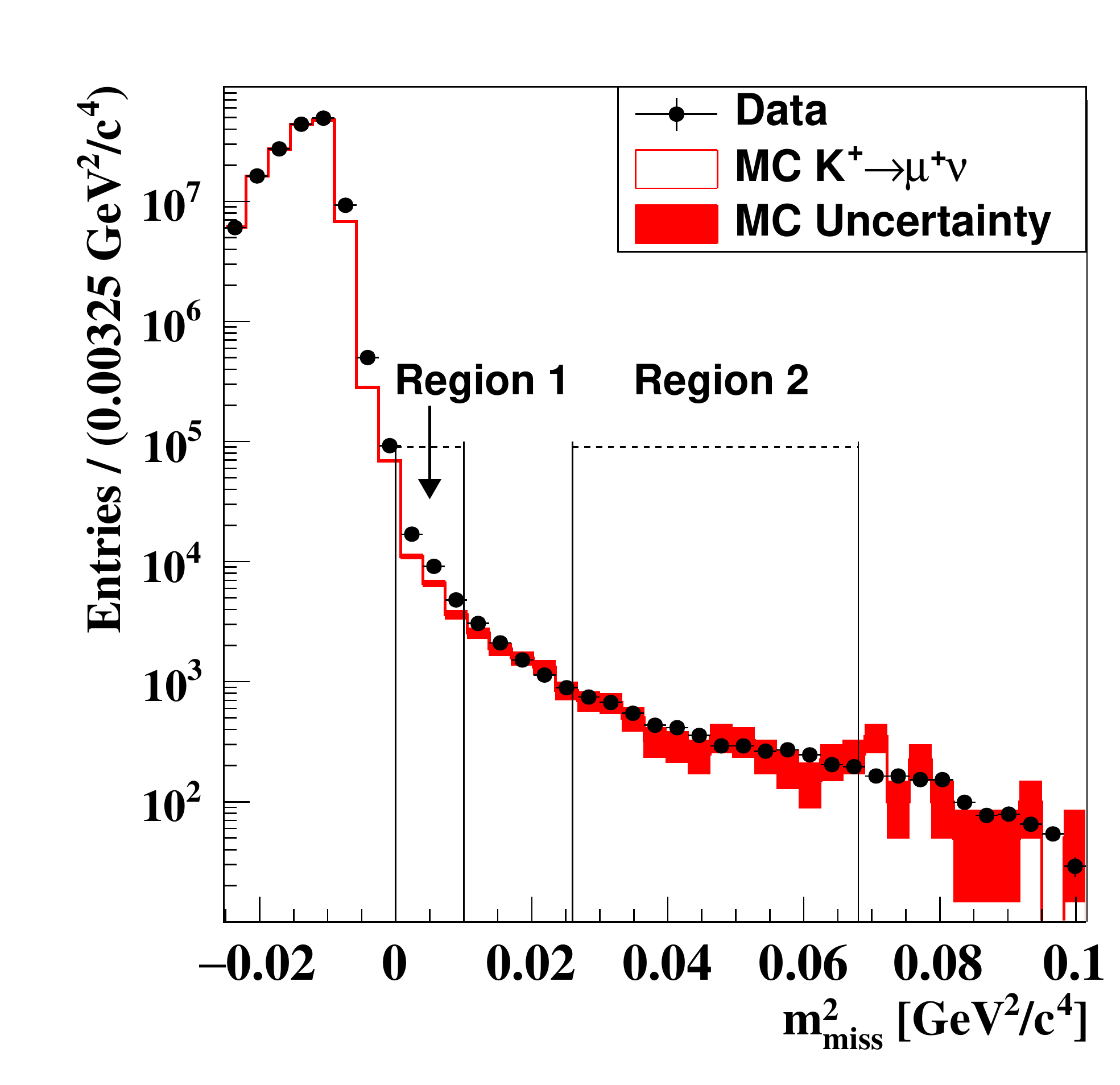}
 \end{minipage}
 \begin{minipage}{18pc}\hspace{-3mm}
 \includegraphics[width=17pc]{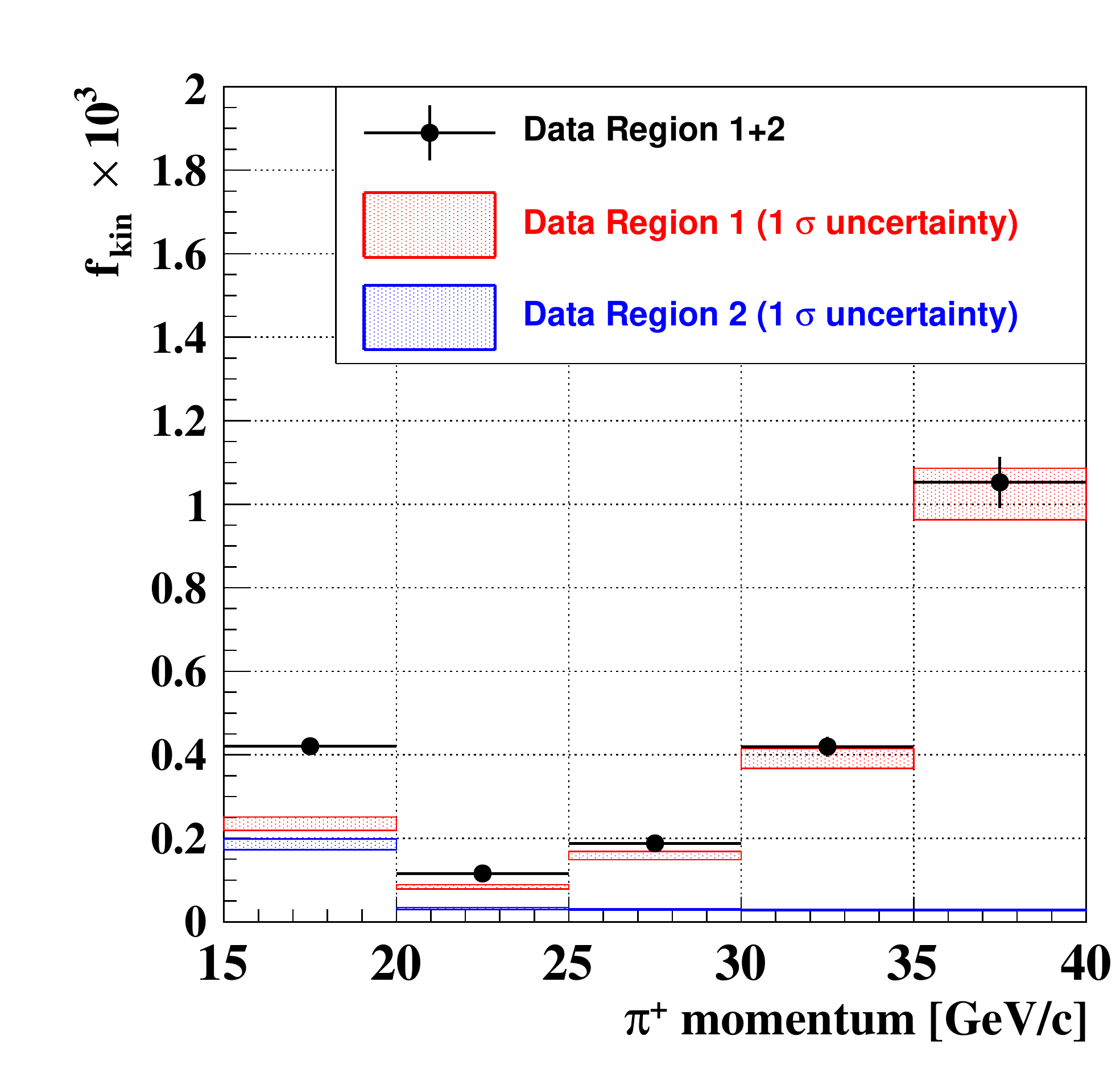}
 \end{minipage}\hspace{7mm}
 \begin{minipage}{17pc}\vspace{1pc}\hspace{4mm}
 \includegraphics[width=16pc]{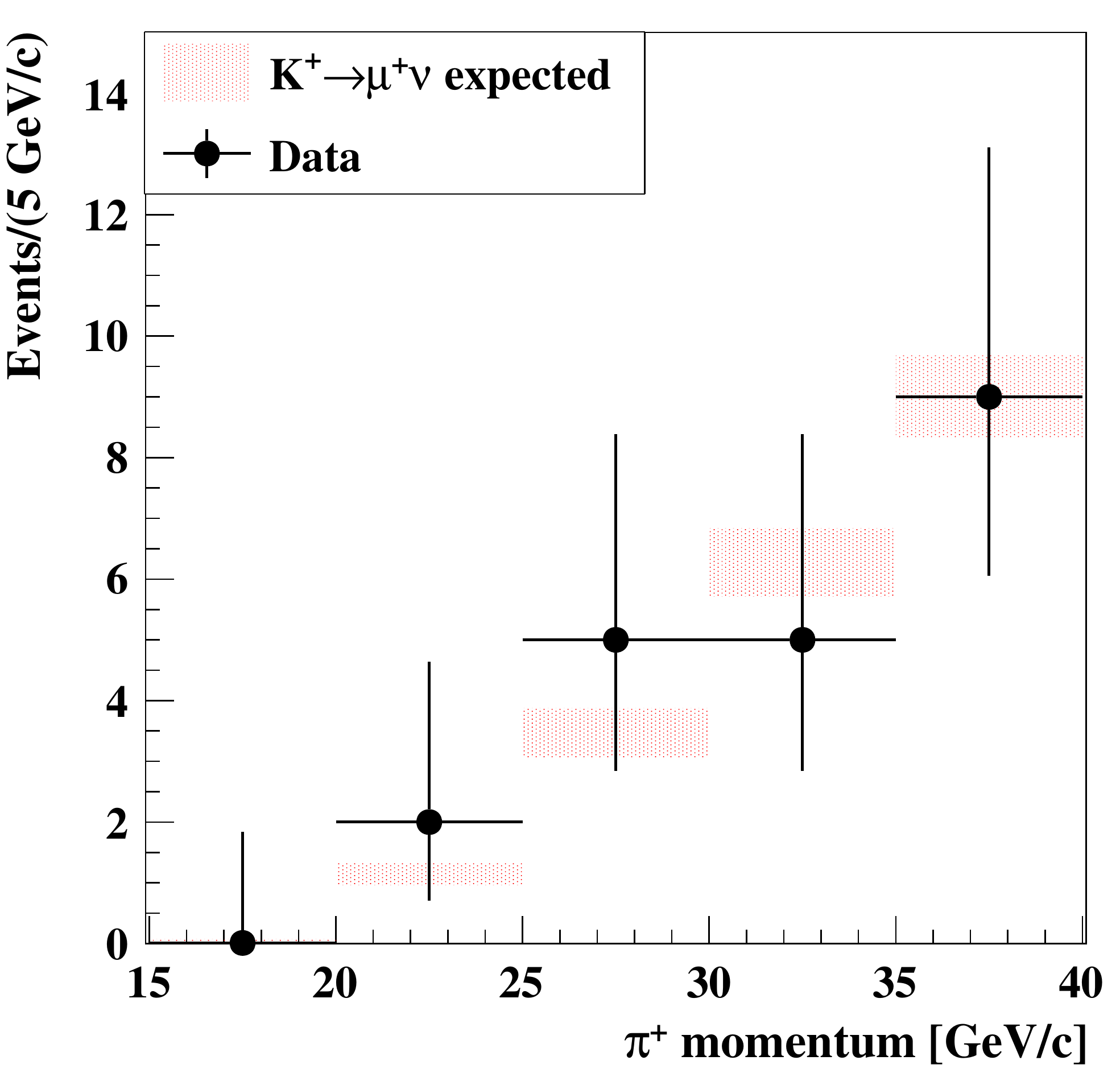}  
 \end{minipage}
  \caption{\label{fig:kmu2ev} 
  {\bf Top left}: distributions of the PNN-triggered events in the  ($\pi^+$ momentum, $\mmis$) plane after the PNN selection in the $\mu\nu$ (red contour) and control $\mmis$ (blue contour) regions. 
  The control region   is used only for validation of the background estimation.
  {\bf Top right}: reconstructed $\mmis$ distribution of the $K^+\to\mu^+\nu$ minimum-bias data events (full symbols, see text for details) integrated over the 15--35~GeV/$c$ momentum range. 
The   distribution of simulated $K^+\to\mu^+\nu$ decays is superimposed. Signal regions 1 and 2 are shown. 
{\bf Bottom left}: the probability $f_{\rm kin}$ measured using the $K^+\to\mu^+\nu$ minimum-bias sample in bins of reconstructed $\pi^+$ momentum, separately for signal region 1 and 2 and combined (black symbols), and  the corresponding statistical uncertainties.
{\bf Bottom right}: expected and observed numbers of background events from $K^+\to\mu^+\nu$ decays in the $\mu\nu$ control region in $\pi^+$ momentum bins. The errors are statistically dominated.}
\end{center}
\end{figure}

After the PNN selection, $N_{\mu\nu}=479$ events from the PNN sample remain in the $\mu\nu$ region. 
The numbers of events in bins of reconstructed $\pi^+$ momentum are presented in Table~\ref{tab:kmu2ev}, and the distribution of these events in the ($\pi^+$ momentum, $\mmis$) plane is shown in Figure~\ref{fig:kmu2ev} (top left). 
The momentum dependence is a consequence of the $K^+\to\mu^+\nu$ kinematics when the $\pi^+$ mass is used to reconstruct $\mmis$ and of the better performance of the RICH in rejecting $\mu^+$ at low momentum. 
The background to $N_{\mu\nu}$ is negligible.

\begin{table}[t]
\begin{center}
 \caption{Number of PNN-triggered events that pass the PNN selection and are reconstructed  in the $\mu\nu$ region, observed in the data in bins of $\pi^+$ momentum.} \label{tab:kmu2ev}
    \begin{tabular}{l|c|c|c|c}
      \toprule
      Momentum bins  (GeV/$c$)& $15-20$ & $20-25$ & $25-30$ & $30-35$ \\\midrule
      Observed events & $1$       & $48$      & $143$    & $287$ \\      \bottomrule
    \end{tabular}
  \end{center}
\end{table}
Two methods are exploited to estimate the $K^+\to\mu^+\nu$ background.

In the first one, the measurement of $f_{\rm kin}$ is based on a $K^+\to\mu^+\nu$ sample selected from minimum-bias data, as described in sections~\ref{sec:pdef} and \ref{sec:kdef}. 
Additionally, the calorimetric BDT probability must be consistent with the identification of a $\mu^+$, while events are discarded if the STRAW track is identified as $\pi^+$ or $e^+$ in the calorimeters.
The decay region is defined as $115<Z_{\rm vertex}<165$~m. 
The rejection of photons and extra charged particles is the same as in the PNN selection (section~\ref{sec:sig}). 
The box cut and the kinematic requirements on $\mmis$ are not applied.
Figure~\ref{fig:kmu2ev} (top right) displays the $\mmis$ spectrum of the $K^+\to\mu^+\nu$ minimum-bias sample used for $f_{\rm kin}$ measurement: $f_{\rm kin}$ is evaluated for each of the signal 
regions 1 and 2 as the ratio of the numbers of events in the signal and $\mu\nu$ regions. 
The signal region definition does not include the cuts on $\mmis$ computed using the momentum evaluated from the RICH information. 
Simulation reproduces the shape of the $\mmis$ spectrum within the statistical uncertainties, and the background is negligible. 
The measured $f_{\rm kin}$ values in bins of reconstructed $\pi^+$ momentum are shown in Figure~\ref{fig:kmu2ev} (bottom left). 
At large momentum, $f_{\rm kin}$ increases because the $\mmis$ of $K^+\rightarrow\mu^+\nu$ events computed assuming the $\pi^+$ mass approaches signal region 1.
Simulations show that the contribution to $f_{\rm kin}$ due to incorrect $K/\pi$ association is sub-dominant with respect to material effects.
The total expected background in the signal region, evaluated with this method by applying Equation~(\ref{eq:bckgest}) in each $\pi^+$ momentum bin, is $N_{\mu\nu}^{\rm exp}=0.14\pm0.007_{\rm stat}\pm0.007_{\rm syst}$. 
The statistical uncertainty is due to $N_{\mu\nu}$.
The systematic uncertainty comes from the stability of the result with the variation of the BDT probability cut, and accounts for a possible bias on $f_{\rm kin}$ due to the $\mu^+$ identification criteria applied in the 
selection of the $K^+\to\mu^+\nu$ minimum-bias sample.
This result relies on the assumption that the $\pi^+$ identification with the RICH and the shape of the $\mmis$ spectrum are uncorrelated. 
This assumption, in principle, is violated because $K^+\to\mu^+\nu$ events may enter the signal region due to track mis-reconstruction, which also affects particle identification with the RICH. 
In addition, the background estimation procedure does not include the cut on $\mmis$ computed using the RICH to measure $f_{\rm kin}$, which can bias the result. 

To investigate the accuracy of these approximations, a second method is employed to evaluate the $K^+\to\mu^+\nu$ background.
In this case the strategy is similar to that previously discussed, but the $\pi^+$ identification by the RICH is removed from the PNN selection used to derive $N_{\mu\nu}$ and added to the selection of the $K^+\to\mu^+\nu$ 
minimum-bias sample used to measure $f_{\rm kin}$. 
In addition, the determination of $f_{\rm kin}$ includes the cuts on $\mmis$ computed using the $\pi^+$ momentum measured from the RICH in the $\pi^+$ hypothesis. 
The $\mu^+$ rejection by the RICH suppresses $f_{\rm kin}$ by two orders of magnitude with respect to that of Figure~\ref{fig:kmu2ev} (bottom left), keeping a similar dependence on the $\pi^+$ momentum. 
This translates into a statistical uncertainty of 20\% on $f_{\rm kin}$. 
The expected background in the signal region is evaluated to be $N_{\mu\nu}^{\rm exp}=0.08\pm0.02_{\rm stat}$. 
This method is free from the bias of the first method, as it relies only on the assumption that $\mu^+$ rejection with the RICH and the calorimeters are uncorrelated. 
Simulations show that this assumption is valid as long as the $\mu^+$ does not decay upstream of the RICH.
\begin{table}[t]
  \begin{center}
  \caption{Expected numbers of $K^+\to\mu^+\nu$ events in signal regions, and numbers of expected and observed events in the control region. The uncertainties are the sums in quadrature of the statistical and systematic ones.}  \label{tab:kmu2result}
    \begin{tabular}{l|c|c|c}
      \toprule
      {Region} & {Expected $K^+\rightarrow\mu^+\nu$} & {Expected $K^+\rightarrow\mu^+\nu$, $\mu^+\rightarrow e^+\nu\bar{\nu}$} & Observed \\\midrule
      Signal region 1 & $0.11\pm0.04$ & $0.04\pm0.02$  & masked \\
      Signal region 2 & $<0.005$         & $<0.005$  & masked \\
      Control region & $10.5\pm1.0$ & $0.5\pm0.2$ & 12 \\
      \bottomrule
    \end{tabular}
  \end{center}
\end{table}

The average of the estimates of $N_{\mu\nu}^{\rm exp}$ from the two methods is used, and a systematic uncertainty equal to half of the difference ($\pm 0.03$) is assigned to account for a possible bias due to the correlation 
between particle identification and shape of the $\mmis$ spectrum. 
This background estimate includes also the contribution from the radiative component of the $K^+\to\mu^+\nu$ decays, as radiative decays enter the $K^+\to\mu^+\nu$ minimum-bias sample used to evaluate $f_{\rm kin}$.

The background from muon decays in flight $\mu^+\to e^+\nu\bar\nu$ is not included in the above estimate (as $f_{\rm kin}$ is measured requiring muon identification), and is determined separately using simulation. The rejection of this background depends on the muon decay position within the detector setup. Decays in the FV affect the kinematics, however, positrons are efficiently rejected by particle identification.
Decays within the STRAW spectrometer 
impact both kinematics and particle identification in the RICH, while decays downstream of the STRAW affect particle identification only.
Simulations show that only $\mu^+$ decays between the third and fourth STRAW chambers are relevant, leading to a worsening of both $K^+\to\mu^+\nu$ kinematics and particle identification.
The background is found to contribute to region 1 only, and is computed to be $0.04\pm0.02$. The uncertainty quoted includes statistical and systematic contributions of similar magnitudes. The latter is evaluated by checks performed on data to validate the simulation of the positron rejection.

The numbers of expected $K^+\to\mu^+\nu$ events in the signal regions are presented in Table~\ref{tab:kmu2result}. The overall background expected  is
\begin{equation}
N^{\rm exp}_{\mu\nu}=0.15\pm0.02_{\rm stat}\pm0.04_{\rm syst},
\end{equation}
with the contributions to the statistical and systematic uncertainties detailed above.

To validate this result, the numbers of expected and observed events are compared in the $\mu\nu$ control region.   
The expected number of events is evaluated with a technique similar to that described above.  This comparison is 
presented in Table~\ref{tab:kmu2result} and 
Figure~\ref{fig:kmu2ev} (bottom right),  showing good agreement between expected and observed numbers of events.

\subsubsection{\boldmath $K^+\to\pi^+\pi^+\pi^-$ decay}
\label{sec:ppp}

\begin{figure}[t]
  \begin{center}
\begin{minipage}{18pc}\vspace{1pc}
 \includegraphics[width=17pc] {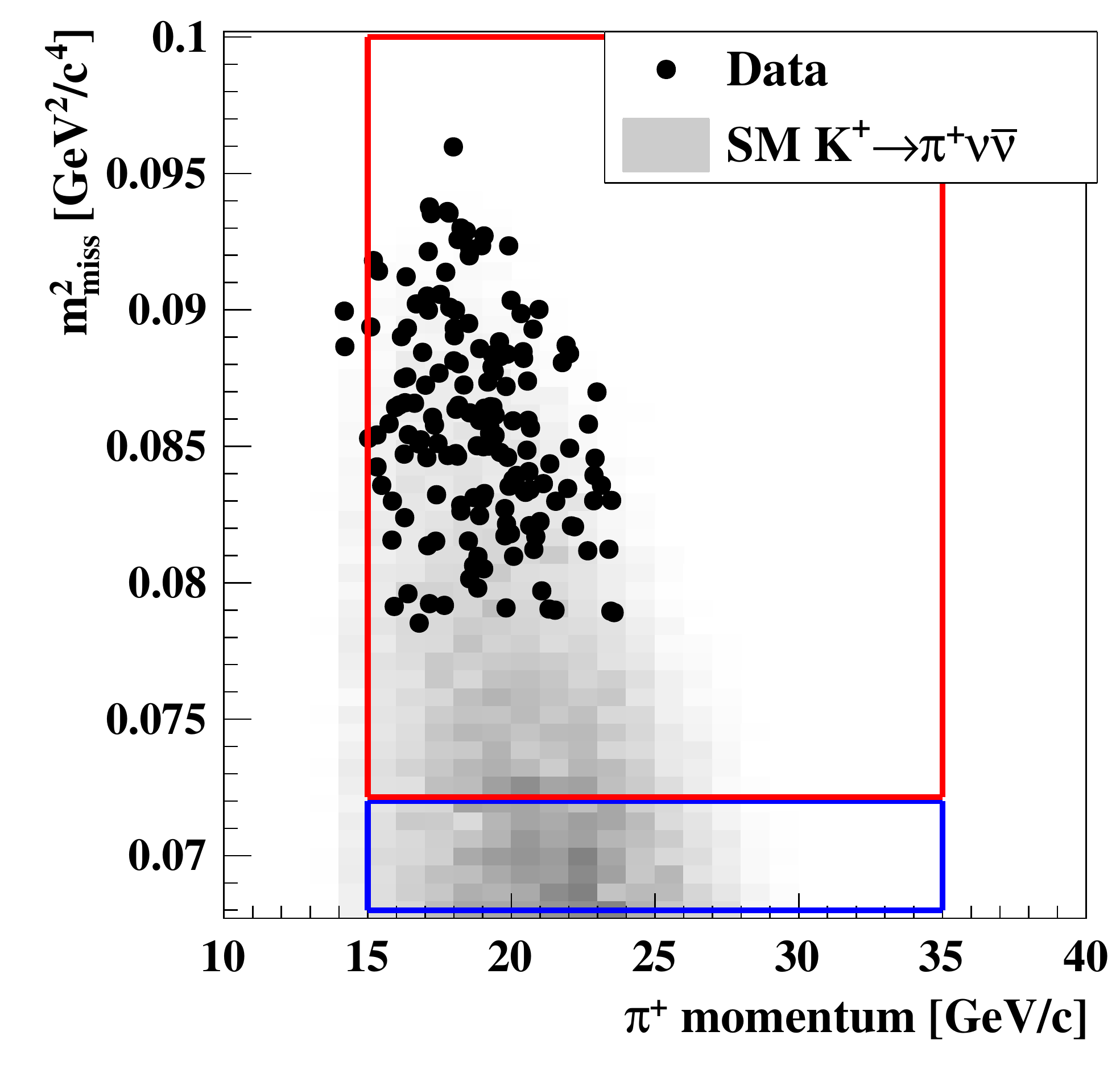}   
 \end{minipage}\hspace{1pc}
 \begin{minipage}{18pc}
 \includegraphics[width=18pc]{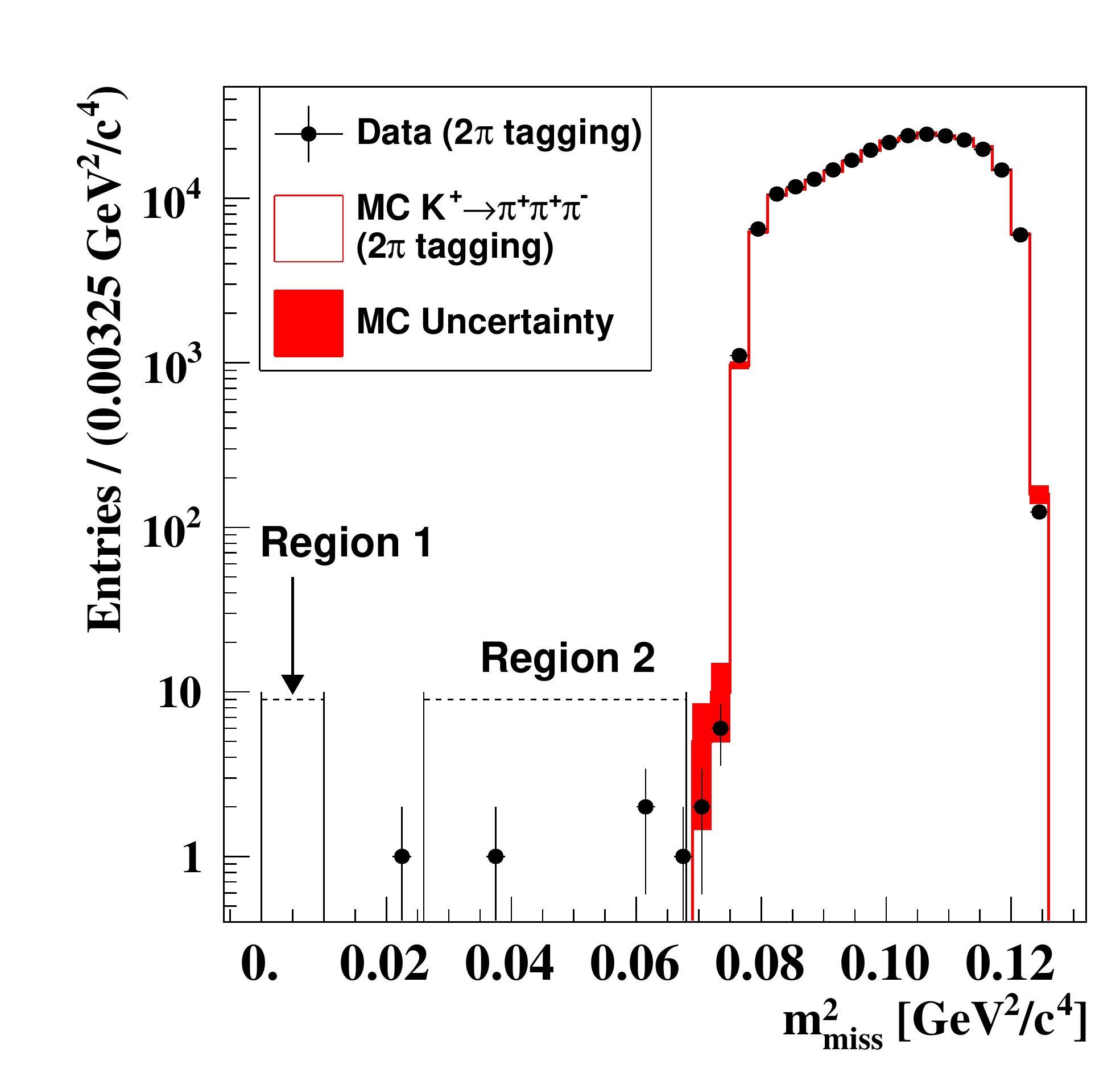}
 \end{minipage}
\caption{\label{fig:3pi} {\bf Left}: distribution  of the PNN-triggered events in the ($\pi^+$ momentum, $\mmis$) plane after the PNN selection in the $3\pi$ (red box) and control $\mmis$ regions (blue box). 
No event is found in the control region.
The shaded grey area represents the distribution of the simulated SM $K^+\to\pi^+\nu\bar\nu$ events (arbitrarily normalized). {\bf Right}: reconstructed $\mmis$ distribution for the unpaired $\pi^+$ from $K^+\to\pi^+\pi^+\pi^-$ decays, obtained using the $2\pi$ tagging method, selected from minimum-bias data and simulated samples. Signal regions 1 and 2   
are shown.}
\end{center}
\end{figure}

After the PNN selection, $N_{3\pi}=161$ events from the PNN sample remain in the $3\pi$ region. The distribution of these events in the  ($\pi^+$ momentum, $\mmis$) plane is shown in Figure~\ref{fig:3pi} (left); the $\pi^+$ momentum is constrained kinematically to the region below 25~GeV/$c$.

The measurement of $f_{\rm kin}$ is based on a $K^+\to\pi^+\pi^+\pi^-$ sample selected from minimum-bias data. For this purpose, a $\pi^+\pi^-$ pair is used to tag the $K^+\to\pi^+\pi^+\pi^-$ decay without biasing the reconstruction of the unpaired $\pi^+$. The $\pi^+$ to be paired to the $\pi^-$ is chosen randomly event by event.
The presence of a two-track vertex in the FV is required, and the quantity $(P_{\pi^+}+P_{\pi^-}-P_{K^+})^2$, where $P_{\pi^\pm}$ are the reconstructed 4-momenta of the $\pi^\pm$ and $P_{K^+}$ is the nominal kaon 4-momentum, must be consistent with the squared  $\pi^+$ mass. The selection proceeds with respect to the unpaired $\pi^+$ as described in sections~\ref{sec:pdef} and \ref{sec:kdef}. Photon veto conditions are applied to the LAV, IRC and SAC only. The box cut, the kinematic cuts on $\mmis$ and the multiplicity rejection are not applied.

The reconstructed $\mmis$ spectra for the tagged unpaired $\pi^+$ for the minimum-bias data and simulated samples are shown in Figure~\ref{fig:3pi}~(right). It is found that $f_{\rm kin}=(1.6^{+2.0}_{-1.0})\times10^{-5}$, where the uncertainties are statistical. Data and simulations are consistent within the uncertainties.

The kinematics of the tagged $\pi^+$ differs from that of the $\pi^+$ remaining after the PNN selection, potentially biasing the $f_{\rm kin}$ measurement. In particular, the $\mmis$ spectrum of the tagged sample does not match the one of the residual events in the $3\pi$ region.
The impact on the $f_{\rm kin}$ measurement is evaluated with simulations, varying the selection criteria. The full PNN selection cannot be applied to the simulated samples due to statistical limitations. Modified PNN selections used for the tests include those without the tagging, and with requirements of at least one and exactly one $\pi^+$ reconstructed in the geometric acceptance. The latter selection is the most PNN-like, and leads to a shape of the $\mmis$ spectrum in the $3\pi$ region matching that of the data events passing the full PNN selection. The values of $f_{\rm kin}$ obtained from simulations with the modified selections are in agreement within the uncertainties quoted above.

A possible bias comes from the dependence of $f_{\rm kin}$ on the $Z$-position of the decay vertex, as the tagging affects the shape of the $Z_{\rm vertex}$ spectrum. To quantify this effect, $f_{\rm kin}$ is evaluated in bins of $Z_{\rm vertex}$ for data and simulated samples.
The variation of $f_{\rm kin}$ across $Z_{\rm vertex}$ bins 
in simulated samples is conservatively considered as a systematic uncertainty. The final result is $f_{\rm kin}=(5\pm5)\times10^{-5}$.

The background computed using Equation~(\ref{eq:bckgest}) is
\begin{equation}
N_{3\pi}=0.008\pm0.008.
\end{equation}
To validate this result, the numbers of expected and observed events are compared in the (unmasked) $3\pi$ control region.  
The expected number of events in the control region is sensitive to the shape of the $\mmis$ spectrum close to the kinematic threshold of the $K^+\to\pi^+\pi^+\pi^-$ decay. Simulation studies lead to a conservative upper limit of $1.5\times10^{-3}$ on $f_{\rm kin}$ in the control region, corresponding to less than 0.24 expected background events. This is consistent with the observation of zero events in the control region.

\subsubsection{\boldmath $K^+\to\pi^+\pi^-e^+\nu$ decay}
\label{sec:ke4}

The $K^+\to\pi^+\pi^-e^+\nu$ decay (denoted $K_{e4}$ below) is characterized by large $\mmis$ and therefore contributes to region 2 only. This background is suppressed by the ${\cal O}(10^{-5})$ branching ratio~\cite{pdg},  the kinematic definition of the signal region, and the multiplicity rejection. The reconstructed $\mmis$ value depends on the kinematics of the undetected charged particles, which impacts the multiplicity rejection.
Because of this correlation,
the $K_{e4}$ background estimation relies on simulation.

\begin{figure}[t]
\begin{center}
\begin{minipage}{18pc}
\includegraphics[width=18pc]{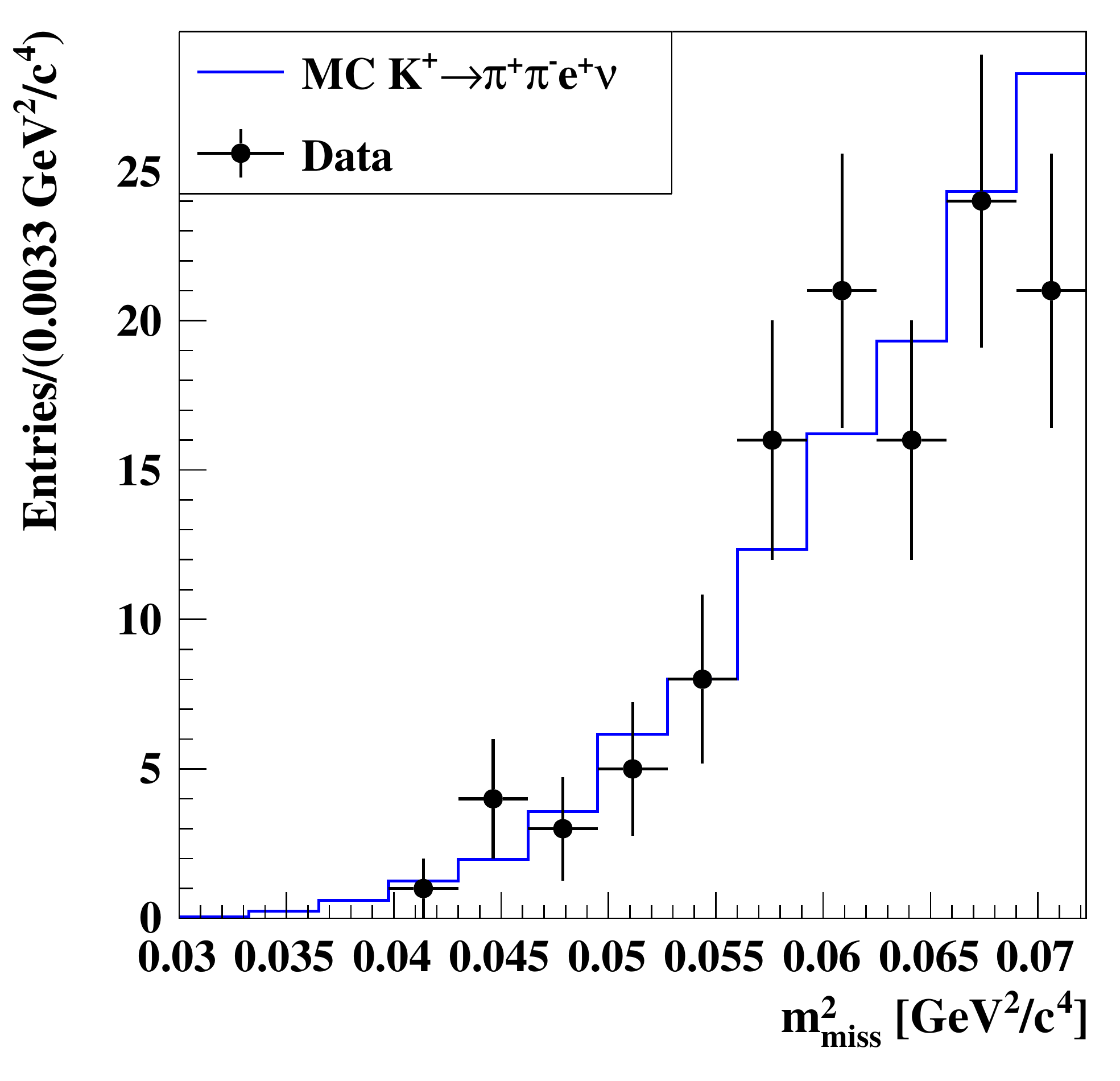}
\end{minipage}\hspace{1pc}
 \begin{minipage}{18pc}
\includegraphics[width=18pc]{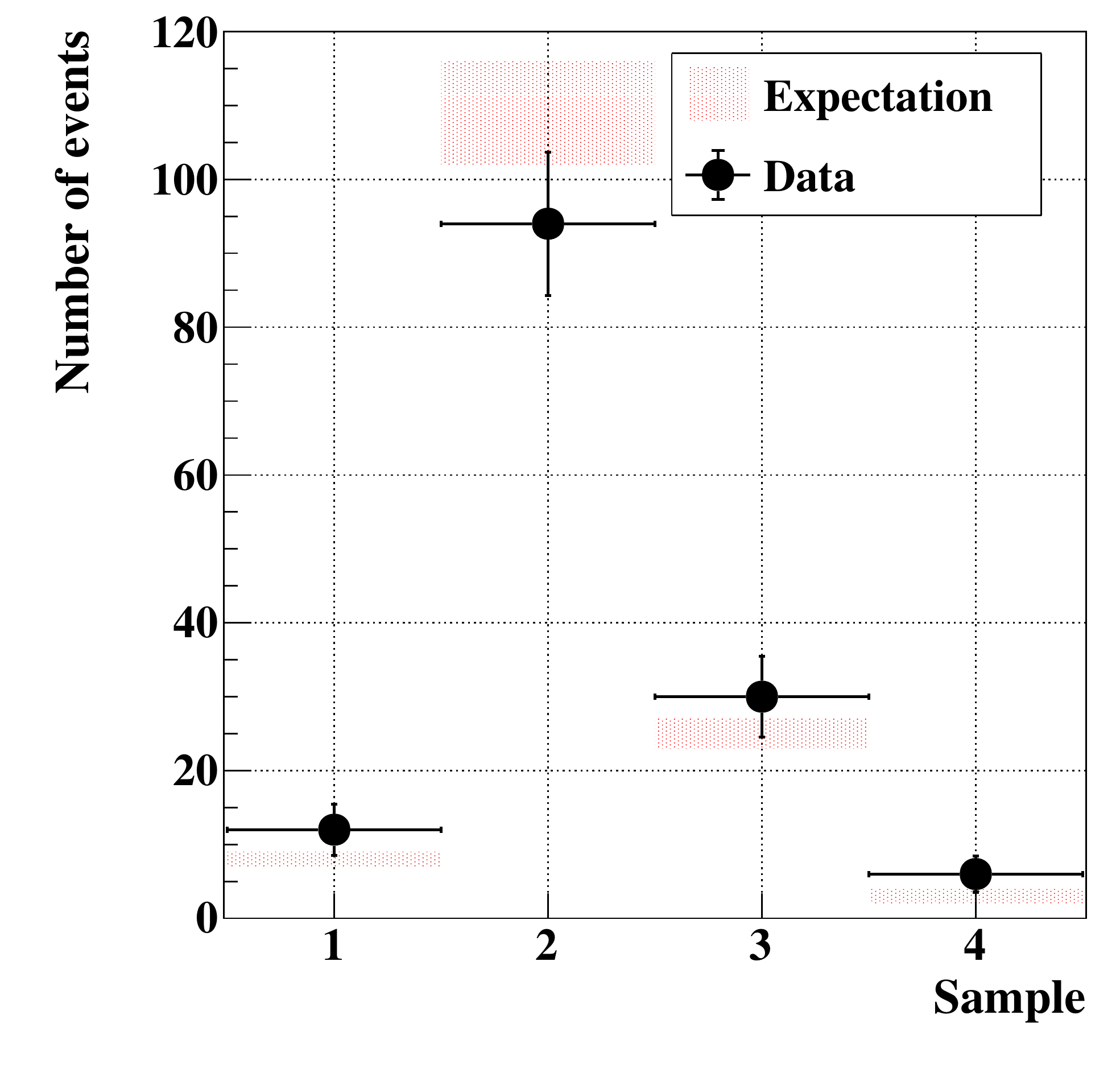}
\end{minipage}
\caption{\label{fig:ke4}{\bf Left}: reconstructed $\mmis$ distribution of the events of sample 2 selected in signal region 2 for data and $K_{e4}$ simulation.
               {\bf Right}: expected number of $K_{e4}$ decays and observed number of data events in region 2 for each of the four samples used to validate the $K_{e4}$ simulation.
               The different samples are defined in the text.}
\end{center}
\end{figure}

The efficiency of the PNN selection evaluated with a sample of $2\times10^9$ simulated $K_{e4}$ decays using the same normalization procedure as for the $SES$ computation is $\varepsilon_{Ke4}=(4\pm2_{\rm stat})\times10^{-9}$. This leads to an estimated background of $N_{Ke4}=(0.12 \pm 0.05_{\rm stat})$ events. To validate this estimate, four modified event selections leading to samples enriched with $K_{e4}$ decays are used:
\begin{enumerate}
\item the PNN selection, with inverted multiplicity conditions in the STRAW;
\item the PNN selection applied to the $\pi^-$ with RICH identification criteria not used, and inverted multiplicity conditions in the STRAW;
\item similar to 2, with the standard STRAW multiplicity conditions used;  and
\item similar to 3, with RICH identification criteria used.
\end{enumerate}
The selection efficiency for $K_{e4}$ decays ranges from $1.1\times10^{-7}$ (selection 4) to $3.7\times10^{-6}$ (selection 2). The corresponding data events entering region 2 are solely $K_{e4}$. The reconstructed $\mmis$ distributions obtained within selection 2 for PNN data and simulated $K_{e4}$ sample show agreement within the statistical uncertainties (Figure~\ref{fig:ke4}~(left)). The expected and observed numbers of events in region 2 within each of the four selections are summarized in Figure~\ref{fig:ke4}~(right). In particular, $3\pm1_{\rm stat}$ events are expected and $6\pm2_{\rm stat}$ events are observed within selection 4 which has the lowest acceptance. This difference is conservatively considered as a systematic uncertainty in $N_{Ke4}^{\rm exp}$, despite the agreement within the statistical uncertainties, leading to the expected background from $K_{e4}$ decays 
\begin{equation}
N_{Ke4}=0.12\pm0.05_{\rm stat}\pm0.06_{\rm syst}.
\end{equation}

\subsubsection{Other {\boldmath $K^+$} decays}
\label{sec:ob}

{\bf Semileptonic decays:} the branching ratios of  the semileptonic decays $K^+\to\pi^0e^+\nu$ and $K^+\to\pi^0\mu^+\nu$ are 5.1\% and 3.4\%, respectively~\cite{pdg}. The presence of the neutrino in the final state prevents kinematic discrimination of these decays from the signal, and the background is suppressed by exploiting
the presence of a $\pi^0$ and a lepton in the final state. The background estimation relies on simulation, with a factorization approach used to overcome unavoidable statistical limitations. Particle identification in the RICH and calorimeters are treated as independent, and the corresponding efficiencies are factored out with respect to the efficiency of the rest of the selection.

The measured muon misidentification probability as  a pion in the RICH detector depends on the particle momentum (section~\ref{sec:pid}).
On average such a probability is about $2\times10^{-3}$ for $K^+\to\pi^0\mu^+\nu$ decays passing the PNN selection; this result is used to validate the simulations.
Positron misidentification probability as a pion in the RICH detector in the 15--35~GeV/$c$ momentum range is evaluated with a simulated sample to be about $10^{-6}$. 
Calorimetric muon and positron misidentification 
probabilities as a pion in this momentum range evaluated with simulations are about $10^{-5}$ and $10^{-3}$, respectively.

The simulation accounts for the joint effect of $\pi^0$ rejection, and the geometric and kinematic acceptances. Simulations show that the former is about $10^{-5}$ (substantially weaker than for $K^+\to\pi^+\pi^0$ decays due to the different photon kinematics), while the latter is about 10\%. The $K^+\to\pi^+\pi^0$ decay is used for normalization. 
This leads to a systematic uncertainty in excess of 10\%, mostly because the particle identification efficiencies do not cancel in the ratio with that of $K^+\to\pi^+\pi^0$. Including the measured random veto and trigger efficiencies, the expected background is found to be less than 0.001 events for both decay modes, and is therefore considered negligible.

 \noindent{\boldmath $K^+\to\pi^+\gamma\gamma$:} the branching ratio of this decay, occurring at the loop level, is  $1.0\times 10^{-6}$~\cite{pdg}. The corresponding background is evaluated with simulations. The decay dynamics favours values of the di-photon invariant mass  above the di-pion threshold, corresponding to $\mmis$ values in the $3\pi$ region. This procedure leads to an overall efficiency of the PNN selection without photon rejection at the 1\% level. The rejection of events in the signal region benefits from the correlation between $\mmis$ and the  photon energy, leading to a photon rejection of order $10^{7}$. The $K^+\to\pi^+\pi^0$ decay is used for normalization. 
  Including the measured random veto and trigger efficiencies, the background is estimated to be $N_{\pi\gamma\gamma}=0.005\pm0.005$, where the conservative uncertainty   accounts for the accuracy of the photon rejection simulation.

\subsection{Upstream background}
\label{sec:ups}

\subsubsection{Background sources}
\label{sec:ups-sources}

Upstream events are defined as interactions or decays of beam particles upstream of the FV. 
An upstream event can mimic a \pnnc decay if:
\begin{itemize}
\item a $\pi^+$ is produced and reaches the downstream detectors; 
\item no additional particles associated to the $\pi^+$ are detected downstream; and
\item a $K^+$ candidate is reconstructed and matched to the $\pi^+$.
\end{itemize}
Based on these conditions, upstream events can be classified as follows: 
\begin{enumerate}
\item \textbf{{Accidental upstream events:}}
   events in which the $\pi^+$ does not originate from the reconstructed $K^+$ candidate.
   In this case the $K^+$ candidate is a pileup GTK track associated accidentally with the $\pi^+$ and tagged as a kaon by the KTAG.  

The mechanisms giving rise to accidental upstream events are the following:
\begin{itemize}
\item[a)] the $\pi^+$ comes from a $K^+$ decaying in the region upstream of GTK3; 
the KTAG signal produced by the parent $K^+$ is associated with a pileup beam $\pi^+$ or proton track,
which is reconstructed as a kaon  in the GTK;
additional particles produced in the decay are absorbed by material in the beam line;
  \item[b)] similar to a), but the matching GTK track belongs to another pileup $K^+$ identified correctly by the KTAG;
  \item[c)] similar to a), but the $\pi^+$ originates from an inelastic interaction of a beam $K^+$ upstream of GTK3;
  \item[d)] similar to b), but the $\pi^+$ originates from an inelastic interaction of a beam $\pi^+$ or proton upstream of GTK3.
  \end{itemize}

\item \textbf{{In-time upstream events:}}
events in which the $\pi^+$ is a primary or a secondary product of an inelastic interaction of a beam $K^+$ in GTK3. In this case, additional particles produced in the interaction must escape detection, as 
no beam line elements can absorb the particles.

Two processes may lead to in-time upstream events:
\begin{itemize}
\item[a)] the interacting $K^+$ produces a prompt $\pi^+$ that reaches the downstream detectors;
\item[b)] the interacting $K^+$ produces a relatively long-lived particle ($K_S$, $K_L$, $K^+$ or $\Lambda$) that decays to a $\pi^+$ in the FV.
  \end{itemize}
\end{enumerate}

\begin{figure}[t]
\begin{center}
\begin{minipage}{.6\textwidth}
\includegraphics[width=1.\textwidth]{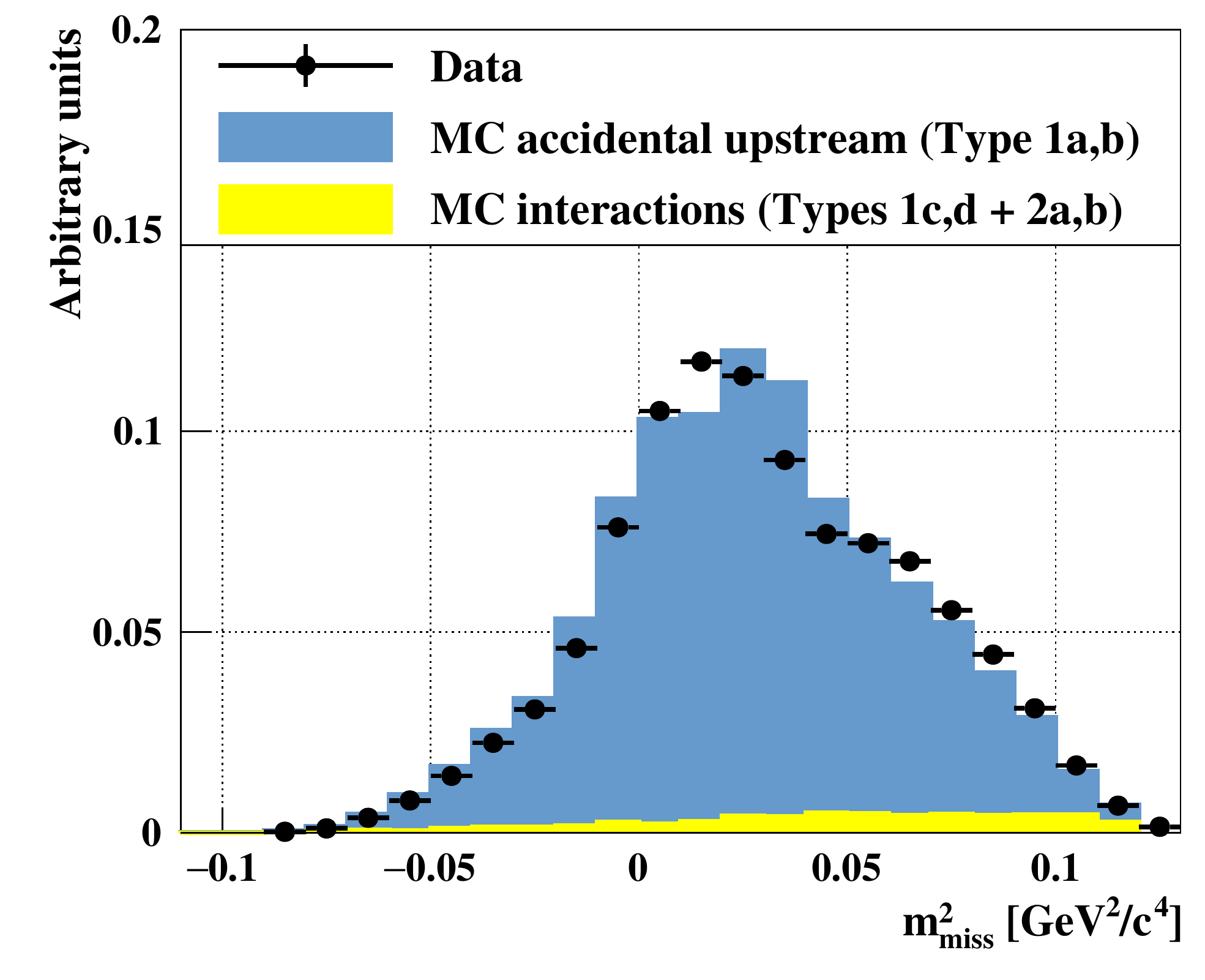}
 \end{minipage}\hfill
\caption{Reconstructed $\mmis$ distributions of PNN data sample and simulated samples obtained from  the upstream event selection described in the text.}
\label{fig:up_mm2_cda}
\end{center}
\end{figure}

\begin{figure}[t]
\begin{minipage}{.49\textwidth}
   \includegraphics[width=1.07\textwidth]{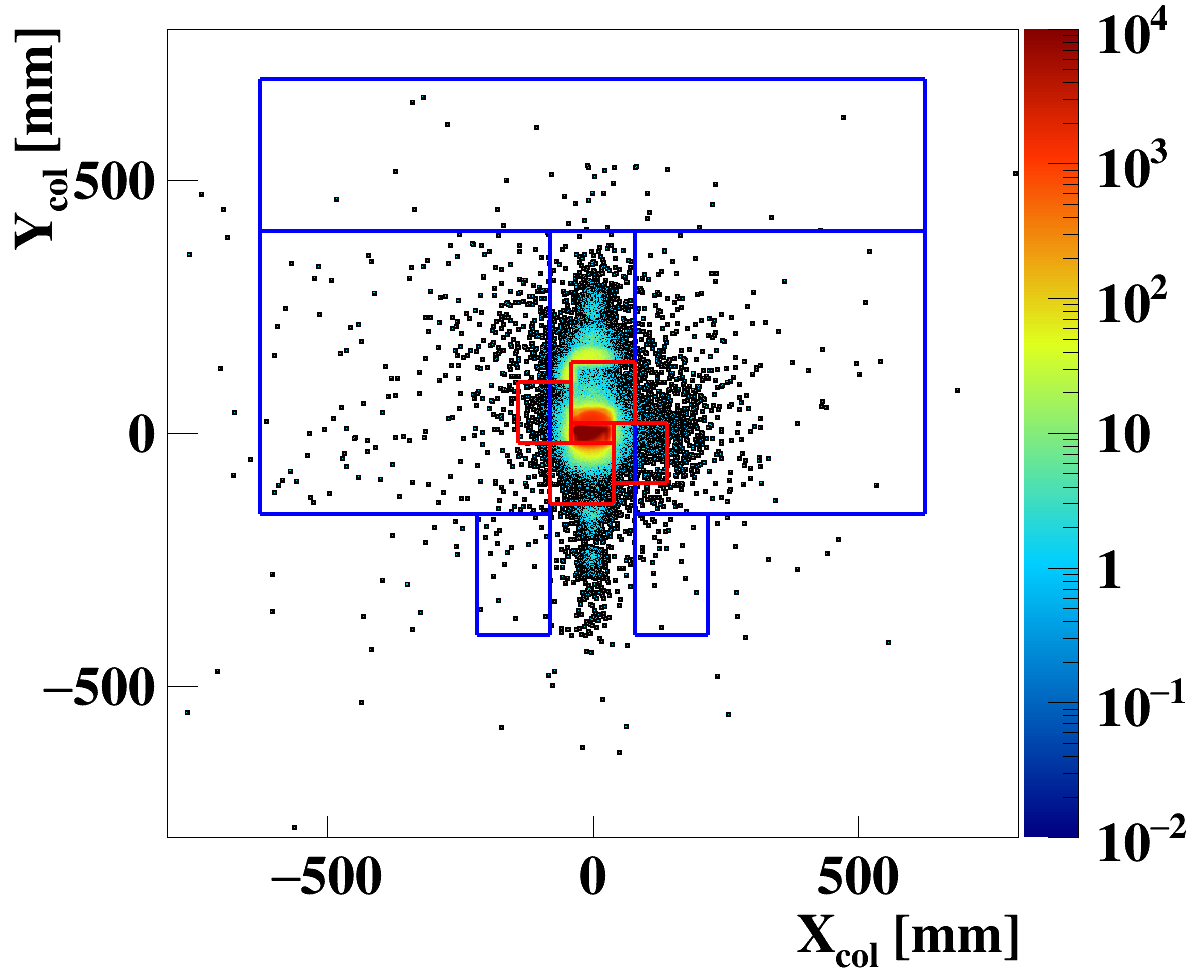} 
    \end{minipage}\hspace{1pc} 
 \begin{minipage}{.49\textwidth}
   \includegraphics[width=0.99\textwidth]{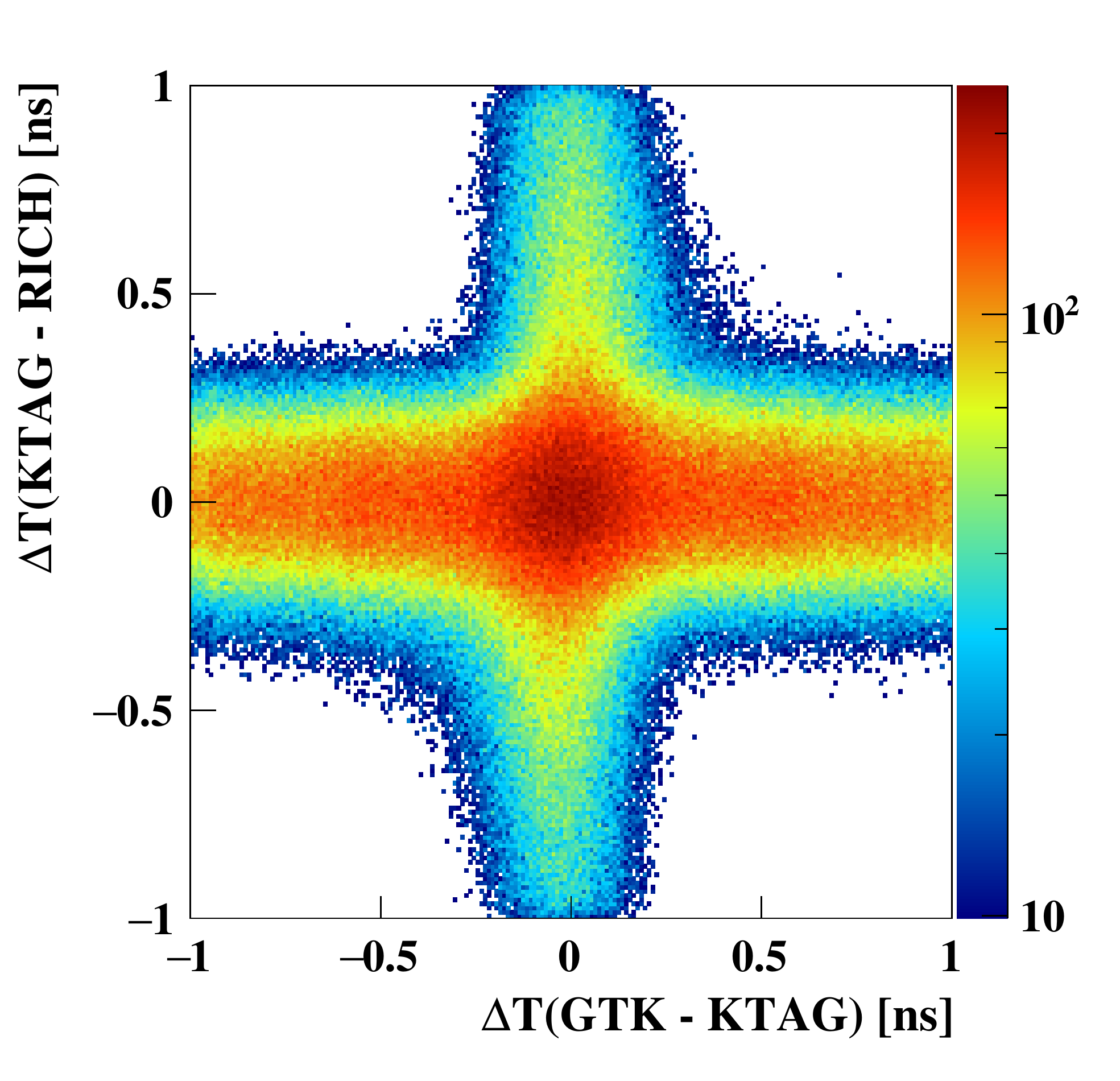}
 \end{minipage}
\caption{{\bf Left}: 
extrapolation of $\pi^+$ tracks in the upstream data sample described in the text to the $(X, Y)$  plane at the $Z$-position of the final collimator.
The blue lines correspond to the last dipole of the second achromat; the contour of the final collimator is shown with a red line. 
{\bf Right}: time difference between KTAG and RICH versus GTK and KTAG for the $\pi^+$s shown on the left plot.}
\label{fig:updt}
\end{figure}

The evidence for the above classification comes from studies based on data and simulated samples. The PNN selection is modified as follows to provide an almost pure sample of upstream data events: the matching conditions for the $K^+$ candidate and the $\pi^+$ track are not applied; no constraints are applied to the reconstructed $Z_{\rm vertex}$; the box cut is not applied; and ${\rm CDA}>4$~mm is required. 
The last condition ensures that the $K/\pi$ matching of the PNN selection is not satisfied, therefore the signal $\mmis$ regions can be explored in the PNN data sample without violating the blind analysis principle. The distribution of $\mmis$ for the selected data and simulated events is shown in Figure~\ref{fig:up_mm2_cda}: simulated upstream events explain the shape of the data. The sample is dominated by $K^+\to\pi^+\pi^+\pi^-$ and $K^+\to\pi^+\pi^0$ decays occurring downstream of the first GTK station (GTK1). 

The $X, Y$ coordinates of the pions selected in the data sample, obtained by extrapolating their tracks to the $ (X, Y)$ plane of the final collimator, 
are shown in Figure~\ref{fig:updt}~(left). In most cases, the pion passes through the beam hole in the final collimator. The shape of the distribution outside of the hole is determined by the material in the beam line: most of the pions outside the hole are contained in the aperture of the last dipole magnet of the beam line. 
Pions from upstream in-time events, originating from GTK3, have an $X, Y$ distribution at the final collimator which overlaps with the $X, Y$ distribution of pions from accidental upstream events. The box cut used in the PNN selection, $|X| < 100$ mm, $|Y| < 500$ mm (section~\ref{sec:sig}) is defined to exclude the whole aperture of the magnet.

The time structure of the selected upstream events is shown in Figure~\ref{fig:updt}~(right).
Accidental coincidence between KTAG and GTK signals is necessary to reconstruct a $K^+$ candidate in events with a $K^+$ decaying or interacting upstream of GTK3. 
On the other hand, the $\pi^+$ in these events produces a RICH signal in time with the KTAG signal of the parent $K^+$. Therefore accidental upstream events of types a) and c) populate the horizontal band in the timing plot. Accidental upstream events of types b) and d) require a pileup $K^+$ in the GTK. 
In this case one of the two KTAG candidates and the GTK track are in time, while the $\pi^+$ signal in the RICH accidentally coincides with the same KTAG candidate.
As a consequence, accidental upstream events of types b) and d) form the vertical band in the timing plot. In-time upstream events populate the central region of the plot.
The distribution of data events in the central region is consistent with that formed by the overlap of the horizontal and vertical bands, and indicates that in-time upstream events account for less than 10\% of the sample, which is in agreement with simulations.

\begin{figure}[t]
\begin{minipage}{1\textwidth}
\centering
 \includegraphics[width=0.7\textwidth]{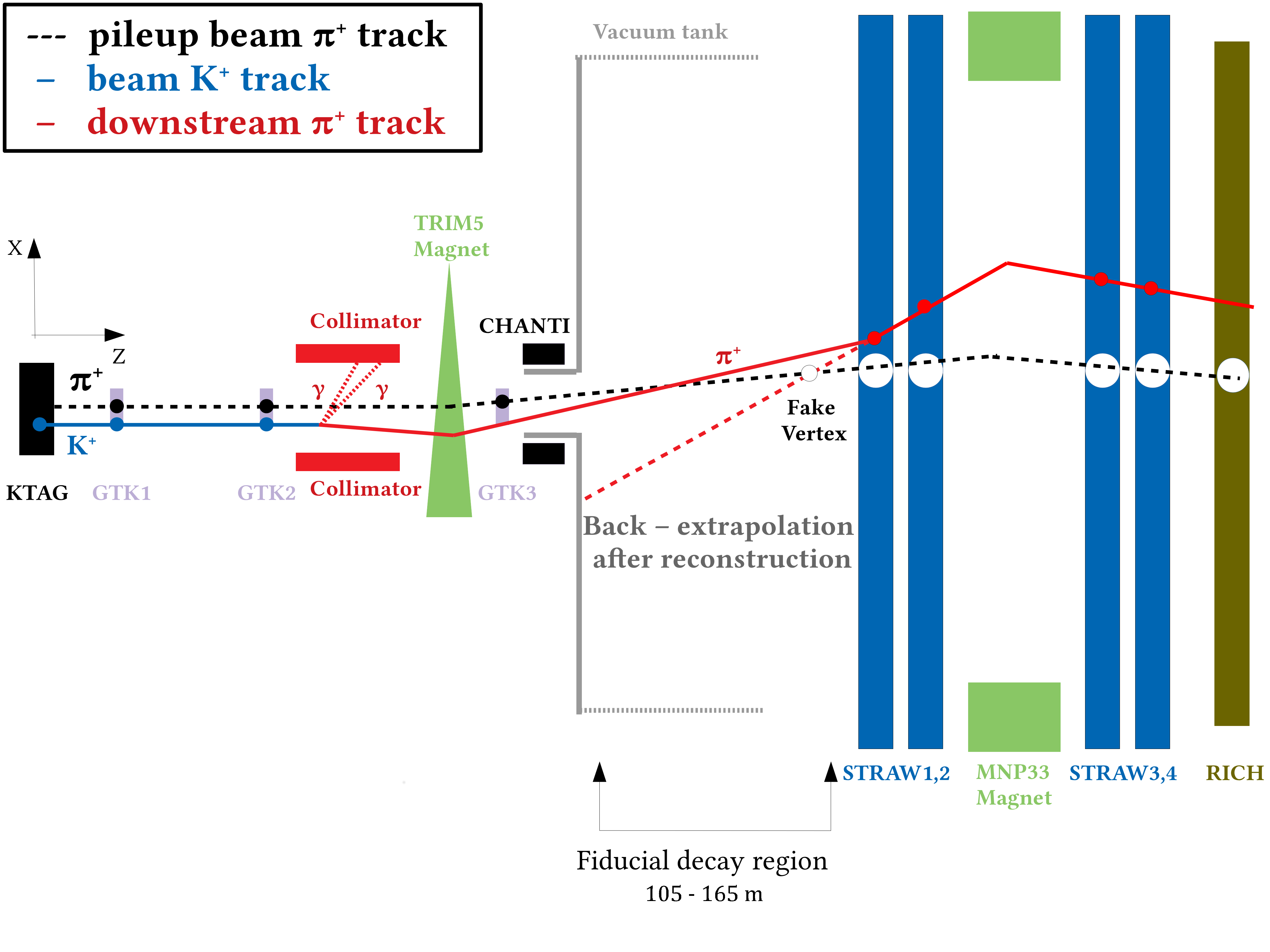}
 \end{minipage}\hfill
\caption{Sketch of an accidental upstream event of type a) in the horizontal plane (not to scale). The GTK stations GTK1, GTK2 and GTK3 are displayed together with the final collimator. The reconstructed $\pi^+$ fails the box cut because of the large-angle scattering at STRAW1.}
\label{fig:sketch}
\end{figure}

The PNN selection criteria mostly effective against accidental upstream background are:
\begin{itemize}
\item the $K/\pi$ association: a coincidence between the two independent particles can only occur accidentally;
\item the $Z_{\rm vertex}$ conditions defining the FV: a $K^+\to\pi^+$ decay vertex
can only be reconstructed in the FV accidentally;
\item the box cut:
         the $\pi^+$ satisfies this condition only if mis-reconstructed or suffering large-angle scattering at STRAW1;
\item the rejection of events with extra hits in at least two GTK stations: the beam particle producing the $\pi^+$ disappears along the beam line in the GTK region; simulations indicate that $\pi^+$ produced upstream of GTK1 cannot reach the FV;
mostly $K^+$ travelling outside the GTK acceptance can pass this condition, as the probability to lose a hit in a GTK station is negligible.  
\end{itemize}
The first three criteria also suppress in-time upstream events along with the CHANTI veto conditions.

An accidental upstream event of type a) contributing to the background is sketched in Figure~\ref{fig:sketch}. The parent kaon decays ($K^+\to\pi^+\pi^0$) downstream of GTK2. The photons from $\pi^0\to\gamma\gamma$ decay are absorbed by the final collimator, while the $\pi^+$ propagates in the magnetic field through the collimator aperture. Finally, the $\pi^+$ direction is modified by large-angle scattering at STRAW1.

\begin{figure}[t]
\begin{minipage}{0.49\textwidth}
  \centering\vspace{-1pc}
   \includegraphics[width=1.06\textwidth]{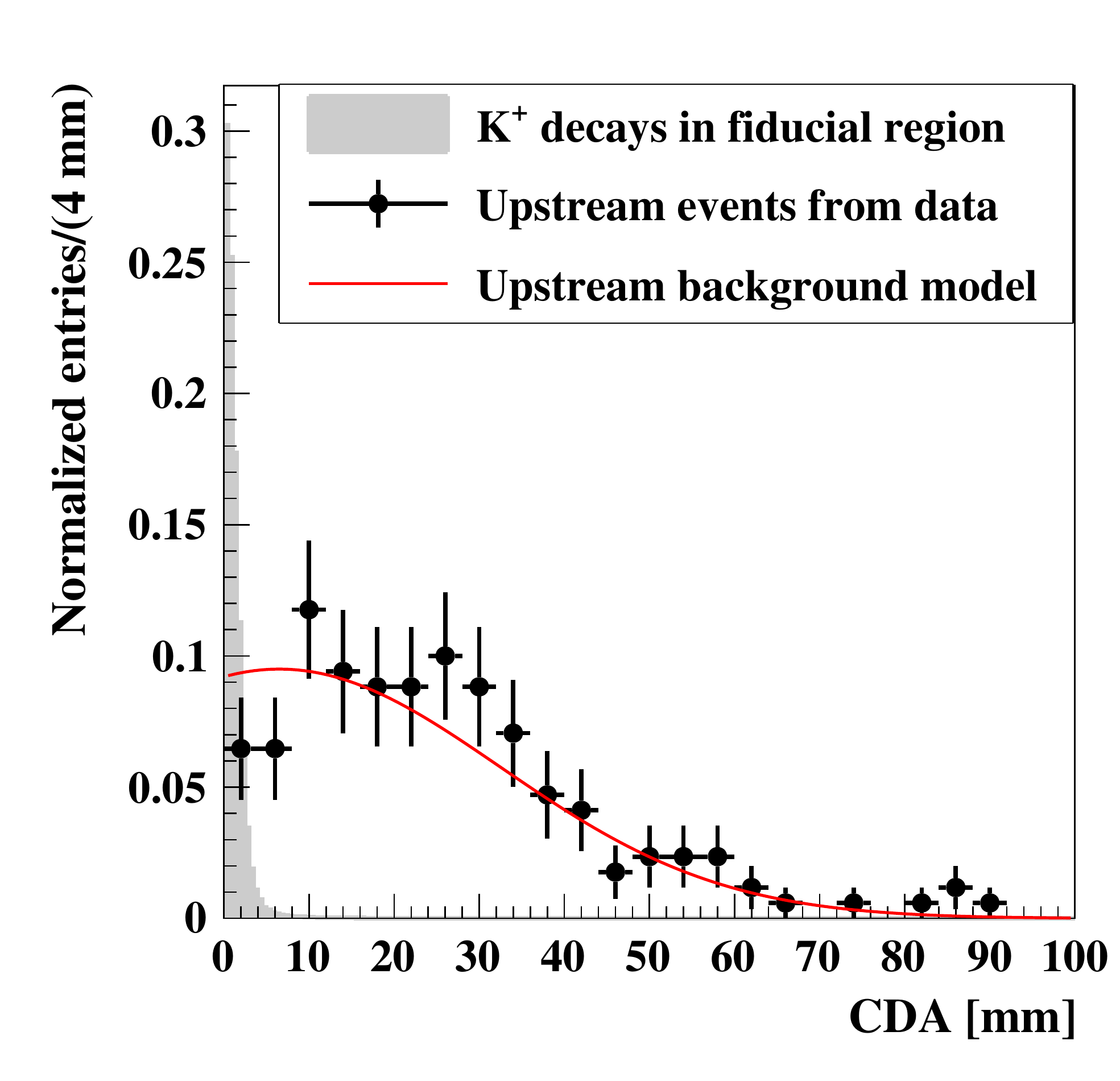}
  \end{minipage}\hspace{1pc}
 \begin{minipage}{.49\textwidth}
  \centering
       \includegraphics[width=1.\textwidth]{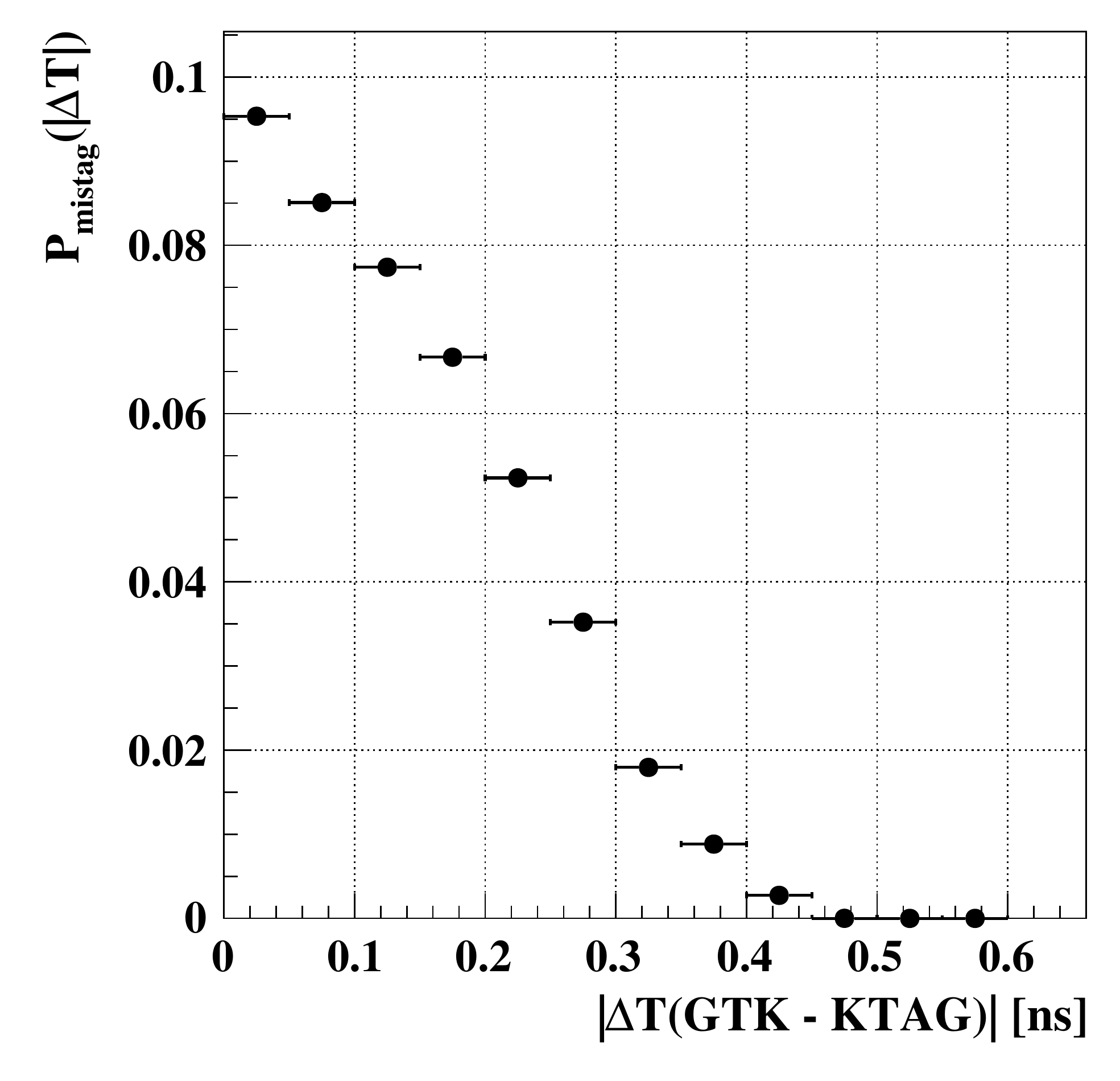}
 \end{minipage}
\caption{{\bf Left}: CDA distribution of upstream events selected as described in the text. The distribution is compared with a model obtained from simulations. The CDA distribution of data events coming from $K^+$ decays in the FV is also shown. {\bf Right}: probability for an upstream event to satisfy the $K/\pi$ matching conditions as a function of $\Delta$T(GTK--KTAG) obtained using the CDA model shown in the left plot.}
\label{fig:pmistag}
\end{figure}

\subsubsection{Upstream background evaluation}
\label{sec:upsval}

The evaluation of the upstream background in the PNN sample does not rely on Monte Carlo simulation, but follows a data-driven approach. A sample of PNN data enriched with upstream events, called the ``upstream sample'' below, is selected using modified PNN criteria: ${\rm CDA}>4$~mm is required instead of the $K/\pi$ matching conditions.
The number of events from the PNN sample passing this selection is $N_{\rm data}=16$. The background from $K^+$ decays in the FV in this sample is estimated to be 0.2 events by analysing background regions of $\mmis$ with the methods described in sections~\ref{sec:pp0} and \ref{sec:kmu2}.

The upstream background is evaluated considering the probability $P_{\rm mistag}$ that an upstream event satisfies the $K/\pi$ matching criteria. This probability depends only on the shape of the CDA distribution and the time difference $\Delta$T(GTK--KTAG) for the events in the horizontal band of Figure~\ref{fig:updt}~(right), and $\Delta$T(KTAG--RICH) for the events in the vertical band. The CDA distribution model is established from simulations of accidental upstream events. This model is validated using a data sample selected similarly to the upstream sample with the following modifications: GTK and CHANTI veto conditions are removed, the condition CDA$>4$~mm is removed, and a timing condition $0.6~{\rm ns}<|$T(KTAG--GTK)$| <3~{\rm ns}$ is used. Data and simulations agree within the statistical uncertainties, as shown in Figure~\ref{fig:pmistag}~(left).
The probability $P_{\rm mistag}$ evaluated with simulations in bins of $\Delta T$
is shown in Figure~\ref{fig:pmistag}~(right). The number of upstream background events is estimated in each of the two bands shown in Figure~\ref{fig:updt}~(right) as
\begin{equation}
\label{eq:ups}
N_{\rm upstream} = f_{\rm scale}\cdot \sum_{i=1}^{12} N_{data}^i P_{\rm mistag}^i   ~~,
\end{equation}
where the sum runs over twelve 100~ps wide $\Delta$T bins covering the $(-0.6, 0.6)$~ns range; 
$N_{\rm data}^i$ is the number of events found in the upstream sample in bin $i$, $P_{\rm mistag}^i$ is the corresponding mis-tagging probability shown in Figure~\ref{fig:pmistag}~(right),
and $f_{\rm scale}=1.06$ accounts for upstream events with ${\rm CDA}\leq4$~mm not included in the $N_{\rm data}$ definition. The last factor is obtained from a study of the T(GTK--KTAG) sidebands, as data and simulations show that the CDA is independent of this quantity.

\begin{figure}[t]
\centering
\includegraphics[width=.5\textwidth]{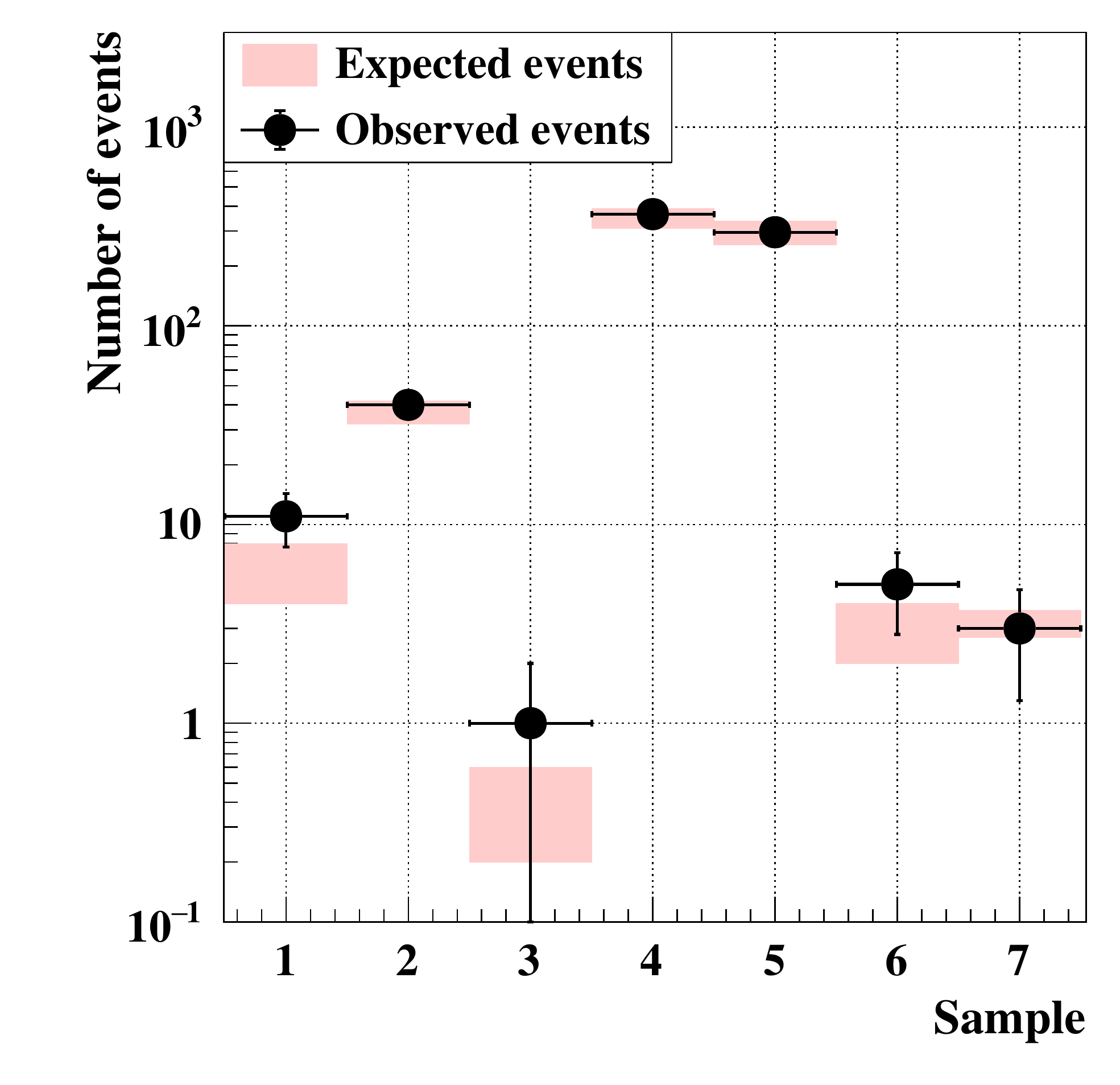}
\caption{Number of expected and observed events in the seven different upstream background validation samples.}
\label{fig:us_val}
\hfill
\end{figure}

The procedure described above is validated using seven different data samples selected modifying the PNN criteria as follows:
\begin{enumerate}
\item $|X_{\rm col}|<100$~mm, $|Y_{\rm col}|<140$~mm for the pion position in the final collimator plane, replacing the box cut;
\item $|X_{\rm col}|<100$~mm, $|Y_{\rm col}|\geq140$~mm, replacing the box cut;
\item $\mmis<-0.05~{\rm GeV}^2/c^4$, replacing the signal region mass definition;
\item as 1), without GTK and CHANTI veto conditions;
\item as 2), without GTK and CHANTI veto conditions;
\item as 3), without GTK and CHANTI veto conditions;
\item GTK and CHANTI veto conditions inverted.
\end{enumerate}
Simulations show that the contributions of the various types of upstream background differ among the samples. The numbers of expected and observed background events in each sample are presented in Figure~\ref{fig:us_val}: they agree within one standard deviation in each sample.
\begin{table}
  \begin{center}
  \caption{Expected numbers of  SM \pnnc decays and of background events in the signal regions.}
   \label{tab:bckgtot}
    \begin{tabular}{l|c}
      \toprule
      {Process} & {Events expected} \\\midrule
       \pnnc (SM) & $2.16\pm0.13_{\rm syst}\pm0.26_{\rm ext}$\\\midrule
      $K^+\rightarrow\pi^+\pi^0(\gamma)$   & $0.29\pm0.03_{\rm stat}\pm0.03_{\rm syst}$ \\
      $K^+\rightarrow\mu^+\nu(\gamma)$    & $0.15\pm0.02_{\rm stat}\pm0.04_{\rm syst}$ \\
      $K^+\rightarrow\pi^+\pi^-e^+\nu$       & $0.12\pm0.05_{\rm stat}\pm0.06_{\rm syst}$ \\
      $K^+\rightarrow\pi^+\pi^+\pi^-$          & $0.008\pm0.008_{\rm syst}$\\
      $K^+\rightarrow\pi^+\gamma\gamma$  & $0.005\pm0.005_{\rm syst}$ \\
      $K^+\rightarrow\pi^0\ell^+\nu~(\ell=\mu,e)$ & $<0.001$ \\
      Upstream background & $0.89\pm0.24_{\rm stat}\pm0.20_{\rm syst}$\\\midrule
      Total background & $1.46\pm0.25_{\rm stat}\pm0.21_{\rm syst}$\\
      \bottomrule
    \end{tabular}
  \end{center}
\end{table}

The number of expected upstream background events is found to be
\begin{equation}
\label{eq:3}
N_{\rm upstream} = 0.89 \pm 0.24_{\rm stat} \pm 0.20_{\rm syst}.
\end{equation}
The statistical uncertainty stems from $N_{\rm data}$. A systematic uncertainty of 12\% is due to the modelling of the CDA distribution, and is derived from the comparison between data and simulations. An additional systematic uncertainty of 20\% is assigned as half of the difference between the expected and observed number of events in sample 6  (with statistics similar to the expected signal). This uncertainty accounts for the accuracy of the assumption 
that all the categories of upstream events have the same CDA distribution.

\subsection{Summary}
The expected backgrounds in signal region are summarized in Table~\ref{tab:bckgtot}. 

As an additional check, the expected and observed numbers of events in the PNN sample are compared in a control region defined by the same $\mmis$ range as the signal regions 1 and 2 but  in the 35--40~GeV$/c$ $\pi^+$ momentum range.
The expected number of background events here is between 0.4 and 0.8 at 90\% CL,
almost equally shared between \kp decays in  the 
FV and upstream events. The corresponding expected number of SM \pnnc events is $0.13\pm0.02$.
One event is observed with a $\pi^+$ momentum of 38 GeV/c and $\mmis\simeq0.03$\,GeV$^2/c^4$, in agreement with the expectation. 


\section{Results}\label{sec:result}
\begin{figure}
  \begin{center}
  \begin{minipage}{28pc}
  \includegraphics[width=28pc]{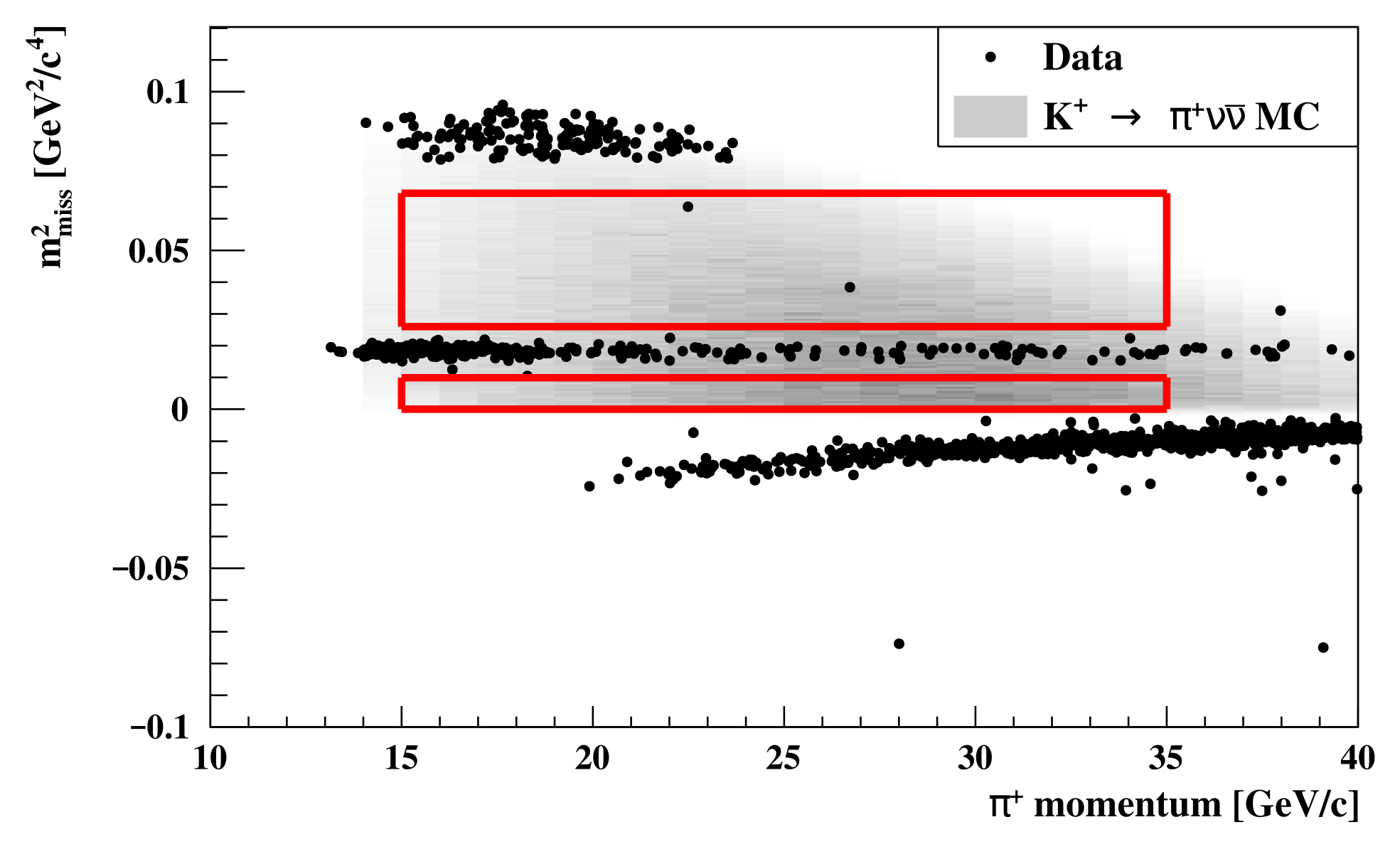}
  \end{minipage}\hspace{1pc}
  \caption{\label{fig:final}
  Reconstructed $\mmis$ as a function of \pic momentum for PNN events (full symbols) satisfying the PNN selection, except the  $\mmis$ and \pic momentum criteria.
   The grey area corresponds to the expected distribution of SM \pnnc MC events (arbitrarily normalized). Red contours define the signal regions. 
   The events observed  
   in the signal regions are shown together with the events found in the background and control regions.}
  \end{center}
\end{figure}

After unmasking the signal regions, two candidate events are found, 
as shown in Figure~\ref{fig:final}. 
\begin{figure}[htb]
  \begin{center}
  \begin{minipage}{20pc}
\includegraphics[width=20pc]{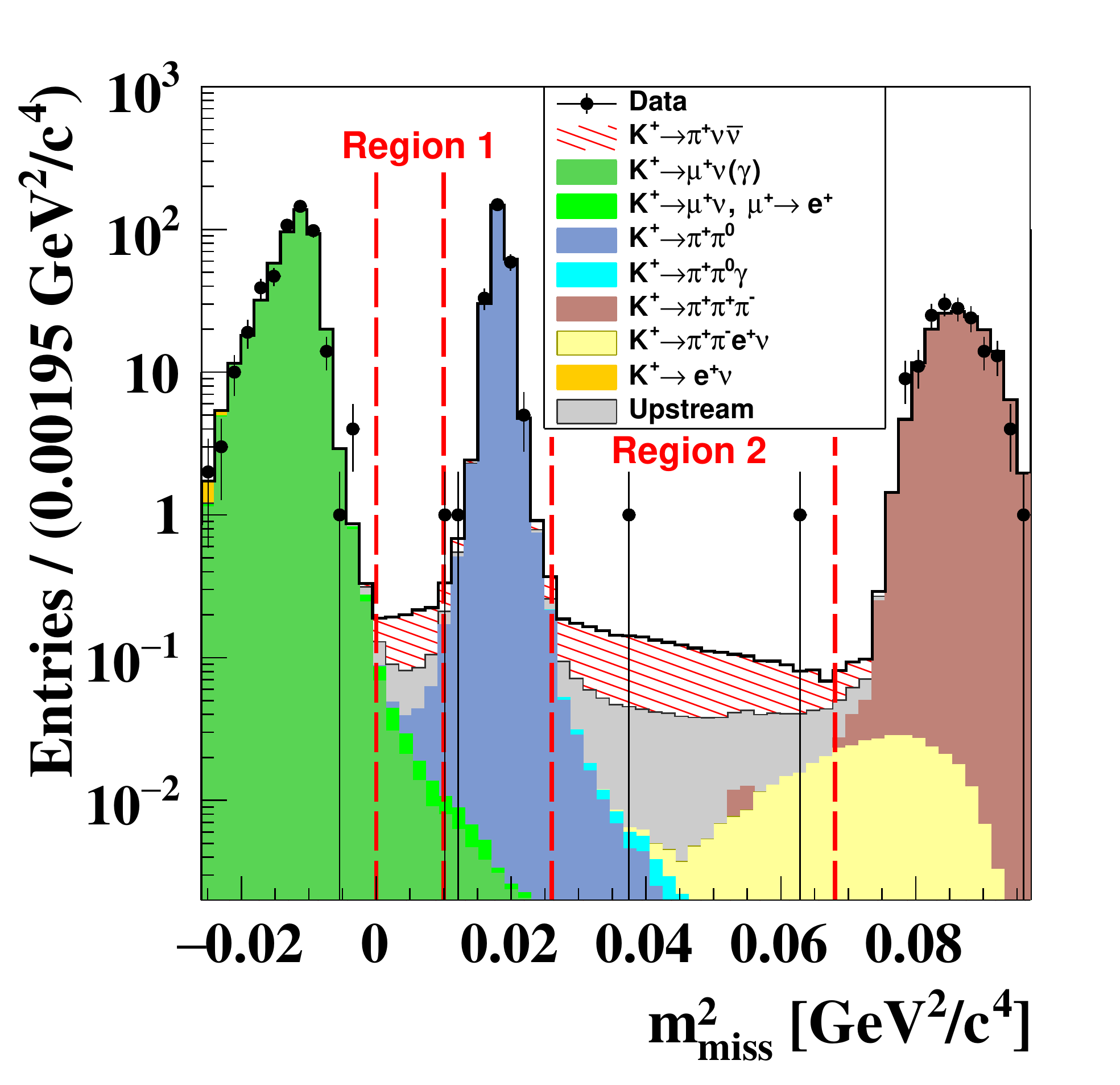}
\end{minipage}\hspace{1pc}
  \caption{\label{fig:final2}Reconstructed $\mmis$ distribution of data events with $\pi^+$ momentum between 15 and 35~GeV/$c$, passing the PNN selection (full symbols). The expected background and SM signal events contributions are superimposed as stacked histograms.
  The $\mmis$  distributions of the $K^+\rightarrow\pi^+\pi^0$,  $K^+\rightarrow\mu^+\nu$ and $K^+\rightarrow\pi^+\pi^+\pi^-$ decays and of the upstream events are extracted from data.  The other contributions are obtained from simulations.}
  \end{center}
\end{figure}
The second and third columns of Table~\ref{tab:events} summarize the characteristics of these events.
\begin{table}[h]
  \begin{center}
    \caption{Observed events in the signal regions after PNN selection.
  Events 1 and 2 come from the analysis of the 2017 data presented here.
  Event 3 comes from the analysis of the 2016 data.}
  \label{tab:events}
    \begin{tabular}{l|c|c|c}
      \toprule
       & {Event 1} & {Event 2} & {Event 3 }\\\midrule
       Year                                  &   2017                                    & 2017                                          &  2016 \\\midrule
       $\mmis$                           &   $0.038\,\text{GeV}^2/c^4$ &  $0.064\,\text{GeV}^2/c^4$      & $0.031\,\text{GeV}^2/c^4$\\
       $\pi^+$ momentum          &   $26.5\,\text{GeV}/c$           &  $22.4\,\text{GeV}/c$ & $15.4\,\text{GeV}/c$\\
       $Z_{vertex}$                    &   $140$\,m                             &  $159$\,m & $146$\,m \\
       $\Delta$T(KTAG - GTK)      &   $-0.171$\,ns                         &  $0.028$\,ns & $0.006$\,ns\\
        $\Delta$T(RICH - KTAG)     &    $-0.082$\,ns                         & $0.209$\,ns & $0.040$\,ns\\
       $(X,Y)$ at final collimator  &    $(228.4,104.1)$\,mm   &  $(189.4,-271.7)$\,mm & $(-372.6,29.8)$\,mm \\ 
     \bottomrule
    \end{tabular}
  \end{center}
\end{table}

Figure~\ref{fig:final2}
shows the $\mmis$ distribution of the events with momentum between 15 and 35~GeV/$c$ passing the PNN selection, compared with that expected from SM \pnnc decays and from the various sources of background.
In this plot the $\mmis$ distribution of the $K^+\rightarrow\pi^+\pi^0$,  $K^+\rightarrow\mu^+\nu$ and $K^+\rightarrow\pi^+\pi^+\pi^-$ decays come from the minimum-bias samples, and normalized to the 
number of events in the corresponding background regions (sections~\ref{sec:pp0}, \ref{sec:kmu2} and \ref{sec:ppp}).
The distribution of the $\mmis$ of the upstream background is extracted from an upstream-event-enriched data sample and is normalized to the number of upstream background events expected in the signal regions.   
The distributions of the other background sources are modelled using MC simulations and normalized to the expected number of events in the signal regions.

The two candidate \pnnc events of this analysis complement the one found by NA62 in the same signal region from the analysis of the 2016 data~\cite{na62pnn1}.
The characteristics of the 2016 candidate are displayed in the fourth column of Table~\ref{tab:events}.
Table~\ref{tab:results} summarizes the numerical results obtained in the \pnnc analysis of the 2017 and 2016 independent data samples.
\begin{table}[h]
  \begin{center}
    \caption{Summary from the \pnnc analyses of the data recorded in  2017 and 2016.}
  \label{tab:results}
    \begin{tabular}{l|c|c}
      \toprule
       & 2017 & 2016 \\\midrule
 Single Event Sensitivity  $SES$    &  $(0.389\pm0.024)\times10^{-10}$  & $(3.15\pm0.24)\times10^{-10}$ \\
   Expected SM \pnnc decays    &   $2.16\pm0.13\pm0.26_{ext}$      & $0.267\pm0.20\pm0.32_{ext}$ \\
   Expected background   $B \pm \delta_B$  &   $1.46\pm 0.30$                               &  $0.15\pm0.093$ \\\midrule
       Observed events                   &   $2$                                                & $1$ \\
     \bottomrule
    \end{tabular}
  \end{center}
\end{table}

The statistical interpretation of the result is obtained from an event counting approach in the full range of the signal region.
The level of the expected background does not allow  
a claim of signal observation nor a claim of inconsistency with the presence of SM \pnnc decays.
Therefore both an upper limit and a measurement of the branching ratio of the \pnnc decay 
are presented.

A fully frequentist hypothesis test, with a profile likelihood ratio as test statistic, is used to combine the results of the 2017 and 2016 analyses.
The parameter of interest is the signal strength $\mu$ defined as the branching ratio in units of the Standard Model one. 
The nuisance parameters are the total expected number of background events in the signal regions  ($B$) and the single event sensitivity ($SES$), obtained separately from the 2016 and 2017 datasets.
Following the method described in~\cite{cranmer} and according to~\cite{onoff}, the number of background events is constrained to follow a Poisson distribution with mean value $(B/\delta_B)^2$ where
$\delta_B$ is the uncertainty of $B$ (Table~\ref{tab:results}).  The mean $(B/\delta_B)^2$ accounts for an equivalent number of events counted in control regions through the auxiliary measurements 
leading to $B$ as described in section~\ref{sec:pnnbckg}.
A log-normal distribution function  is used to constrain the $SES$ around the measured value.

\begin{figure}[h]
  \begin{center}
  \begin{minipage}{30pc}
  \includegraphics[width=30pc]{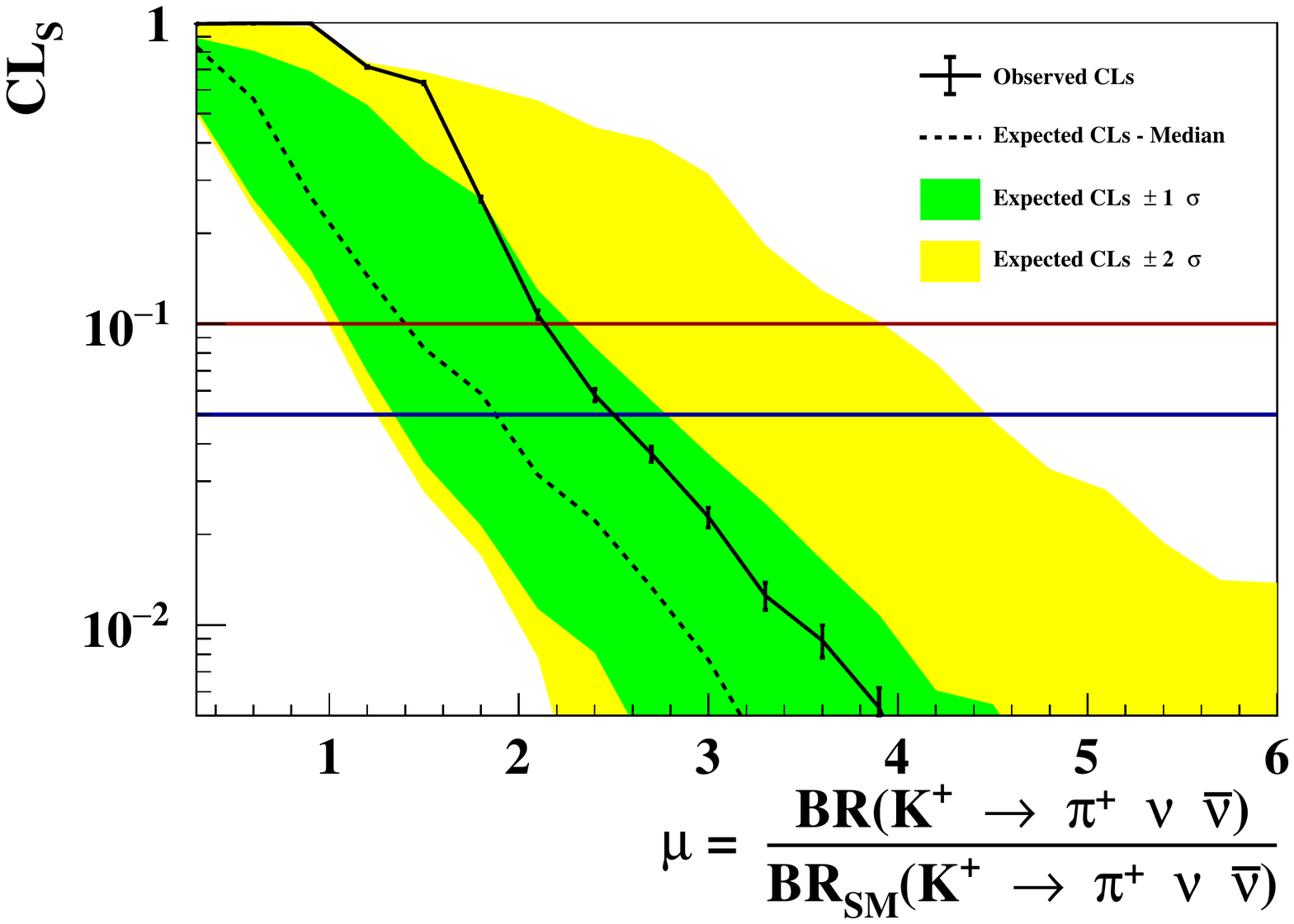}
  \end{minipage}
  \caption{\label{fig:ul} CL$_\text{S}$ $p$-values as a function of the branching ratio of the \pnnc decay expressed in units of the Standard Model value.
  The red (blue) line corresponds to the 90\% (95\%) CL.}
  \end{center}
\end{figure}

The likelihood functions of the results of the 2016 and 2017 analyses are multiplied to form a single combined function, which is profiled with respect to the nuisance parameters.  
The upper limit on the branching ratio of the \pnnc decay is obtained using a CL$_{\text S}$ method~\cite{cls} for several values of the signal strength $\mu$ (Figure~\ref{fig:ul}).
The 90\% CL expected upper limit is BR$(K^+\rightarrow\pi^+\nu\bar{\nu}) <1.24\times10^{-10}$ and the observed one is:
\begin{equation}
\text{BR}(K^+\rightarrow\pi^+\nu\bar{\nu})<1.78 \times10^{-10}.    
\end{equation}
This result translates to a  Grossman-Nir limit~\cite{gnb} of the SM \pnnz 
branching ratio equal to $7.8\times10^{-10}$.

The data also allow  the setting of a 68\% CL interval on the SM branching ratio of the \pnnc decay.
Using the prescriptions of~\cite{FC} and~\cite{RL}, the measured \pnnc branching ratio is:
\begin{equation}
\text{BR}(K^+\rightarrow\pi^+\nu\bar{\nu})=(0.48^{+0.72}_{-0.48})\times10^{-10}.
\end{equation}

\section{Conclusions}\label{sec:end}
An investigation of \pnnc has been performed using the data collected by the NA62 experiment at CERN in 2017.
The experiment has reached the best single event sensitivity so far in this decay mode, corresponding to
$(0.389\pm0.024_{syst})\times10^{-10}$. 
This translates into  an expectation of $(2.16\pm0.13_{syst}\pm0.26_{ext})$  
\pnnc events 
in the signal regions, assuming the Standard Model  BR of ($8.4\pm 1.0) \times 10^{-11}$. 
A further  1.5 background events are expected in the same signal regions, 
mainly due to a single $\pi^+$ produced along  the beam line  upstream of the \kp decay volume and 
accidentally matched to a beam kaon. 
Using a blind analysis procedure, two candidate events 
have been observed in the signal regions, consistent with expectation. 
These two candidates,  together with the single candidate observed from the analysis of the 2016 data, lead to the most stringent upper limit on the  branching ratio
BR$(K^+\rightarrow\pi^+\nu\bar{\nu})<1.78\times10^{-10}$ at 90\%\,CL and set the Grossman-Nir limit 
on BR$(K_L\rightarrow\pi^0\nu\bar{\nu})$ to $7.8\times10^{-10}$. 
The corresponding 68\%\,CL measurement of the \pnnc branching ratio is $(0.48^{+0.72}_{-0.48})\times10^{-10}$.
This result constrains some New Physics models that can predict large enhancements previously allowed by the measurements published by the E787 and E949 BNL experiments~\cite{pnnlh,pnnrs,pnnzz,pnnzp,pnnlu,pnnlq}.
The NA62 experiment has collected and is now analysing almost twice as much data in 2018 as that reported upon here, and further optimization of the analysis strategy is expected significantly to reduce the uncertainty in the measured BR of the \pnnc decay. 
\clearpage

\section*{Acknowledgements}
It is a pleasure to express our appreciation to the staff of the CERN laboratory and the technical
staff of the participating laboratories and universities for their efforts in the operation of the
experiment and data processing.

The cost of the experiment and its auxiliary systems was supported by the funding agencies of 
the Collaboration Institutes. We are particularly indebted to: 
F.R.S.-FNRS (Fonds de la Recherche Scientifique - FNRS), Belgium;
BMES (Ministry of Education, Youth and Science), Bulgaria;
NSERC (Natural Sciences and Engineering Research Council), funding SAPPJ-2018-0017 Canada;
NRC (National Research Council) contribution to TRIUMF, Canada;
MEYS (Ministry of Education, Youth and Sports),  Czech Republic;
BMBF (Bundesministerium f\"{u}r Bildung und Forschung) contracts 05H12UM5, 05H15UMCNA and 05H18UMCNA, Germany;
INFN  (Istituto Nazionale di Fisica Nucleare),  Italy;
MIUR (Ministero dell'Istruzione, dell'Univer\-sit\`a e della Ricerca),  Italy;
CONACyT  (Consejo Nacional de Ciencia y Tecnolog\'{i}a),  Mexico;
IFA (Institute of Atomic Physics) Romanian CERN-RO No.1/16.03.2016 and Nucleus Programme PN 19 06 01 04,  Romania;
INR-RAS (Institute for Nuclear Research of the Russian Academy of Sciences), Moscow, Russia; 
JINR (Joint Institute for Nuclear Research), Dubna, Russia; 
NRC (National Research Center)  ``Kurchatov Institute'' and MESRF (Ministry of Education and Science of the Russian Federation), Russia; 
MESRS  (Ministry of Education, Science, Research and Sport), Slovakia; 
CERN (European Organization for Nuclear Research), Switzerland; 
STFC (Science and Technology Facilities Council), United Kingdom;
NSF (National Science Foundation) Award Numbers 1506088 and 1806430,  U.S.A.;
ERC (European Research Council)  ``UniversaLepto'' advanced grant 268062, ``KaonLepton'' starting grant 336581, Europe.

Individuals have received support from:
Charles University Research Center (UNCE/SCI/ 013), Czech Republic;
Ministry of Education, Universities and Research (MIUR  ``Futuro in ricerca 2012''  grant RBFR12JF2Z, Project GAP), Italy;
Russian Foundation for Basic Research  (RFBR grants 18-32-00072, 18-32-00245), Russia; 
Russian Science Foundation (RSF 19-72-10096), Russia;
the Royal Society  (grants UF100308, UF0758946), United Kingdom;
STFC (Rutherford fellowships ST/J00412X/1, ST/M005798/1), United Kingdom;
ERC (grants 268062,  336581 and  starting grant 802836 ``AxScale'');
EU Horizon 2020 (Marie Sk\l{}odowska-Curie grants 701386, 842407, 893101).

\end{document}